\documentclass[twoside]{article}
\usepackage{qic,epsfig}
\usepackage{mathrsfs}
\usepackage{eucal}
\usepackage{amssymb}
\usepackage{cite}
\usepackage{dsfont}
\usepackage{threeparttable}
\usepackage[table]{xcolor}
\usepackage[all]{xy}
\usepackage{multirow}
\newcommand{\be}{\begin{equation}}
\newcommand{\ee}{\end{equation}}
\newcommand{\brr}{\begin{eqnarray}}
\newcommand{\err}{\end{eqnarray}}
\newcommand{\nn}{\nonumber}
\newcommand{\bd}{\begin{displaymath}}
\newcommand{\ed}{\end{displaymath}}

\newcommand{\bib}{\bibitem}
\newcommand{\col}{\mathrel{\mathop:}=}
\def\alf{\alpha}
\def\bet{\beta}
\def\gam{\gamma}

\def\lam{\lambda}
\def\om{\omega}
\def\eps{\epsilon}

\def\rpar{\right)}
\def\lpar{\left(}
\def\rbk{\right]}
\def\lbk{\left[}
\def\rbr{\right\}}
\def\lbr{\left\{}
\def\lb{\label}
\def\im{{\rm i}}

\def\tr{\mbox{${\rm Tr}$}}
\def\ro{\mbox{$\hat{\rho}$}}
\def\vro{\mbox{$\hat{\varrho}$}}

\def\ima{\mbox{${\mathtt{Im}}$}}
\def\re{\mbox{${\mathtt{Re}}$}}
\def\rg{\rangle}
\def\lg{\langle}
\def\half{\frac{1}{2}}
\def\lb{\label}
\def\ftn{\footnote}
\begin{document}
\setlength{\textheight}{8.0truein}    
\runninghead{Representations of two-qubit and ququart states via discrete Wigner functions}
            {Marchiolli and Galetti}
\normalsize\textlineskip
\thispagestyle{empty}
\setcounter{page}{1}
\copyrightheading{0}{0}{2019}{000--000}
\vspace*{0.88truein}
\alphfootnote
\fpage{1}
\centerline{\bf REPRESENTATIONS OF TWO-QUBIT AND QUQUART STATES}
\vspace*{0.035truein}
\centerline{\bf VIA DISCRETE WIGNER FUNCTIONS}
\vspace*{0.37truein}
\centerline{\footnotesize MARCELO A. MARCHIOLLI}
\vspace*{0.015truein}
\centerline{\footnotesize\it Avenida General Os\'{o}rio 414, centro, 14.870-100 Jaboticabal, S\~{a}o Paulo, Brazil}
\vspace*{0.015truein}
\centerline{\footnotesize{\it E-mail}: \texttt{marcelo$\_$march@bol.com.br}}
\vspace*{10pt}
\centerline{\footnotesize DI\'{O}GENES GALETTI}
\vspace*{0.015truein}
\centerline{\footnotesize\it Instituto de F\'{\i}sica Te\'{o}rica, Universidade Estadual Paulista, 
Rua Dr. Bento Teobaldo Ferraz 271}
\baselineskip=10pt
\centerline{\footnotesize\it Bloco II, Barra Funda, 01140-070 S\~{a}o Paulo, S\~{a}o Paulo, Brazil}
\vspace*{0.015truein}
\centerline{\footnotesize{\it E-mail}: \texttt{diogaletti@hotmail.com}}
\vspace*{0.225truein}
\publisher{(received date)}{(revised date)}
\vspace*{0.21truein}
\abstracts{By means of a well-grounded mapping scheme linking Schwinger unitary operators and generators of the special unitary 
group $\mathrm{SU(N)}$, it is possible to establish a self-consistent theoretical framework for finite-dimensional discrete phase
spaces which has the discrete $\mathrm{SU(N)}$ Wigner function as a legitimate by-product. In this paper, we apply these results
with the aim of putting forth a detailed study on the discrete $\mathrm{SU(2)} \otimes \mathrm{SU(2)}$ and $\mathrm{SU(4)}$ Wigner
functions, in straight connection with experiments involving, among other things, the tomographic reconstruction of density
matrices related to the two-qubit and ququart states. Next, we establish a formal correspondence between both the descriptions
that allows us to visualize the quantum correlation effects of these states in finite-dimensional discrete phase spaces. Moreover,
we perform a theoretical investigation on the two-qubit X-states, which combines discrete Wigner functions and their respective 
marginal distributions in order to obtain a new function responsible for describing qualitatively the quantum correlation effects.
To conclude, we also discuss possible extensions to the discrete Husimi and Glauber-Sudarshan distribution functions, as well as 
future applications on spin chains.}{}{}
\vspace*{10pt}
\keywords{Finite-dimensional Discrete Phase Spaces, Discrete Wigner Function, \\
\hspace*{1.4cm} Two-qubit States, Ququart States, Two-qubit X-states, Entanglement}
\vspace*{3pt}
\communicate{to be filled by the Editorial}
\vspace*{1pt}\textlineskip    
\section{Introduction}
\lb{s1}        

Despite the heated debates occurred in 1935 on the ``spooky" feature of quantum theory \cite{EPR,S1935}, nowadays it is 
well-accepted that entanglement essentially corresponds to a holistic property of multipartite quantum systems whose genuinely
quantum correlations represent not only an important physical resource for many quantum processes (such as, for instance, quantum
cryptography, quantum teleportation, and dense coding), but also a fundamental feature for our understanding on how Nature behaves
at the microscopic and/or mesoscopic levels. Due its nontrivial structure, quantum entanglement suffers of certain inexorable
effects (with deleterious consequences) and intrinsic limitations: it is highly sensitive to environment and does not increase on
average when systems are distributed in spatially separated regions. Despite these inherent difficulties, both the theoretical 
advances \cite{RPMK2009,Acin2013,Aol2015,Chi2018,Li-Qiao,Witten2018} and the outstanding experimental achievements\cite{Nagali,
Resh,Gedik,Kues,Guo,NC2019,Tison} in the last decades reveal a very promising future for new quantum technologies based on the 
entanglement effects.

One of the fundamental concerns of quantum information theory, in particular the chapter associated with quantum entanglement, 
refers to characterize, control and quantify entanglement \cite{RPMK2009,MC,Vedral,Plenio2007,Beng,Barnett,EMM,Marinescu,Wilde}.
In particular, let us focus on bipartite systems constituted by subsystems of low dimensionality in Hilbert space, where some
well-known entanglement measures (namely, concurrence, negativity, and relative entropy of entanglement) are widely used to
characterize the underlying quantum states \cite{Mend-1,MMH}. Moreover, let us also adopt the theoretical framework established
in \cite{MG2019} for finite-dimensional discrete phase spaces, which permits us to construct discrete Wigner functions related
to the special unitary group $\mathrm{SU(N)}$. So, a pertinent question then naturally emerges from these considerations: ``Can 
the discrete Wigner function be used as an effective mathematical tool in the analysis of quantum correlation effects in bipartite
systems or even in single systems described by a finite number of levels?"

In this paper we introduce both the discrete $\mathrm{SU(2)} \otimes \mathrm{SU(2)}$ and $\mathrm{SU(4)}$ Wigner functions 
associated with spin representations of two two-level systems and a single four-level system, respectively, which lead us to
describe general two-qubit and ququart states. The theoretical background used to obtain these discrete distribution functions
is based on a well-grounded mapping scheme between Schwinger unitary operators and generators of $\mathrm{SU(N)}$, which provides
a sound pathway to the formulation of a genuinely discrete Wigner function for arbitrary quantum systems described essentially by
finite-dimensional state vector spaces. This particular mapping scheme is performed by means of the $\mathit{mod(N)}$-invariant
operator basis introduced in \cite{GP1992} and whose mathematical properties were extensively explored in \cite{GM1996}, which
allowed us to formulate a finite-dimensional phase space labelled by genuinely discrete variables \cite{AG1990}.

Initially, we adopt the well-established Fano's prescription for two-qubit states as pairs of two-level systems \cite{Fano1957,
Fano1983} to obtain the corresponding discrete Wigner function that obeys the criterion `easy-to-handle'. As a first immediate
by-product the discrete Wigner functions associated with the reduced density operators are promptly determined. For the sake of
simplicity, we briefly introduce the two-qubit density matrix in the computational-basis representation and, as a consequence, we
derive its discrete Wigner function expressed in terms of the matrix elements. Furthermore, we apply these results to the Bell
states \cite{MC,Vedral,Plenio2007,Beng,Barnett,EMM,Marinescu} and Werner states \cite{Werner} in order to establish the first
compact expressions for the respective discrete Wigner functions. However, these functions suffer a minor problem: they cannot be
visualized in a unique phase space due to the initial decomposition of the mapping kernel in two kernels, each one responsible for
a particular subsystem. To conclude, we discuss possible experimental scenarios where the discrete Wigner functions can be used 
as an effective theoretical tool to monitorate the entanglement between both the qubits.

Following, we focus on the generators of $\mathrm{SU(4)}$ and their connections with the Schwinger unitary operators, where it 
is possible to show that all the fifteen generators can be expressed as specific combinations of these unitary operators in a
one-to-one correspondence. Hence, the discrete $\mathrm{SU(4)}$ Wigner function obtained from this mathematical procedure is quite
general since it is written as a function of the density-matrix elements associated with an arbitrary four-level system. As an
interesting application, we investigate a recent experiment involving Nuclear Magnetic Ressonance (NMR) techniques used to 
implement an oracle based quantum algorithm that solves a black-box problem faster than any classical counterpart by means of a
single ququart \cite{Gedik}. Here, we show that the discrete Wigner functions related to each stage of the experiment are not mere
figurative mathematical tools, but rather valuable and important theoretical instruments which allow to expand our knowledge on
the physical processes involved. Next, we adopt the theoretical prescription described in Refs. \cite{Resh,Manko} to establish an
isomorphic correspondence between ququart and two-qubit states, which allows us to rewrite the Fano's decomposition for two-qubit
density matrix in terms of the $\mathrm{SU(4)}$ generators by means of an adequate change of basis. This procedure leads us to
solve the previous problem associated with the visualization of two-qubit discrete Wigner functions in finite-dimensional discrete
phase spaces. To corroborate these results, we revisit once again the two-qubit Bell and Werner states with the aim of searching
for a noticeable signature of the entanglement effect through their respective discrete $\mathrm{SU(4)}$ Wigner functions.
Besides, the two-qubit X-states \cite{Mend-1} are also investigated in this manuscript, where the corresponding discrete 
$\mathrm{SU(2)} \otimes \mathrm{SU(2)}$ and $\mathrm{SU(4)}$ Wigner functions reveal new quite interesting results: it is possible 
to define a functional onto this particular finite phase space, which is responsible for recognizing the quantum correlations 
present in these states. Some relevant points associated with possible extensions to the discrete Husimi and Glauber-Sudarshan 
quasiprobability distribution functions, as well as potential applications on spin chains and definitions of \textsl{fidelity},
were also included in our discussions. 

This paper is structured as follows. In Section \ref{s2}, we fix the quantum-algebraic framework that paves the way to establish
an important set of solid mathematical results associated with the discrete $\mathrm{SU(N)}$ Wigner functions. In Section 
\ref{s3}, we present two different but complementary group-theoretical approaches which lead us to describe both the two-qubit and
ququart states through the discrete $\mathrm{SU(2)} \otimes \mathrm{SU(2)}$ and $\mathrm{SU(4)}$ Wigner functions. In addition,
we also apply these results to study the entanglement effect in the two-qubit Bell and Werner states, as well as to analyse all
the experimental stages of a particular NMR experiment involving ququarts via discrete Wigner functions. Next, Section \ref{s4} is
dedicated to the study of two-qubit X-states, whose results allow to define a functional on the corresponding discrete phase space
responsible for recognizing the quantum correlations exhibited by the X-states under scrutiny. Finally, in Section \ref{s5} we
summarize our main results and discuss some possible avenues for future research. To conclude, Appendix A contains technical
details on the $\mathrm{SU(4)}$ generators, their relations with the Schwinger unitary operators, and the respective mapped 
expressions in the corresponding finite-dimensional discrete phase space.

\section{Preliminaries on the discrete Wigner functions for SU(N)}
\lb{s2}

Let us initially introduce the density-matrix space for $N$-level quantum systems, here related to the Hilbert space
$\mathcal{H}_{N}$, through the mathematical definition \cite{Kimura2003,BK2003}
\bd
\mathscr{L}_{+,1}(\mathcal{H}_{N}) = \lbr \ro \in \mathscr{L}(\mathcal{H}_{N}) \; | \; \tr [ \ro ] = 1 , \; \ro = \ro^{\dagger} ,
\; \rho_{\ell} \geq 0 \;\; (\ell = 1,\ldots,N) \rbr
\ed
which embodies, in principle, three important requirements on the matrix density $\ro$:
\begin{itemlist}
 \item $\tr [ \ro ] = 1$: its normalization is preserved;
 \item $\ro = \ro^{\dagger}$: by definition, $\ro$ must be a Hermitian matrix; and finally,
 \item $\rho_{\ell} \in \mathbb{R}_{+} (\forall \ell)$: it must be positive semidefinite \cite{Horn} (i.e., all its eigenvalues
 are positive).
\end{itemlist}
This last requirement has become of vital importance in quantum information theory \cite{Kraus}, since it permits to identify
entangled states \cite{Peres1996,HHH1996} and classify quantum channels \cite{HSR2003,Ruskai,JP2018} by means of positive and
completely positive maps. The further requirement $\tr [ \ro^{2} ] \leq 1$ can be interpreted as a direct consequence of 
$\ro=\ro^{\dagger}$ for any $\ro \in \mathscr{L}_{+,1}(\mathcal{H}_{N})$, being the saturation reached in this case only for pure 
states. In fact, this condition characterizes the Bloch-vector space to be in a ball in $\mathbb{R}^{N^{2}-1}$ \cite{Kimura2003}. 
Furthermore, the notation $\mathscr{L}(\mathcal{H}_{N})$ represents the set of linear operators on $\mathcal{H}_{N}$.

The next step consists in considering the complete orthonormal operator basis constituted by $N^{2}-1$ generators 
$\{ \hat{g}_{i} \}_{i=1,\ldots,N^{2}-1}$ associated with the special unitary group $\mathrm{SU(N)}$, which are characterized by
$N \times N$ skew-Hermitian matrices satisfying a special set of mathematical relations \cite{HE1981,Pfeifer,MW1998}:
\begin{itemlist}
 \item Basic rules
 \bd
 \mathtt{(i)} \; \hat{g}_{i} = \hat{g}_{i}^{\dagger} , \quad \mathtt{(ii)} \; \tr [ \hat{g}_{i} ] = 0 , \quad 
 \mathtt{(iii)} \; \tr [ \hat{g}_{i} \hat{g}_{j} ] = 2 \delta_{ij} ;
 \ed
 \item Commutation relation
 \bd
 \mathtt{(iv)} \; [ \hat{g}_{i},\hat{g}_{j} ] = 2 \im \sum_{k=1}^{N^{2}-1} \mathscr{F}_{ijk} \hat{g}_{k} \; \Rightarrow \;
 \mathscr{F}_{ijk} = - \frac{\im}{4} \tr [ [ \hat{g}_{i},\hat{g}_{j} ] \hat{g}_{k} ] \;\; (\mbox{antisymmetric tensor}) ;
 \ed
 \item Anticommutation relation
 \bd
 \mathtt{(v)} \; \{ \hat{g}_{i},\hat{g}_{j} \} = \frac{4}{N} \delta_{ij} \hat{\mathds{I}}_{N} + 2 \sum_{k=1}^{N^{2}-1} 
 \mathscr{D}_{ijk} \hat{g}_{k} \; \Rightarrow \; 
 \mathscr{D}_{ijk} = \frac{1}{4} \tr [ \{ \hat{g}_{i},\hat{g}_{j} \} \hat{g}_{k} ] \;\; (\mbox{symmetric tensor}) ,
 \ed
 where $\hat{\mathds{I}}_{N}$ denotes the $N$-dimensional unit matrix;
 \item Jacobi identity
 \brr
 & & \mathtt{(vi)} \; [ \hat{g}_{i}, [ \hat{g}_{j},\hat{g}_{k} ] ] + [ \hat{g}_{j}, [ \hat{g}_{k},\hat{g}_{i} ] ] + 
 [ \hat{g}_{k}, [ \hat{g}_{i},\hat{g}_{j} ] ] = 0 , \nn \\
 & & \mathtt{(vii)} \; [ \hat{g}_{i}, \{ \hat{g}_{j},\hat{g}_{k} \} ] + [ \hat{g}_{j}, \{ \hat{g}_{k},\hat{g}_{i} \} ] + 
 [ \hat{g}_{k}, \{ \hat{g}_{i},\hat{g}_{j} \} ] = 0 ; \nn
 \err
 \item Trace of products
 \brr
 & & \mathtt{(viii)} \; \tr [ \hat{g}_{i} \hat{g}_{j} \hat{g}_{k} ] = 2 \mathscr{J}_{ijk} \quad \mbox{with} \quad 
 \mathscr{J}_{ijk} = \mathscr{D}_{ijk} + \im \mathscr{F}_{ijk} , \nn \\
 & & \mathtt{(ix)} \; \tr [ \hat{g}_{i} \hat{g}_{j} \hat{g}_{k} \hat{g}_{l} ] = \frac{4}{N} \delta_{ij} \delta_{kl} + 
 2 \sum_{p=1}^{N^{2}-1} \mathscr{J}_{ijp} \mathscr{J}_{pkl} . \nn
 \err
\end{itemlist}
Note that $\mathscr{F}_{ijk}$ and $\mathscr{D}_{ijk}$ are well-known constants in literature \cite{MW1998,AL1987} and whose 
respective values can be found in tabulated form for different values of $N$. Therefore, any linear operator $\hat{O}$ can be
decomposed in terms of the elements arising from such operator basis,
\be
\lb{eq1}
\hat{O} = \frac{1}{N} \tr [ \hat{O} ] \mathds{I}_{N} + \half \sum_{i=1}^{N^{2}-1} \mathcal{O}_{i} \hat{g}_{i} ,
\ee
with $\mathcal{O}_{i} \equiv \tr [ \hat{g}_{i} \hat{O} ]$ representing the $N^{2}-1$ coefficients of the expansion. In particular,
if one considers the $N$-level quantum systems, the associated density matrix $\ro \in \mathscr{L}_{+,1}(\mathcal{H}_{N})$ admits
the following expression \cite{HE1981}:
\be
\lb{eq2}
\ro = \frac{1}{N} \mathds{I}_{N} + \half \sum_{i=1}^{N^{2}-1} \lg \hat{g}_{i} \rg \hat{g}_{i} ,
\ee
where the mean values $\lg \hat{g}_{i} \rg \equiv \tr [ \hat{g}_{i} \ro ]$ are the components of the generalized Bloch vector 
\bd
\mathbf{g} = \lpar \lg \hat{g}_{1} \rg,\ldots,\lg \hat{g}_{N^{2}-1} \rg \rpar \in \mathbb{R}^{N^{2}-1} .
\ed
In this algebraic approach, the condition
\be
\lb{eq3}
\tr [ \ro^{2} ] = \frac{1}{N} + \half | \mathbf{g} |^{2} \leq 1
\ee
allows us to determine if a given density matrix (\ref{eq2}) describes pure or mixed states: indeed, from the experimental point 
of view, it is sufficient to measure the length of the generalized Bloch vector in such a case \cite{KK2005}. The constructive
aspects of the generators of $\mathrm{SU(N)}$ can be properly found in Refs. \cite{HE1981,MW1998}.

Now, let us introduce the $\mathit{mod(N)}$-invariant unitary operator basis \cite{GP1992}
\be
\lb{eq4}
\hat{G}(\mu,\nu) \col \frac{1}{\sqrt{N}} \sum_{\eta,\xi=0}^{N-1} \om^{-(\mu \eta + \nu \xi)} \om^{\half N \Phi(\eta,\xi;N)}
\hat{S}_{\mathrm{S}}(\eta,\xi) \quad (\mu,\nu = 0,\ldots,N-1)
\ee
written in terms of the discrete Fourier transform of the symmetrized basis
\bd
\hat{S}_{\mathrm{S}}(\eta,\xi) = \frac{1}{\sqrt{N}} \om^{\half \eta \xi} \hat{U}^{\eta} \hat{V}^{\xi} ,
\ed
where $\hat{U}$ and $\hat{V}$ correspond to the Schwinger unitary operators \cite{Schwinger} defined in an $N$-dimensional state
vector space, whose mathematical properties were extensively explored in Refs. \cite{GM1996,MM2013}. It is worth stressing that
the extra phase $\Phi(\eta,\xi;N) = N I_{\eta}^{N} I_{\xi}^{N} - \eta I_{\xi}^{N} - \xi I_{\eta}^{N}$ appearing in (\ref{eq4}) is
responsible for the $\mathit{mod(N)}$-invariance property of this operator basis, $I_{\varepsilon}^{N} = \lbk \frac{\varepsilon}
{N} \rbk$ being the integer part of $\varepsilon$ with respect to $N$. Thus, as expected from a well-grounded unitary operator basis, the decomposition of any linear operator can also be promptly established in this case, that is
\be
\lb{eq5}
\hat{O} = \frac{1}{N} \sum_{\mu,\nu=0}^{N-1} \mathrm{O}(\mu,\nu) \hat{G}(\mu,\nu) .
\ee
The coefficients $\mathrm{O}(\mu,\nu) = \tr [ \hat{G}^{\dagger}(\mu,\nu) \hat{O} ]$ show a one-to-one correspondence between
operators and functions belonging to an $N^{2}$-dimensional discrete phase space. For $\hat{O} \equiv \ro \in \mathscr{L}_{+,1}
(\mathcal{H}_{N})$, this particular decomposition assumes the compact form
\be
\lb{eq6}
\ro = \frac{1}{N} \sum_{\mu,\nu=0}^{N-1} W(\mu,\nu) \hat{G}(\mu,\nu) 
\ee
with $W(\mu,\nu) \col \tr [ \hat{G}^{\dagger}(\mu,\nu) \ro ]$ formally defining the discrete Wigner function \cite{GP1992}. 
Henceforth, the \textsl{finite-dimensional discrete phase space} will represent a finite mesh with $N^{2}$ points labelled by 
genuinely discrete variables \cite{AG1990}.

\vspace*{4pt}
\begin{table}[!t]
\tcaption{All the possible values of discrete Wigner function (\ref{eq8}) for $0 \leq \mu,\nu \leq 1$.}
\centerline{\smalllineskip
\begin{tabular}{l c c c c}
$\mu$ & $\nu$ & w.r.t. $(P_{x},P_{y},P_{z})$  & w.r.t. $(\rho_{11},\rho_{12},\rho_{22})$ \\
\hline \\
0     & 0     & $\half \lpar 1 + P_{x} - P_{y} + P_{z} \rpar$ & $\rho_{11} + \re (\rho_{12}) + \ima (\rho_{12})$ \\ \\
0     & 1     & $\half \lpar 1 - P_{x} + P_{y} + P_{z} \rpar$ & $\rho_{11} - \re (\rho_{12}) - \ima (\rho_{12})$ \\ \\
1     & 0     & $\half \lpar 1 + P_{x} + P_{y} - P_{z} \rpar$ & $\rho_{22} + \re (\rho_{12}) - \ima (\rho_{12})$ \\ \\
1     & 1     & $\half \lpar 1 - P_{x} - P_{y} - P_{z} \rpar$ & $\rho_{22} - \re (\rho_{12}) + \ima (\rho_{12})$ \\ \\
\hline \\
\end{tabular}}
\lb{tab1}
\end{table}
How the discrete Wigner function for $\mathrm{SU(N)}$ can be constructed out from these theoretical frameworks? This question was
properly answered in \cite{MG2019} through the connections established between generators of the group $\mathrm{SU(N)}$ and
Schwinger unitary operators via $\mathit{mod(N)}$-invariant operator basis, that is, for each generator $\hat{g}_{i}$ exists a
one-to-one correspondence with a given decomposition of unitary operators reached, by its turns, through the discrete mapping
kernel (\ref{eq4}). Pursuing this guideline, the discrete Wigner function for $\mathrm{SU(N)}$ is defined as follows:
\be
\lb{eq7}
W(\mu,\nu) = \frac{1}{N} + \half \sum_{i=1}^{N^{2}-1} \lg \hat{g}_{i} \rg \lpar \hat{g}_{i} \rpar \! (\mu,\nu) ,
\ee
where $\lpar \hat{g}_{i} \rpar \! (\mu,\nu) = \tr [ \hat{G}^{\dagger}(\mu,\nu) \hat{g}_{i} ]$ corresponds to the representatives
in the finite-dimensional discrete phase space of the generators $\{ \hat{g}_{i} \}_{i=1,\ldots,N^{2}-1}$. This elegant and
compact mathematical result represents an alternative approach to those recently proposed by Tilma, Everitt \textit{et al.}
\cite{TN2012,PRL2016} for Wigner functions with continuous representations (through coherent states or Euler angles): in fact,
Eq. (\ref{eq7}) describes the Wigner function defined upon a finite-dimensional phase space labelled by genuine discrete variables
associated with spin representations.

To illustrate such argument, let us now consider, from the non-relativistic quantum theory point of view, the group 
$\mathrm{SU(2)}$ and its corresponding generators $\{ \hat{\sigma}_{i} \}_{i=x,y,z}$, where $\hat{\sigma}_{i}$ denotes the Pauli 
matrices. Thus, it is possible to demonstrate that $\hat{\sigma}_{i} = \hat{V} \delta_{ix} - \im \hat{U} \hat{V} \delta_{iy} + 
\hat{U} \delta_{iz}$ indeed establishes the correspondence, once both the operators share the same set of orthonormal 
eigenvectors\ftn{Similar relations were already obtained for the Gell-Mann matrices $\lam$'s associated with the group 
$\mathrm{SU(3)}$ -- see Ref. \cite{MG2019} for technical details.}. In this particular example, the discrete Wigner function is
given by
\be
\lb{eq8}
W(\mu,\nu) = \half \lbk 1 + (-1)^{\nu} P_{x} + (-1)^{\mu + \nu +1} P_{y} + (-1)^{\mu} P_{z} \rbk ,
\ee
with $P_{i} = \tr [ \ro \hat{\sigma}_{i} ] \in [-1,1]$ for $i=x,y,z$ corresponding to the polarization-vector components which 
obey the relation $P_{x}^{2} + P_{y}^{2} + P_{z}^{2} \leq 1$ (the saturation occurs only for pure states). Table \ref{tab1} shows
all the possible values of $W(\mu,\nu)$ in the finite-dimensional phase space here labelled by the discrete variables $0 \leq \mu,
\nu \leq 1$ with respect to (w.r.t.) $(P_{x},P_{y},P_{z})$ and also as a function of the matrix elements $(\rho_{11},\rho_{12},
\rho_{22})$, once that $(P_{x},P_{y},P_{z}) = \lpar 2 \re (\rho_{12}), - 2 \ima (\rho_{12}), \rho_{11} - \rho_{22} \rpar$. As
expected, the absolute minimum values reached by (\ref{eq8}) happen for pure states.

\section{Description of two-qubit and ququart states via discrete Wigner functions}
\lb{s3}

In this section, we establish two different but complementary group-theoretical approaches to describe two-qubit and ququart
states through discrete Wigner functions: the first approach considers the Klein's group $\mathrm{SU(2)} \otimes \mathrm{SU(2)}$ 
and encompasses the Fano's description for two-qubit states as pairs of two-level systems \cite{Fano1957,Fano1983}, whereas the
second one embodies the group $\mathrm{SU(4)}$ in order to describe ququart states as a single four-level system. Besides, we
obtain the respective exact discrete Wigner functions associated with a finite-dimensional discrete phase space.

\subsection{The Klein's group}

Initially, let us mention that $\{ \hat{\mathds{I}}_{2}^{(1)}, \hat{\sigma}_{x}^{(1)}, \hat{\sigma}_{y}^{(1)}, 
\hat{\sigma}_{z}^{(1)} \} \otimes \{ \hat{\mathds{I}}_{2}^{(2)}, \hat{\sigma}_{x}^{(2)}, \hat{\sigma}_{y}^{(2)}, 
\hat{\sigma}_{z}^{(2)} \}$ represents the operator basis used to describe a completely general two-qubit state, where the
superscripts $(1)$ and $(2)$ correspond to the qubits $1$ and $2$. Following, let us adopt the Fano's prescription for the density
matrix \cite{Fano1983}
\be
\lb{eq9}
\ro = \frac{1}{4} \lbk \hat{\mathds{I}}_{4} + \sum_{i=x,y,z} a_{i} \hat{\sigma}_{i}^{(1)} \otimes \hat{\mathds{I}}_{2}^{(2)} +
\sum_{j=x,y,z} b_{j} \hat{\mathds{I}}_{2}^{(1)} \otimes \hat{\sigma}_{j}^{(2)} + \sum_{i,j=x,y,z} c_{ij} \hat{\sigma}_{i}^{(1)}
\otimes \hat{\sigma}_{j}^{(2)} \rbk
\ee
with real coefficients\ftn{Although this operator basis is expressed as a tensor product, this fact does not imply that $\ro$ 
can be decomposed as $\ro^{(1)} \otimes \ro^{(2)}$: indeed, only for $a_{i} = P_{i}^{(1)}$, $b_{j} = P_{j}^{(2)}$ and $c_{ij} = 
P_{i}^{(1)} P_{j}^{(2)}$, the condition $\ro = \ro^{(1)} \otimes \ro^{(2)}$ is verified; otherwise, we obtain $\ro \neq \ro^{(1)}
\otimes \ro^{(2)}$. This genuinely quantum property is essential for characterizing, in such a case, the bipartite states in 
separable states $(\ro = \ro^{(1)} \otimes \ro^{(2)})$ and entangled states $(\ro \neq \ro^{(1)} \otimes \ro^{(2)})$, with 
immediate implications in quantum mechanics and quantum information theory \cite{RPMK2009,Li-Qiao}.}, where $\ro \in
\mathscr{L}_{+,1}(\mathcal{H}_{2} \otimes \mathcal{H}_{2})$. Note that the positivity of all four eigenvalues (necessary condition 
to ensure that $\ro$ is positive semidefinite) is reached through the solution of the following nontrivial set of inequalities
\cite{BK2003}:
\brr
\lb{eq10}
\left\{ \begin{array}{lll}
  \tr [ \ro^{2} ] \leq 1 , \\
  \tr [ \ro^{3} ] \geq \frac{3}{2} \tr [ \ro^{2} ] - \half , \\
  \tr [ \ro^{4} ] \leq \frac{1}{6} - \tr [ \ro^{2} ] + \half \lpar \tr [ \ro^{2} ] \rpar^{2} + \frac{4}{3} \tr [ \ro^{3} ] .
        \end{array}
\right.
\err
For the sake of completeness, the first inequality
\bd
\tr [ \ro^{2} ] = \frac{1}{4} \lbk 1 + \sum_{i=x,y,z} \lpar a_{i}^{2} + b_{i}^{2} \rpar + \sum_{i,j=x,y,z} c_{ij} \rbk \leq 1
\ed
consists of a mathematical condition that distinguishes between mixed and pure states, with the saturation occurring only for pure
states. In addition, the reduced density matrix related to the qubit $1(2)$ is obtained from $\ro$ by taking the partial trace
over the subspace of the qubit $2(1)$, namely, 
\be
\lb{eq11}
\ro_{\mathtt{R}}^{(1)} \col \tr_{2} [ \ro ] = \half \lbk \hat{\mathds{I}}_{2}^{(1)} + \sum_{i=x,y,z} a_{i} \hat{\sigma}_{i}^{(1)}
\rbk 
\ee
and
\be
\lb{eq12}
\ro_{\mathtt{R}}^{(2)} \col \tr_{1} [ \ro ] = \half \lbk \hat{\mathds{I}}_{2}^{(2)} + \sum_{j=x,y,z} b_{j} \hat{\sigma}_{j}^{(2)} 
\rbk .
\ee
These results determine only the partial information of the bipartite system $\ro$ under scrutiny since they see only the quantum 
state of a given subsystem \cite{Vedral}.

Now, let us calculate the discrete Wigner function associated with the density matrix (\ref{eq9}). For such a particular task,
we initially consider the mapping kernel $\hat{G}^{\dagger}(\mu_{a},\nu_{a})$, here related to the subspaces $a=1,2$ of each 
qubit, in order to obtain a first expression for
\bd
W(\mu_{1},\nu_{1},\mu_{2},\nu_{2}) \col \tr [ \hat{G}^{\dagger}(\mu_{1},\nu_{1}) \otimes \hat{G}^{\dagger}(\mu_{2},\nu_{2}) \ro ]
\qquad (0 \leq \mu_{1},\nu_{1},\mu_{2},\nu_{2} \leq 1). 
\ed
Next, substituting Eq. (\ref{eq9}) in this definition, we get the intermediate result
\brr
& & W(\mu_{1},\nu_{1},\mu_{2},\nu_{2}) = \frac{1}{4} \left[ 1 + \sum_{i=x,y,z} a_{i} \tr [ \hat{G}^{\dagger}(\mu_{1},\nu_{1})
\hat{\sigma}_{i}^{(1)} ] + \sum_{j=x,y,z} b_{j} \tr [ \hat{G}^{\dagger}(\mu_{2},\nu_{2}) \hat{\sigma}_{j}^{(2)} ] \right. \nn \\
& & \hspace{3.8cm} + \left. \sum_{i,j=x,y,z} c_{ij} \tr [ \hat{G}^{\dagger}(\mu_{1},\nu_{1}) \hat{\sigma}_{i}^{(1)} ] \,
\tr [ \hat{G}^{\dagger}(\mu_{2},\nu_{2}) \hat{\sigma}_{j}^{(2)} ] \right] , \nn
\err
where the terms $\tr [ \hat{G}^{\dagger}(\mu_{a},\nu_{a}) \hat{\sigma}_{\ell}^{(a)} ]$ for $a=1,2$ and $\ell=x,y,z$ are determined
as follows:
\bd
\tr [ \hat{G}^{\dagger}(\mu_{a},\nu_{a}) \hat{\sigma}_{\ell}^{(a)} ] = (-1)^{\nu_{a}} \delta_{\ell x} + (-1)^{\mu_{a}+\nu_{a}+1}
\delta_{\ell y} + (-1)^{\mu_{a}} \delta_{\ell z} .
\ed
So, after a few algebraic manipulations, the discrete Wigner function $W(\mu_{1},\nu_{1},\mu_{2},\nu_{2})$ achieves the general
expression
\brr
\lb{eq13}
& & W(\mu_{1},\nu_{1},\mu_{2},\nu_{2}) = \frac{1}{4} \bigl[ 1 + (-1)^{\nu_{1}} a_{x} + (-1)^{\mu_{1}+\nu_{1}+1} a_{y} + 
(-1)^{\mu_{1}} a_{z} + (-1)^{\nu_{2}} b_{x} \nn \\
& & \hspace{1cm} + (-1)^{\mu_{2}+\nu_{2}+1} b_{y} + (-1)^{\mu_{2}} b_{z} + (-1)^{\nu_{1}+\nu_{2}} c_{xx} + 
(-1)^{\nu_{1}+\mu_{2}+\nu_{2}+1} c_{xy} + (-1)^{\nu_{1}+\mu_{2}} c_{xz} \nn \\
& & \hspace{1cm} + (-1)^{\mu_{1}+\nu_{1}+\nu_{2}+1} c_{yx} + (-1)^{\mu_{1}+\nu_{1}+\mu_{2}+\nu_{2}} c_{yy} + 
(-1)^{\mu_{1}+\nu_{1}+\mu_{2}+1} c_{yz} \nn \\
& & \hspace{1cm} + (-1)^{\mu_{1}+\nu_{2}} c_{zx} + (-1)^{\mu_{1}+\mu_{2}+\nu_{2}+1} c_{zy} + (-1)^{\mu_{1}+\mu_{2}} c_{zz} 
\bigr] ,
\err
whose normalization condition
\bd
\frac{1}{4} \sum_{\mu_{1},\nu_{1},\mu_{2},\nu_{2}} W(\mu_{1},\nu_{1},\mu_{2},\nu_{2}) = 1
\ed
can be promptly verified through the results showed in Table \ref{tab2}. However, Eq. (\ref{eq13}) presents an apparent
disadvantage: the visualization of this function in the finite-dimensional phase space labelled by the discrete variables 
$(\mu_{1},\nu_{1},\mu_{2},\nu_{2})$ is not fully functional since only bidimensional projections are easily manageable.
\vspace*{4pt}
\begin{table}[!ht]
\tcaption{Discrete Wigner function (\ref{eq13}) in terms of the coefficients $a_{i}$, $b_{j}$, and $c_{ij}$ for $i,j=x,y,z$.}
\centerline{\footnotesize\smalllineskip
\begin{tabular}{l c c c c c}
$\mu_{1}$ & $\nu_{1}$ & $\mu_{2}$ & $\nu_{2}$ & $4 W(\mu_{1},\nu_{1},\mu_{2},\nu_{2})$  \\
\hline
0 & 0 & 0 & 0 & $1 + a_{x} - a_{y} + a_{z} + b_{x} - b_{y} + b_{z} + c_{xx} - c_{xy} + c_{xz} - c_{yx} + c_{yy} - c_{yz} + 
c_{zx} - c_{zy} + c_{zz}$ \\
0 & 0 & 1 & 0 & $1 + a_{x} - a_{y} + a_{z} + b_{x} + b_{y} - b_{z} + c_{xx} + c_{xy} - c_{xz} - c_{yx} - c_{yy} + c_{yz} + 
c_{zx} + c_{zy} - c_{zz}$ \\
0 & 0 & 0 & 1 & $1 + a_{x} - a_{y} + a_{z} - b_{x} + b_{y} + b_{z} - c_{xx} + c_{xy} + c_{xz} + c_{yx} - c_{yy} - c_{yz} - 
c_{zx} + c_{zy} + c_{zz}$ \\
0 & 0 & 1 & 1 & $1 + a_{x} - a_{y} + a_{z} - b_{x} - b_{y} - b_{z} - c_{xx} - c_{xy} - c_{xz} + c_{yx} + c_{yy} + c_{yz} - 
c_{zx} - c_{zy} - c_{zz}$ \\
1 & 0 & 0 & 0 & $1 + a_{x} + a_{y} - a_{z} + b_{x} - b_{y} + b_{z} + c_{xx} - c_{xy} + c_{xz} + c_{yx} - c_{yy} + c_{yz} -
c_{zx} + c_{zy} - c_{zz}$ \\
1 & 0 & 1 & 0 & $1 + a_{x} + a_{y} - a_{z} + b_{x} + b_{y} - b_{z} + c_{xx} + c_{xy} - c_{xz} + c_{yx} + c_{yy} - c_{yz} - 
c_{zx} - c_{zy} + c_{zz}$ \\
1 & 0 & 0 & 1 & $1 + a_{x} + a_{y} - a_{z} - b_{x} + b_{y} + b_{z} - c_{xx} + c_{xy} + c_{xz} - c_{yx} + c_{yy} + c_{yz} + 
c_{zx} - c_{zy} - c_{zz}$ \\
1 & 0 & 1 & 1 & $1 + a_{x} + a_{y} - a_{z} - b_{x} - b_{y} - b_{z} - c_{xx} - c_{xy} - c_{xz} - c_{yx} - c_{yy} - c_{yz} + 
c_{zx} + c_{zy} + c_{zz}$ \\
0 & 1 & 0 & 0 & $1 - a_{x} + a_{y} + a_{z} + b_{x} - b_{y} + b_{z} - c_{xx} + c_{xy} - c_{xz} + c_{yx} - c_{yy} + c_{yz} + 
c_{zx} - c_{zy} + c_{zz}$ \\
0 & 1 & 1 & 0 & $1 - a_{x} + a_{y} + a_{z} + b_{x} + b_{y} - b_{z} - c_{xx} - c_{xy} + c_{xz} + c_{yx} + c_{yy} - c_{yz} + 
c_{zx} + c_{zy} - c_{zz}$ \\
0 & 1 & 0 & 1 & $1 - a_{x} + a_{y} + a_{z} - b_{x} + b_{y} + b_{z} + c_{xx} - c_{xy} - c_{xz} - c_{yx} + c_{yy} + c_{yz} - 
c_{zx} + c_{zy} + c_{zz}$ \\
0 & 1 & 1 & 1 & $1 - a_{x} + a_{y} + a_{z} - b_{x} - b_{y} - b_{z} + c_{xx} + c_{xy} + c_{xz} - c_{yx} - c_{yy} - c_{yz} -
c_{zx} - c_{zy} - c_{zz}$ \\
1 & 1 & 0 & 0 & $1 - a_{x} - a_{y} - a_{z} + b_{x} - b_{y} + b_{z} - c_{xx} + c_{xy} - c_{xz} - c_{yx} + c_{yy} - c_{yz} - 
c_{zx} + c_{zy} - c_{zz}$ \\
1 & 1 & 1 & 0 & $1 - a_{x} - a_{y} - a_{z} + b_{x} + b_{y} - b_{z} - c_{xx} - c_{xy} + c_{xz} - c_{yx} - c_{yy} + c_{yz} - 
c_{zx} - c_{zy} + c_{zz}$ \\
1 & 1 & 0 & 1 & $1 - a_{x} - a_{y} - a_{z} - b_{x} + b_{y} + b_{z} + c_{xx} - c_{xy} - c_{xz} + c_{yx} - c_{yy} - c_{yz} + 
c_{zx} - c_{zy} - c_{zz}$ \\
1 & 1 & 1 & 1 & $1 - a_{x} - a_{y} - a_{z} - b_{x} - b_{y} - b_{z} + c_{xx} + c_{xy} + c_{xz} + c_{yx} + c_{yy} + c_{yz} + 
c_{zx} + c_{zy} + c_{zz}$ \\
\hline \\
\end{tabular}}
\lb{tab2}
\end{table}

As a further remark, let us now determine the discrete Wigner functions associated with the reduced density matrices (\ref{eq11})
and (\ref{eq12}), i.e.,
\brr
& & \mathcal{W}_{\mathtt{R}}(\mu_{1},\nu_{1}) \col \tr [ \hat{G}^{\dagger}(\mu_{1},\nu_{1}) \ro_{\mathtt{R}}^{(1)} ] = \half 
\lbk 1 + (-1)^{\nu_{1}} a_{x} + (-1)^{\mu_{1}+\nu_{1}+1} a_{y} + (-1)^{\mu_{1}} a_{z} \rbk , \nn \\
& & \mathcal{W}_{\mathtt{R}}(\mu_{2},\nu_{2}) \col \tr [ \hat{G}^{\dagger}(\mu_{2},\nu_{2}) \ro_{\mathtt{R}}^{(2)} ] = \half 
\lbk 1 + (-1)^{\nu_{2}} b_{x} + (-1)^{\mu_{2}+\nu_{2}+1} b_{y} + (-1)^{\mu_{2}} b_{z} \rbk . \nn
\err
These results present, as expected, a particular connection with the partial sums of Eq. (\ref{eq13}) through the relations
\bd
\mathcal{W}_{\mathtt{R}}(\mu_{1},\nu_{1}) = \half \sum_{\mu_{2},\nu_{2}} W(\mu_{1},\nu_{1},\mu_{2},\nu_{2}) \quad \mbox{and}
\quad \mathcal{W}_{\mathtt{R}}(\mu_{2},\nu_{2}) = \half \sum_{\mu_{1},\nu_{1}} W(\mu_{1},\nu_{1},\mu_{2},\nu_{2}) .
\ed
The restriction $c_{ij} = a_{i} b_{j}$ for $i,j=x,y,z$ implies that $W(\mu_{1},\nu_{1},\mu_{2},\nu_{2})$ splits into the product
$\mathcal{W}_{\mathtt{R}}(\mu_{1},\nu_{1}) \mathcal{W}_{\mathtt{R}}(\mu_{2},\nu_{2})$, this condition being directly related to 
the separable states.

\subsubsection{The computational basis}

In general, an arbitrary two-qubit state is defined as a linear superposition of the (orthonormalized) computational-basis states
\cite{Marinescu}
\bd
\bigl\{ | 0_{1} 0_{2} \rg, | 0_{1} 1_{2} \rg, | 1_{1} 0_{2} \rg, | 1_{1} 1_{2} \rg \; : \; | i_{1} \rg \otimes | j_{2} \rg \equiv
| i_{1} j_{2} \rg \;\; (i,j=0,1) \bigr\} ,
\ed
namely,
\be
\lb{eq14}
| \Psi \rg = \alf | 0_{1} 0_{2} \rg + \bet | 0_{1} 1_{2} \rg + \gam | 1_{1} 0_{2} \rg + \delta | 1_{1} 1_{2} \rg
\ee
with $\alf,\bet,\gam,\delta \in \mathbb{C}$ and $| \alf |^{2} + | \bet |^{2} + | \gam |^{2} + | \delta |^{2} = 1$. From an 
operational point of view, it is common to deal with the matrix representation for $\ro = | \Psi \rg \lg \Psi |$ in the 
computacional basis, which leads us to obtain the positive semidefinite $4 \times 4$ Hermitian matrix
\be
\lb{eq15}
\ro = \left( \begin{array}{cccc}
\rho_{11}        & \rho_{12}        & \rho_{13}        & \rho_{14} \\
\rho_{12}^{\ast} & \rho_{22}        & \rho_{23}        & \rho_{24} \\
\rho_{13}^{\ast} & \rho_{23}^{\ast} & \rho_{33}        & \rho_{34} \\
\rho_{14}^{\ast} & \rho_{24}^{\ast} & \rho_{34}^{\ast} & \rho_{44}
\end{array} \right) = \left( \begin{array}{cccc}
| \alf |^{2}       & \alf \bet^{\ast}   & \alf \gam^{\ast}   & \alf \delta^{\ast} \\
\alf^{\ast} \bet   & | \bet |^{2}       & \bet \gam^{\ast}   & \bet \delta^{\ast} \\
\alf^{\ast} \gam   & \bet^{\ast} \gam   & | \gam |^{2}       & \gam \delta^{\ast} \\
\alf^{\ast} \delta & \bet^{\ast} \delta & \gam^{\ast} \delta & | \delta |^{2}
\end{array} \right)
\ee
such that $\tr [ \ro ] = 1$. Note that Eqs. (\ref{eq9}) and (\ref{eq15}) are connected by means of a change of basis: indeed, for
$\hat{\mathds{I}}_{2}^{(a)} = | 0_{a} \rg \lg 0_{a} | + | 1_{a} \rg \lg 1_{a} |$, $\hat{\sigma}_{x}^{(a)} = | 0_{a} \rg \lg 
1_{a} | + | 1_{a} \rg \lg 0_{a} |$, $\hat{\sigma}_{y}^{(a)} = - \im ( | 0_{a} \rg \lg 1_{a} | - | 1_{a} \rg \lg 0_{a} | )$, and
$\hat{\sigma}_{z}^{(a)} = | 0_{a} \rg \lg 0_{a} | - | 1_{a} \rg \lg 1_{a} |$ (with $a=1,2$), the first one can be written in a 
similar fashion as the second one or vice-versa, where now the elements $\rho_{ij}$ for $i \leq j$ are evaluated as follows:
\brr
& & \rho_{11} = \frac{1}{4} ( 1 + a_{z} + b_{z} + c_{zz} ) , \;\; \rho_{12} = \frac{1}{4} [ b_{x} + c_{zx} - \im ( b_{y} + 
c_{zy} ) ] , \nn \\
& & \rho_{13} = \frac{1}{4} [ a_{x} + c_{xz} - \im ( a_{y} + c_{yz} ) ] , \;\; \rho_{14} = \frac{1}{4} [ c_{xx} - c_{yy} - \im
( c_{xy} + c_{yx} ) ] , \nn \\
& & \rho_{22} = \frac{1}{4} ( 1 + a_{z} - b_{z} - c_{zz} ) , \;\; \rho_{23} = \frac{1}{4} [ c_{xx} + c_{yy} + \im ( c_{xy} -
c_{yx} ) ] , \nn \\
& & \rho_{24} = \frac{1}{4} [ a_{x} - c_{xz} - \im ( a_{y} - c_{yz} ) ] , \;\; \rho_{33} = \frac{1}{4} ( 1 - a_{z} + b_{z} - 
c_{zz} ) , \nn \\
& & \rho_{34} = \frac{1}{4} [ b_{x} - c_{zx} - \im ( b_{y} - c_{zy} ) ] , \;\; \rho_{44} = \frac{1}{4} ( 1 - a_{z} - b_{z} +
c_{zz} ) . \nn
\err
This system of linear equations is invertible and this fact allows us to rewrite $ W(\mu_{1},\nu_{1},\mu_{2},\nu_{2})$ as a 
function of the matrix elements appeared in (\ref{eq15}),
\brr
\lb{eq16}
& & W(\mu_{1},\nu_{1},\mu_{2},\nu_{2}) = \frac{1}{4} \Bigl\{ 1 + \Gamma_{11}(\mu_{1},\mu_{2}) + \Gamma_{22}(\mu_{1},\mu_{2}) +
\Gamma_{33}(\mu_{1},\mu_{2}) + \Gamma_{44}(\mu_{1},\mu_{2}) \nn \\
& & \hspace{2cm} + 2 (-1)^{\nu_{1}} \lbk \Gamma_{13}(\mu_{1},\mu_{2}) + \Gamma_{24}(\mu_{1},\mu_{2}) \rbk + 2 (-1)^{\nu_{2}} \lbk
\Gamma_{12}(\mu_{1},\mu_{2}) + \Gamma_{34}(\mu_{1},\mu_{2}) \rbk \nn \\
& & \hspace{2cm} + 2 (-1)^{\nu_{1}+\nu_{2}} \lbk \Gamma_{14}(\mu_{1},\mu_{2}) + \Gamma_{23}(\mu_{1},\mu_{2}) \rbk \Bigr\}
\err
where
\brr
& & \Gamma_{11}(\mu_{1},\mu_{2}) = \lbk (-1)^{\mu_{1}} + (-1)^{\mu_{2}} + (-1)^{\mu_{1}+\mu_{2}} \rbk \rho_{11} , \nn \\
& & \Gamma_{22}(\mu_{1},\mu_{2}) = \lbk (-1)^{\mu_{1}} - (-1)^{\mu_{2}} - (-1)^{\mu_{1}+\mu_{2}} \rbk \rho_{22} , \nn \\
& & \Gamma_{33}(\mu_{1},\mu_{2}) = \lbk -(-1)^{\mu_{1}} + (-1)^{\mu_{2}} - (-1)^{\mu_{1}+\mu_{2}} \rbk \rho_{33} , \nn \\
& & \Gamma_{44}(\mu_{1},\mu_{2}) = \lbk -(-1)^{\mu_{1}} - (-1)^{\mu_{2}} + (-1)^{\mu_{1}+\mu_{2}} \rbk \rho_{44} , \nn \\
& & \Gamma_{12}(\mu_{1},\mu_{2}) = \lbk 1 + (-1)^{\mu_{1}} \rbk \lbk \mathtt{Re}(\rho_{12}) + (-1)^{\mu_{2}} \mathtt{Im}
(\rho_{12}) \rbk , \nn \\
& & \Gamma_{13}(\mu_{1},\mu_{2}) = \lbk 1 + (-1)^{\mu_{2}} \rbk \lbk \mathtt{Re}(\rho_{13}) + (-1)^{\mu_{1}} \mathtt{Im}
(\rho_{13}) \rbk , \nn \\
& & \Gamma_{14}(\mu_{1},\mu_{2}) = \lbk 1 - (-1)^{\mu_{1}+\mu_{2}} \rbk \mathtt{Re}(\rho_{14}) + \lbk (-1)^{\mu_{1}} +
(-1)^{\mu_{2}} \rbk \mathtt{Im}(\rho_{14}) , \nn \\
& & \Gamma_{23}(\mu_{1},\mu_{2}) = \lbk 1 + (-1)^{\mu_{1}+\mu_{2}} \rbk \mathtt{Re}(\rho_{23}) + \lbk (-1)^{\mu_{1}} -
(-1)^{\mu_{2}} \rbk \mathtt{Im}(\rho_{23}) , \nn \\
& & \Gamma_{24}(\mu_{1},\mu_{2}) = \lbk 1 - (-1)^{\mu_{2}} \rbk \lbk \mathtt{Re}(\rho_{24}) + (-1)^{\mu_{1}} \mathtt{Im}
(\rho_{24}) \rbk , \nn \\
& & \Gamma_{34}(\mu_{1},\mu_{2}) = \lbk 1 - (-1)^{\mu_{1}} \rbk \lbk \mathtt{Re}(\rho_{34}) + (-1)^{\mu_{2}} \mathtt{Im}
(\rho_{34}) \rbk . \nn 
\err
Table \ref{tab3} shows all the possible values of Eq. (\ref{eq16}) in the finite-dimensional discrete phase space labelled by 
$0 \leq \mu_{1},\nu_{1},\mu_{2},\nu_{2} \leq 1$. Such a compilation of results allows us to demonstrate that this discrete
quasiprobability distribution function obeys the criterion easy-to-handle.
\vspace*{4pt}
\begin{table}[!t]
\tcaption{Discrete Wigner function (\ref{eq16}) as a function of the matrix elements (\ref{eq15}).}
\centerline{\footnotesize\smalllineskip
\begin{tabular}{l c c c c c}
$\mu_{1}$ & $\nu_{1}$ & $\mu_{2}$ & $\nu_{2}$ & $W(\mu_{1},\nu_{1},\mu_{2},\nu_{2})$  \\
\hline
0 & 0 & 0 & 0 & $\rho_{11} + \mathtt{Re}(\rho_{12} + \rho_{13} + \rho_{23}) + \mathtt{Im}(\rho_{12} + \rho_{13} + \rho_{14})$ \\
0 & 0 & 1 & 0 & $\rho_{22} + \mathtt{Re}(\rho_{12} + \rho_{14} + \rho_{24}) - \mathtt{Im}(\rho_{12} - \rho_{23} - \rho_{24})$ \\
0 & 0 & 0 & 1 & $\rho_{11} - \mathtt{Re}(\rho_{12} - \rho_{13} + \rho_{23}) - \mathtt{Im}(\rho_{12} - \rho_{13} + \rho_{14})$ \\
0 & 0 & 1 & 1 & $\rho_{22} - \mathtt{Re}(\rho_{12} + \rho_{14} - \rho_{24}) + \mathtt{Im}(\rho_{12} - \rho_{23} + \rho_{24})$ \\
1 & 0 & 0 & 0 & $\rho_{33} + \mathtt{Re}(\rho_{13} + \rho_{14} + \rho_{34}) - \mathtt{Im}(\rho_{13} + \rho_{23} - \rho_{34})$ \\
1 & 0 & 1 & 0 & $\rho_{44} + \mathtt{Re}(\rho_{23} + \rho_{24} + \rho_{34}) - \mathtt{Im}(\rho_{14} + \rho_{24} + \rho_{34})$ \\
1 & 0 & 0 & 1 & $\rho_{33} + \mathtt{Re}(\rho_{13} - \rho_{14} - \rho_{34}) - \mathtt{Im}(\rho_{13} - \rho_{23} + \rho_{34})$ \\
1 & 0 & 1 & 1 & $\rho_{44} - \mathtt{Re}(\rho_{23} - \rho_{24} + \rho_{34}) + \mathtt{Im}(\rho_{14} - \rho_{24} + \rho_{34})$ \\
0 & 1 & 0 & 0 & $\rho_{11} + \mathtt{Re}(\rho_{12} - \rho_{13} - \rho_{23}) + \mathtt{Im}(\rho_{12} - \rho_{13} - \rho_{14})$ \\
0 & 1 & 1 & 0 & $\rho_{22} + \mathtt{Re}(\rho_{12} - \rho_{14} - \rho_{24}) - \mathtt{Im}(\rho_{12} + \rho_{23} + \rho_{24})$ \\
0 & 1 & 0 & 1 & $\rho_{11} - \mathtt{Re}(\rho_{12} + \rho_{13} - \rho_{23}) - \mathtt{Im}(\rho_{12} + \rho_{13} - \rho_{14})$ \\
0 & 1 & 1 & 1 & $\rho_{22} - \mathtt{Re}(\rho_{12} - \rho_{14} + \rho_{24}) + \mathtt{Im}(\rho_{12} + \rho_{23} - \rho_{24})$ \\
1 & 1 & 0 & 0 & $\rho_{33} - \mathtt{Re}(\rho_{13} + \rho_{14} - \rho_{34}) + \mathtt{Im}(\rho_{13} + \rho_{23} + \rho_{34})$ \\
1 & 1 & 1 & 0 & $\rho_{44} - \mathtt{Re}(\rho_{23} + \rho_{24} - \rho_{34}) + \mathtt{Im}(\rho_{14} + \rho_{24} - \rho_{34})$ \\
1 & 1 & 0 & 1 & $\rho_{33} - \mathtt{Re}(\rho_{13} - \rho_{14} + \rho_{34}) + \mathtt{Im}(\rho_{13} - \rho_{23} - \rho_{34})$ \\
1 & 1 & 1 & 1 & $\rho_{44} + \mathtt{Re}(\rho_{23} - \rho_{24} - \rho_{34}) - \mathtt{Im}(\rho_{14} - \rho_{24} - \rho_{34})$ \\
\hline \\
\end{tabular}}
\lb{tab3}
\end{table}

It is worth stressing that Eq. (\ref{eq14}) does not necessarily imply the decomposition
\bd
\underbrace{\lpar a_{1} | 0_{1} \rg + b_{1} | 1_{1} \rg \rpar}_{| \psi^{(1)} \rg} \otimes \underbrace{\lpar a_{2} | 0_{2} \rg +
b_{2} | 1_{2} \rg \rpar}_{| \psi^{(2)} \rg} = a_{1} a_{2} | 0_{1} 0_{2} \rg + a_{1} b_{2} | 0_{1} 1_{2} \rg + b_{1} a_{2} | 1_{1} 
0_{2} \rg + b_{1} b_{2} | 1_{1} 1_{2} \rg ,
\ed
which describes a separable pure two-qubit state. Indeed, this case occurs only for the coefficients $\alf = a_{1} a_{2}$, 
$\bet = a_{1} b_{2}$, $\gam = b_{1} a_{2}$, and $\delta = b_{1} b_{2}$; otherwise, $| \Psi \rg$ characterizes an entangled
two-qubit state (that is, $| \Psi \rg \neq | \psi^{(1)} \rg \otimes | \psi^{(2)} \rg$). In addition, the associated density matrix
(\ref{eq15}) leads us to obtain 
\bd
\tr [ \ro^{2} ] = \rho_{11}^{2} + \rho_{22}^{2} + \rho_{33}^{2} + \rho_{44}^{2} + 2 \lpar | \rho_{12} |^{2} + | \rho_{13} |^{2} +
| \rho_{14} |^{2} + | \rho_{23} |^{2} + | \rho_{24} |^{2} + | \rho_{34} |^{2} \rpar \leq 1 ,
\ed
the saturation arising only for pure states -- note that both the discrete Wigner functions (\ref{eq13}) and (\ref{eq16}) can also 
be used to evaluate the expression for $\tr [ \ro^{2} ]$ since that
\bd
\tr [ \ro^{2} ] = \frac{1}{4} \sum_{\mu_{1},\nu_{1},\mu_{2},\nu_{2}} W^{2}(\mu_{1},\nu_{1},\mu_{2},\nu_{2}) \leq 1 .
\ed
Summarizing, there are four different possibilities of describing an arbitrary bipartite quantum state \cite{Peres1996,HHH1996},
being the two-qubit quantum state a legitimate representative of this general case: entangled pure state, entangled mixed state, 
separable pure state, and finally, separable mixed state. Next, we will illustrate these cases through well-established examples
in the literature with the help of the associated discrete Wigner functions.

\subsubsection{Applications}

The first example of entangled pure two-qubit states are the Bell states $| \Psi_{\pm} \rg$ and $| \Phi_{\pm} \rg$, defined by
\cite{Vedral,Marinescu}
\bd
| \Psi_{\pm} \rg = \frac{1}{\sqrt{2}} \lpar | 0_{1} 1_{2} \rg \pm | 1_{1} 0_{2} \rg \rpar \quad \mbox{and} \quad
| \Phi_{\pm} \rg = \frac{1}{\sqrt{2}} \lpar | 0_{1} 0_{2} \rg \pm | 1_{1} 1_{2} \rg \rpar ,
\ed
whose respective density operators assume the following forms:
\brr
\lb{eq17}
\ro_{\Psi_{\pm}} &=& \half \lpar | 0_{1} 1_{2} \rg \lg 0_{1} 1_{2} | \pm | 0_{1} 1_{2} \rg \lg 1_{1} 0_{2} | \pm | 1_{1} 0_{2} \rg
\lg 0_{1} 1_{2} | + | 1_{1} 0_{2} \rg \lg 1_{1} 0_{2} | \rpar , \nn \\
&=& \frac{1}{4} \lpar \hat{\mathds{I}}_{4} \pm \hat{\sigma}_{x}^{(1)} \otimes \hat{\sigma}_{x}^{(2)} \pm \hat{\sigma}_{y}^{(1)}
\otimes \hat{\sigma}_{y}^{(2)} - \hat{\sigma}_{z}^{(1)} \otimes \hat{\sigma}_{z}^{(2)} \rpar , \\
\lb{eq18}
\ro_{\Phi_{\pm}} &=& \half \lpar | 0_{1} 0_{2} \rg \lg 0_{1} 0_{2} | \pm | 0_{1} 0_{2} \rg \lg 1_{1} 1_{2} | \pm | 1_{1} 1_{2} \rg
\lg 0_{1} 0_{2} | + | 1_{1} 1_{2} \rg \lg 1_{1} 1_{2} | \rpar , \nn \\
&=& \frac{1}{4} \lpar \hat{\mathds{I}}_{4} \pm \hat{\sigma}_{x}^{(1)} \otimes \hat{\sigma}_{x}^{(2)} \mp \hat{\sigma}_{y}^{(1)}
\otimes \hat{\sigma}_{y}^{(2)} + \hat{\sigma}_{z}^{(1)} \otimes \hat{\sigma}_{z}^{(2)} \rpar .
\err
Once the coefficients of (\ref{eq17}) and (\ref{eq18}) are completely determined, it turns immediate to obtain the corresponding
discrete Wigner functions for each set of Bell states,
\brr
\lb{eq19}
W_{\Psi_{\pm}}(\mu_{1},\nu_{1},\mu_{2},\nu_{2}) = \frac{1}{4} \lbr 1 - (-1)^{\mu_{1}+\mu_{2}} \pm (-1)^{\nu_{1}+\nu_{2}} \lbk
1 + (-1)^{\mu_{1}+\mu_{2}} \rbk \rbr , \\
\lb{eq20}
W_{\Phi_{\pm}}(\mu_{1},\nu_{1},\mu_{2},\nu_{2}) = \frac{1}{4} \lbr 1 + (-1)^{\mu_{1}+\mu_{2}} \pm (-1)^{\nu_{1}+\nu_{2}} \lbk
1 - (-1)^{\mu_{1}+\mu_{2}} \rbk \rbr .
\err
By using the expressions obtained in the right-hand-side of Table \ref{tab3}, it is possible to show that Eqs. (\ref{eq19}) and
(\ref{eq20}) assume only two unique values for any $0 \leq \mu_{1},\nu_{1},\mu_{2},\nu_{2} \leq 1$: $+ \half$ and $- \half$. In
addition, note that $| \Psi_{\pm} \rg$ and $| \Phi_{\pm} \rg$ are indeed maximally entangled pure states since their associated
reduced density matrices present the same value $\half \hat{\mathds{I}}_{2}$, that is, the states $| 0_{1(2)} \rg$ and $| 1_{1(2)}
\rg$ are equally likely to be found with the same probability $\half$. In this regard, the discrete Wigner functions established
in our analysis possess an important rule: they permit to show that the relations\ftn{Originally, these relations for the discrete
Wigner functions are associated with the two-qubit density-matrix decomposition $\ro = \ro^{(1)} \otimes \ro^{(2)}+\hat{\Delta}$,
such that $\tr_{1} [ \hat{\Delta} ] = \tr_{2} [ \Delta ] = 0 \, \hat{\mathds{I}}_{2}$, where $\hat{\Delta}$ contains, in
principle, all the possible classical and quantum correlations admitted by $\ro^{(1)}$ and $\ro^{(2)}$ \cite{Baio2019}.}
\bd
\Delta_{\Psi_{\pm}}(\mu_{1},\nu_{1},\mu_{2},\nu_{2}) = W_{\Psi_{\pm}}(\mu_{1},\nu_{1},\mu_{2},\nu_{2}) - 
\mathcal{W}_{\mathtt{R},\Psi_{\pm}}(\mu_{1},\nu_{1}) \mathcal{W}_{\mathtt{R},\Psi_{\pm}}(\mu_{2},\nu_{2}) = - \frac{3}{4} \;
\mbox{or} \; \frac{1}{4}
\ed
and 
\bd
\Delta_{\Phi_{\pm}}(\mu_{1},\nu_{1},\mu_{2},\nu_{2}) = W_{\Phi_{\pm}}(\mu_{1},\nu_{1},\mu_{2},\nu_{2}) - 
\mathcal{W}_{\mathtt{R},\Phi_{\pm}}(\mu_{1},\nu_{1}) \mathcal{W}_{\mathtt{R},\Phi_{\pm}}(\mu_{2},\nu_{2}) = - \frac{3}{4} \;
\mbox{or} \; \frac{1}{4}
\ed
prevail for all the finite-dimensional discrete phase space, which represent a genuine signature of the states under scrutiny
\cite{Galvao}.

As an interesting second example, let us consider the Werner states \cite{Werner}
\be
\lb{eq21}
\ro_{\mathtt{W}} = \mathcal{F} \ro_{\Psi_{-}} + \frac{1-\mathcal{F}}{3} \bigl( \ro_{\Psi_{+}} + \ro_{\Phi_{+}} + \ro_{\Phi_{-}} 
\bigr) \qquad ( 0 \leq \mathcal{F} \leq 1 )
\ee
which are defined as mixtures of the Bell states, where the parameter $\mathcal{F}$ determines the degree of mixing for these
states. Substituting Eqs. (\ref{eq17}) and (\ref{eq18}) in this definition, it is quite easy to show that
\be
\lb{eq22}
\ro_{\mathtt{W}} = \frac{1}{4} \lbk \hat{\mathds{I}}_{4} + \frac{1-4\mathcal{F}}{3} \lpar \hat{\sigma}_{x}^{(1)} \otimes
\hat{\sigma}_{x}^{(2)} + \hat{\sigma}_{y}^{(1)} \otimes \hat{\sigma}_{y}^{(2)} + \hat{\sigma}_{z}^{(1)} \otimes 
\hat{\sigma}_{z}^{(2)} \rpar \rbk ,
\ee
whose reduced density matrices coincide with $\half \hat{\mathds{I}}_{2}$ and do not depend on $\mathcal{F}$. So, the expression
for $\tr [ \ro_{\mathtt{W}}^{2} ] = \frac{1}{3} ( 1 - 2 \mathcal{F} + 4 \mathcal{F}^{2} ) \leq 1$ leads us, in principle, to 
characterize this state as follows:
\begin{itemize}
\item $\mathcal{F} = 1$: $\ro_{\mathtt{W}}$ describes a maximally entangled pure state;
\item $\half < \mathcal{F} < 1$: according to the Peres-Horodecki criterion \cite{Peres1996,HHH1996}, $\ro_{\mathtt{W}}$ basically
depicts entangled mixed states;
\item $\mathcal{F} = \half$: $\ro_{\mathtt{W}}$ corresponds to a separable mixed state \cite{Vedral}; and finally,
\item $\mathcal{F} = \frac{1}{4}$: the states $| 0_{1} 0_{2} \rg$, $| 0_{1} 1_{2} \rg$, $| 1_{1} 0_{2} \rg$, and $| 1_{1} 1_{2}
\rg$ are equally likely to be found (in this case, with the same probability $\frac{1}{4}$), since $\ro_{\mathtt{W}}$ is given
by $\frac{1}{4} \hat{\mathds{I}}_{4}$.
\end{itemize}
With respect to its discrete Wigner function
\be
\lb{eq23}
W_{\mathtt{W}}(\mu_{1},\nu_{1},\mu_{2},\nu_{2}) = \frac{1}{4} \lbr 1 + \frac{1-4\mathcal{F}}{3} \lbk (-1)^{\mu_{1} + \mu_{2}} +
(-1)^{\mu_{1} + \nu_{1} + \mu_{2} + \nu_{2}} + (-1)^{\nu_{1} + \nu_{2}} \rbk \rbr ,
\ee
it assumes two distinct values, $\frac{1}{6} + \frac{\mathcal{F}}{3}$ and $\half - \mathcal{F}$, which change according to
$\mathcal{F} \in [0,1]$ and $0 \leq \mu_{1},\nu_{1},\mu_{2},\nu_{2} \leq 1$: for instance, $\mathcal{F}=0$ gives 
$W_{\mathtt{W}} = \frac{1}{6}$ or $\half$, while $\mathcal{F}=1$ leads to $W_{\mathtt{W}} = - \half$ or $\half$; however, if one
considers $\mathcal{F}=\half$, we obtain $W_{\mathtt{W}} = 0$ or $\frac{1}{3}$. Moreover, it is also important to observe that
$\Delta_{\mathtt{W}}(\mu_{1},\nu_{1},\mu_{2},\nu_{2})$ admits the following situations: 
\begin{itemize}
\item $\mathcal{F}=1$: $\Delta_{\mathtt{W}} = - \frac{3}{4}$ or $\frac{1}{4}$ (see previous example);
\item $\mathcal{F}=\half$: $\Delta_{\mathtt{W}} = - \frac{1}{4}$ or $\frac{1}{12}$ (separable mixed state);
\item $\mathcal{F}=\frac{1}{4}$: $\Delta_{\mathtt{W}}=0$ ($W_{\mathtt{W}}$ constant and equal to $\frac{1}{4}$);
\item $\mathcal{F}=0$: $\Delta_{\mathtt{W}} = - \frac{1}{12}$ or $\frac{1}{4}$; otherwise, $\Delta_{\mathtt{W}} =
\frac{\mathcal{F}}{3} - \frac{1}{12}$ or $\frac{1}{4} - \mathcal{F}$. 
\end{itemize}
Further results related to the Werner states in connection with its local and nonlocal properties can be promptly found in Ref.
\cite{IH2000}.

Similar analysis via discrete Wigner functions can also be applied to the isotropic states \cite{HH1999} and maximally entangled
mixed states \cite{IH-2,VAM2001,MJWK2001,Wei2003}, or even in the investigation on the connection between entangled states and the
closest disentangled states \cite{Is2003}. On the other hand, we left behind the strong visual appeal associated with the discrete
Wigner functions in this Klein's group approach, at the cost of obtaining a theoretical framework completely compatible with the
Fano's prescription for two-qubit density matrix (\ref{eq9}). This disadvantage does not represent a complicated problem for our 
considerations, since it can be apparently solved through the use of Eq. (\ref{eq7}) and the generators of $\mathrm{SU(4)}$. 

From a practical point of view, Nuclear Magnetic Ressonance (NMR) actually corresponds to one of the possible experimental
tecniques \cite{NMR} which can be used for reconstructing both the density matrices (\ref{eq9}) and (\ref{eq15}) associated with
two-qubit states and consequently, to directly determine the respective discrete Wigner functions (\ref{eq13}) and (\ref{eq16}).
Recently, Micadei \textit{et al.} \cite{NC2019} conducted an important experimental investigation on the reversal of heat flow
between two initially quantum-correlated qubits prepared in local thermal states at different temperatures, basically employing
the aforementioned experimental technique. In fact, the authors employed the quantum-state tomography \cite{NMR} in order to
reconstruct the global two-qubit density matrix and then calculate the changes of internal energies of each qubit during the 
thermal contact. In this experimental approach, we argue that discrete Wigner functions should be used as an effective theoretical
tool to monitorate the pre-existing correlations between both the qubits.

\subsection{The group $\mathrm{SU(4)}$}

Initially, let us consider the discrete Wigner function (\ref{eq7}) with $N=4$ fixed, which is equivalent to construct the 
function
\be
\lb{eq24}
W(\mu,\nu) = \frac{1}{4} + \half \sum_{i} \lg \hat{g}_{i} \rg \lpar \hat{g}_{i} \rpar \! (\mu,\nu) \qquad (i=1,\ldots,15)
\ee
for a given four-level quantum system, which is described by the density matrix $\vro \in \mathscr{L}_{+,1}(\mathcal{H}_{4})$
explicitly presented in appendix A. Note that $\{ \hat{g}_{i} \}_{i=1,\ldots,15}$ denote the generators of the special unitary 
group $\mathrm{SU(4)}$ and represent the building blocks of this fundamental process. In addition, $\lg \hat{g}_{i} \rg$ and 
$\lpar \hat{g}_{i} \rpar \! (\mu,\nu)$ were previously defined in section \ref{s2} and promptly calculated in appendix A. These
results provide a completely general discrete Wigner function for $\mathrm{SU(4)}$, that is
\brr
\lb{eq25}
W(\mu,\nu) &=& \frac{1}{4} + \frac{1}{4} \lpar 3 \delta_{\mu,0}^{[4]} - \delta_{\mu,1}^{[4]} - \delta_{\mu,2}^{[4]} -
\delta_{\mu,3}^{[4]} \rpar \varrho_{11} - \frac{1}{4} \lpar \delta_{\mu,0}^{[4]} - 3 \delta_{\mu,1}^{[4]} + \delta_{\mu,2}^{[4]} +
\delta_{\mu,3}^{[4]} \rpar \varrho_{22} \nn \\
& & - \, \frac{1}{4} \lpar \delta_{\mu,0}^{[4]} + \delta_{\mu,1}^{[4]} - 3 \delta_{\mu,2}^{[4]} + \delta_{\mu,3}^{[4]} \rpar 
\varrho_{33} - \frac{1}{4} \lpar \delta_{\mu,0}^{[4]} + \delta_{\mu,1}^{[4]} + \delta_{\mu,2}^{[4]} - 3 \delta_{\mu,3}^{[4]} \rpar
\varrho_{44} \nn \\
& & + \, \half \frac{\sin \lbk \lpar \mu - \half \rpar \pi \rbk}{\sin \lbk \lpar \mu - \half \rpar \frac{\pi}{4} \rbk} \lbk \cos 
\lpar \frac{\nu \pi}{2} \rpar \re ( \varrho_{12} ) - \sin \lpar \frac{\nu \pi}{2} \rpar \ima ( \varrho_{12} ) \rbk \nn \\
& & + \, 2 \delta_{\mu,1}^{[4]} \lbk \cos ( \nu \pi ) \re ( \varrho_{13} ) - \sin ( \nu \pi ) \ima ( \varrho_{13} ) \rbk \nn \\
& & + \, \half \frac{\sin \lbk \lpar \mu - \frac{3}{2} \rpar \pi \rbk}{\sin \lbk \lpar \mu - \frac{3}{2} \rpar \frac{\pi}{4} \rbk}
(-1)^{\nu} \lbk \cos \lpar \frac{\nu \pi}{2} \rpar \re ( \varrho_{14} ) - \sin \lpar \frac{\nu \pi}{2} \rpar \ima ( \varrho_{14} )
\rbk \nn \\
& & + \, \half \frac{\sin \lbk \lpar \mu - \frac{3}{2} \rpar \pi \rbk}{\sin \lbk \lpar \mu - \frac{3}{2} \rpar \frac{\pi}{4} \rbk}
\lbk \cos \lpar \frac{\nu \pi}{2} \rpar \re ( \varrho_{23} ) - \sin \lpar \frac{\nu \pi}{2} \rpar \ima ( \varrho_{23} ) \rbk 
\nn \\
& & + \, 2 \delta_{\mu,2}^{[4]} \lbk \cos ( \nu \pi ) \re ( \varrho_{24} ) - \sin ( \nu \pi ) \ima ( \varrho_{24} ) \rbk \nn \\
& & + \, \half \frac{\sin \lbk \lpar \mu - \frac{5}{2} \rpar \pi \rbk}{\sin \lbk \lpar \mu - \frac{5}{2} \rpar \frac{\pi}{4} \rbk}
\lbk \cos \lpar \frac{\nu \pi}{2} \rpar \re ( \varrho_{34} ) - \sin \lpar \frac{\nu \pi}{2} \rpar \ima ( \varrho_{34} ) \rbk , 
\err
where the superscript $[4]$ on the Kronecker deltas denotes that this function is different from zero when its labels are
$\mathit{mod}(4)$-congruent. Table \ref{tab-su4} presents all the possible values that Eq. (\ref{eq25}) assumes in the 
finite-dimensional discrete phase space.
\vspace*{4pt}
\begin{table}[!t]
\tcaption{All possible values of the discrete Wigner function (\ref{eq25}) in terms of the density-matrix elements associated 
with a four-level system for $0 \leq \mu,\nu \leq 3$. It is worth stressing that by means of quadrupolar NMR techniques applied on
four-level systems (or ququarts), the reconstruction process of this function is completely feasible from the experimental point
of view \cite{Gedik}.}
\centerline{\footnotesize\smalllineskip
\begin{tabular}{l c c c}
$\mu$ & $\nu$ & $W(\mu,\nu)$ \\
\hline
$0$ & $0$ & $\varrho_{11} + \sqrt{\frac{2+\sqrt{2}}{2}} \, \re ( \varrho_{12} ) - \sqrt{\frac{2-\sqrt{2}}{2}} \, 
\re ( \varrho_{14} + \varrho_{23} - \varrho_{34} )$ \\
$0$ & $1$ & $\varrho_{11} - \sqrt{\frac{2+\sqrt{2}}{2}} \, \ima ( \varrho_{12} ) - \sqrt{\frac{2-\sqrt{2}}{2}} \, 
\ima ( \varrho_{14} - \varrho_{23} + \varrho_{34} )$ \\
$0$ & $2$ & $\varrho_{11} - \sqrt{\frac{2+\sqrt{2}}{2}} \, \re ( \varrho_{12} ) + \sqrt{\frac{2-\sqrt{2}}{2}} \, 
\re ( \varrho_{14} + \varrho_{23} - \varrho_{34} )$ \\
$0$ & $3$ & $\varrho_{11} + \sqrt{\frac{2+\sqrt{2}}{2}} \, \ima ( \varrho_{12} ) + \sqrt{\frac{2-\sqrt{2}}{2}} \, 
\ima ( \varrho_{14} - \varrho_{23} - \varrho_{34} )$ \\
$1$ & $0$ & $\varrho_{22} + \sqrt{\frac{2+\sqrt{2}}{2}} \, \re ( \varrho_{12} + \varrho_{14} + \varrho_{23} ) + 
2 \, \re ( \varrho_{13} ) - \sqrt{\frac{2-\sqrt{2}}{2}} \, \re ( \varrho_{34} )$ \\
$1$ & $1$ & $\varrho_{22} - \sqrt{\frac{2+\sqrt{2}}{2}} \, \ima ( \varrho_{12} - \varrho_{14} + \varrho_{23} ) - 
2 \, \re ( \varrho_{13} ) + \sqrt{\frac{2-\sqrt{2}}{2}} \, \ima ( \varrho_{34} )$ \\
$1$ & $2$ & $\varrho_{22} - \sqrt{\frac{2+\sqrt{2}}{2}} \, \re ( \varrho_{12} + \varrho_{14} + \varrho_{23} ) + 
2 \, \re ( \varrho_{13} ) + \sqrt{\frac{2-\sqrt{2}}{2}} \, \re ( \varrho_{34} )$ \\
$1$ & $3$ & $\varrho_{22} + \sqrt{\frac{2+\sqrt{2}}{2}} \, \ima ( \varrho_{12} - \varrho_{14} + \varrho_{23} ) - 
2 \, \re ( \varrho_{13} ) - \sqrt{\frac{2-\sqrt{2}}{2}} \, \ima ( \varrho_{34} )$ \\
$2$ & $0$ & $\varrho_{33} - \sqrt{\frac{2-\sqrt{2}}{2}} \, \re ( \varrho_{12} ) + \sqrt{\frac{2+\sqrt{2}}{2}} \, 
\re ( \varrho_{14} + \varrho_{23} + \varrho_{34} ) + 2 \, \re ( \varrho_{24} )$ \\
$2$ & $1$ & $\varrho_{33} + \sqrt{\frac{2-\sqrt{2}}{2}} \, \ima ( \varrho_{12} ) + \sqrt{\frac{2+\sqrt{2}}{2}} \, 
\ima ( \varrho_{14} - \varrho_{23} - \varrho_{34} ) - 2 \, \re ( \varrho_{24} )$ \\
$2$ & $2$ & $\varrho_{33} + \sqrt{\frac{2-\sqrt{2}}{2}} \, \re ( \varrho_{12} ) - \sqrt{\frac{2+\sqrt{2}}{2}} \, 
\re ( \varrho_{14} + \varrho_{23} + \varrho_{34} ) + 2 \, \re ( \varrho_{24} )$ \\
$2$ & $3$ & $\varrho_{33} - \sqrt{\frac{2-\sqrt{2}}{2}} \, \ima ( \varrho_{12} ) - \sqrt{\frac{2+\sqrt{2}}{2}} \, 
\ima ( \varrho_{14} - \varrho_{23} - \varrho_{34} ) - 2 \, \re ( \varrho_{24} )$ \\
$3$ & $0$ & $\varrho_{44} + \sqrt{\frac{2-\sqrt{2}}{2}} \, \re ( \varrho_{12} - \varrho_{14} - \varrho_{23} ) + 
\sqrt{\frac{2+\sqrt{2}}{2}} \, \re ( \varrho_{34} )$ \\
$3$ & $1$ & $\varrho_{44} - \sqrt{\frac{2-\sqrt{2}}{2}} \, \ima ( \varrho_{12} + \varrho_{14} - \varrho_{23} ) - 
\sqrt{\frac{2+\sqrt{2}}{2}} \, \ima ( \varrho_{34} )$ \\
$3$ & $2$ & $\varrho_{44} - \sqrt{\frac{2-\sqrt{2}}{2}} \, \re ( \varrho_{12} - \varrho_{14} - \varrho_{23} ) - 
\sqrt{\frac{2+\sqrt{2}}{2}} \, \re ( \varrho_{34} )$ \\
$3$ & $3$ & $\varrho_{44} + \sqrt{\frac{2-\sqrt{2}}{2}} \, \ima ( \varrho_{12} + \varrho_{14} - \varrho_{23} ) + 
\sqrt{\frac{2+\sqrt{2}}{2}} \, \ima ( \varrho_{34} )$ \\
\hline \\
\end{tabular}}
\lb{tab-su4}
\end{table}

Recently, Gedik \textit{et al.} \cite{Gedik} showed that a single ququart is enough to implement an oracle based quantum algorithm
that solves a black-box problem faster than any classical algorithm. In this experimental approach, the main idea is to determine
the parity of a cyclic permutation of the elements $\{ | 0 \rg, | 1 \rg, | 2 \rg, | 3 \rg \}$ through a single evaluation of the 
permutation function using as initial state the ququart $| \psi_{1} \rg = \hat{\mathfrak{F}} | 1 \rg = \half \lpar | 0 \rg + \im |
1 \rg - | 2 \rg - \im | 3 \rg \rpar$, where
\bd
\hat{\mathfrak{F}} \col \half \lpar \begin{array}{cccc}
1 & 1    & 1  & 1 \\
1 & \im  & -1 & -\im \\
1 & -1   & 1  & -1 \\
1 & -\im & -1 & \im \\
\end{array} \rpar
\ed
denotes the discrete Fourier operator for $\mathrm{SU(4)}$ such that $\hat{\mathfrak{F}}^{4} = \hat{\mathds{I}}_{4}$ and
$\hat{\mathfrak{F}} \hat{\mathfrak{F}}^{\dagger} = \hat{\mathfrak{F}}^{\dagger} \hat{\mathfrak{F}} = \hat{\mathds{I}}_{4}$. Now,
let us discuss such experimental approach by means of the schematic diagram exhibited below:
\xymatrix{
& & \hat{\mathcal{U}}_{6} | \psi_{1} \rg \ar[r]^-{\hat{\mathfrak{F}}^{\dagger}} & \hat{\mathfrak{F}}^{\dagger} 
\hat{\mathcal{U}}_{6} | \psi_{1} \rg = \hat{\mathfrak{F}}^{\dagger} \hat{\mathcal{U}}_{6} \hat{\mathfrak{F}} | 1 \rg 
\ar@{|->>}[r] & | 3 \rg \lg 3 | \; (\texttt{negative}) \\
| 1 \rg \ar[r]^-{\hat{\mathfrak{F}}} & \hat{\mathfrak{F}} | 1 \rg \equiv | \psi_{1} \rg \ar[ru]^-{\hat{\mathcal{U}}_{6}} 
\ar[rd]^-{\hat{\mathcal{U}}_{2}} & \\
& & \hat{\mathcal{U}}_{2} | \psi_{1} \rg \ar[r]^-{\hat{\mathfrak{F}}^{\dagger}} & \hat{\mathfrak{F}}^{\dagger} 
\hat{\mathcal{U}}_{2} | \psi_{1} \rg = \hat{\mathfrak{F}}^{\dagger} \hat{\mathcal{U}}_{2} \hat{\mathfrak{F}} | 1 \rg 
\ar@{|->>}[r] & | 1 \rg \lg 1 | \; (\texttt{positive}) }
\begin{description}
\item[step 1.] Basically, this first step is responsible for creating the initial state $| 1 \rg$;
\item[step 2.] the next one applies the discrete Fourier operator $\hat{\mathfrak{F}}$ upon the initial state $| 1 \rg$ in order
to obtain $| \psi_{1} \rg = \hat{\mathfrak{F}} | 1 \rg$; following,
\item[step 3.] two different pulses are then applied on $| \psi_{1} \rg$ with the aim of producing the unitary matrices
\bd
\hat{\mathcal{U}}_{2} = \lpar \begin{array}{cccc}
0 & 0 & 0 & 1 \\
1 & 0 & 0 & 0 \\
0 & 1 & 0 & 0 \\
0 & 0 & 1 & 0 \\
\end{array} \rpar \qquad \mbox{and} \qquad
\hat{\mathcal{U}}_{6} = \lpar \begin{array}{cccc}
0 & 0 & 1 & 0 \\
0 & 1 & 0 & 0 \\
1 & 0 & 0 & 0 \\
0 & 0 & 0 & 1 \\
\end{array} \rpar ,
\ed
i.e., for $\hat{\mathcal{U}}_{2} | \psi_{1} \rg$ we obtain $- \im | \psi_{1} \rg$, while $\hat{\mathcal{U}}_{6} | \psi_{1} \rg$
gives $- \half \lpar | 0 \rg - \im | 1 \rg - | 2 \rg + \im | 3 \rg \rpar$;
\item[step 4.] as a subsequent step we apply, once again, the operator $\hat{\mathfrak{F}}^{\dagger}$ on each one of the resulting
states, which implies in $\hat{\mathfrak{F}}^{\dagger} \hat{\mathcal{U}}_{2} | \psi_{1} \rg = - \im | 1 \rg$ and
$\hat{\mathfrak{F}}^{\dagger} \hat{\mathcal{U}}_{6} | \psi_{1} \rg = - | 3 \rg$; finally,
\item[step 5.] we measure both the possibilities through its respective density matrices. In particular, the measurements of 
$| 1 \rg \lg 1|$ and $| 3 \rg \lg 3 |$ correspond to the positive and negative cyclic permutations.
\end{description}
In fact, all the measurements performed in the experiment are associated with the tomographic reconstruction of the density
matrix for each aforementioned step with errors always smaller than $6\%$. In order to illustrate part of these experimental 
steps, we have performed numerical calculations that lead us to show three-dimensional plots of the discrete Wigner function
(\ref{eq25}) for all the range $0 \leq \mu,\nu \leq 3$ -- see Figure \ref{fig1}. These pictures clearly demonstrate the relevant
role of discrete $\mathrm{SU(4)}$ Wigner functions in the comprehension of the physical processes involved in each step: for
instance, the effects of unitary transformations $\hat{\mathfrak{F}}$ and $\hat{\mathcal{U}}_{6}$ on the respective states
$| 1 \rg$ and $| \psi_{1} \rg$ are quite significant and visually different, i.e., the first unitary operation shuffles the
state $| 1 \rg$ (thereby producing state $| \psi_{1} \rg$) -- see picture (b) -- while the second one promotes the exchange of
states by means of displacements in the discrete phase space -- see picture (c). In this context, these functions are not mere
figurative mathematical tools, but rather valuable theoretical instruments that allow to increase our knowledge on the physical
processes involved. 
\begin{figure}[t]
\begin{minipage}[b]{0.4\linewidth}
\includegraphics[width=\linewidth]{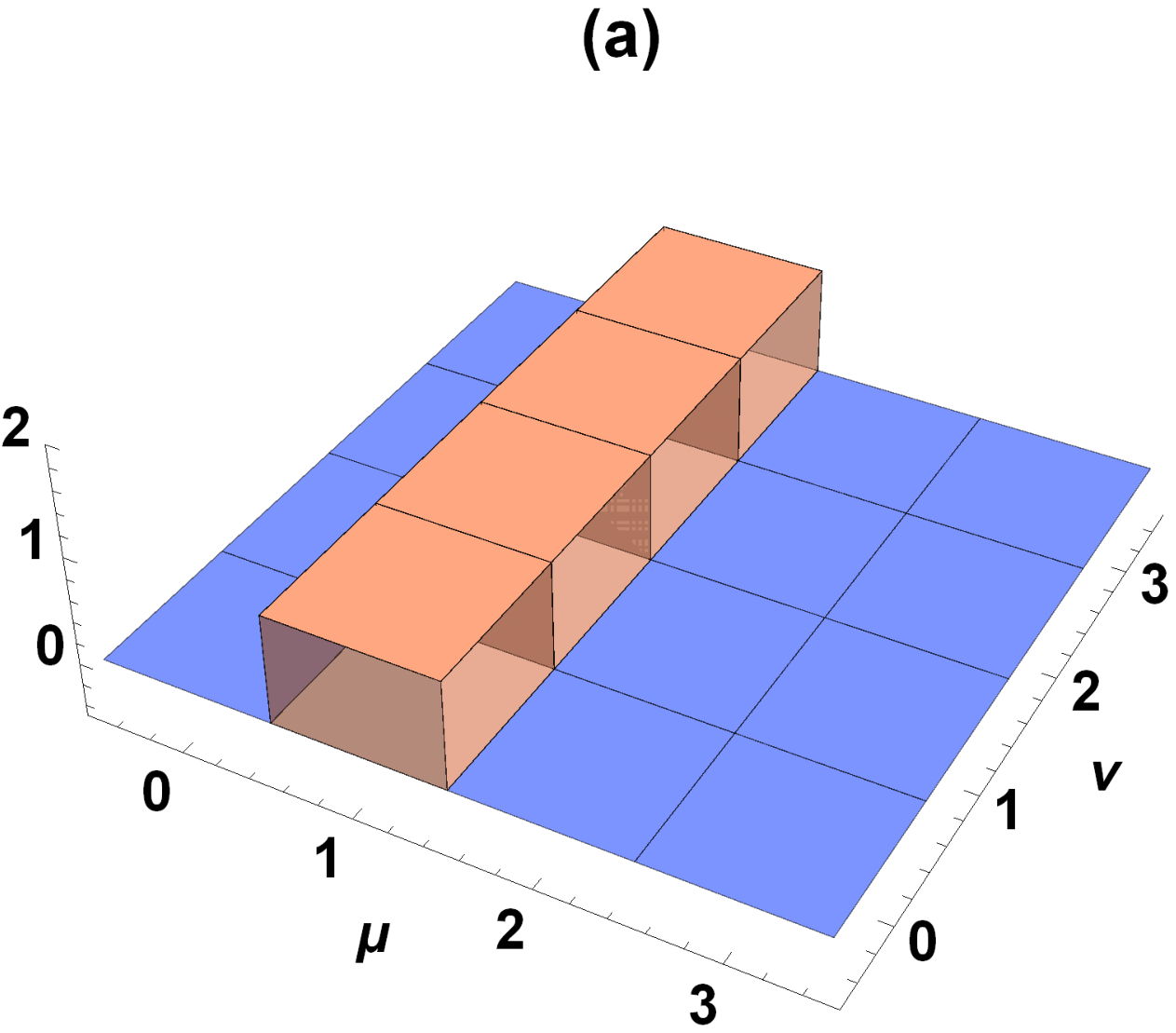}
\end{minipage} \hfill
\begin{minipage}[b]{0.4\linewidth}
\includegraphics[width=\linewidth]{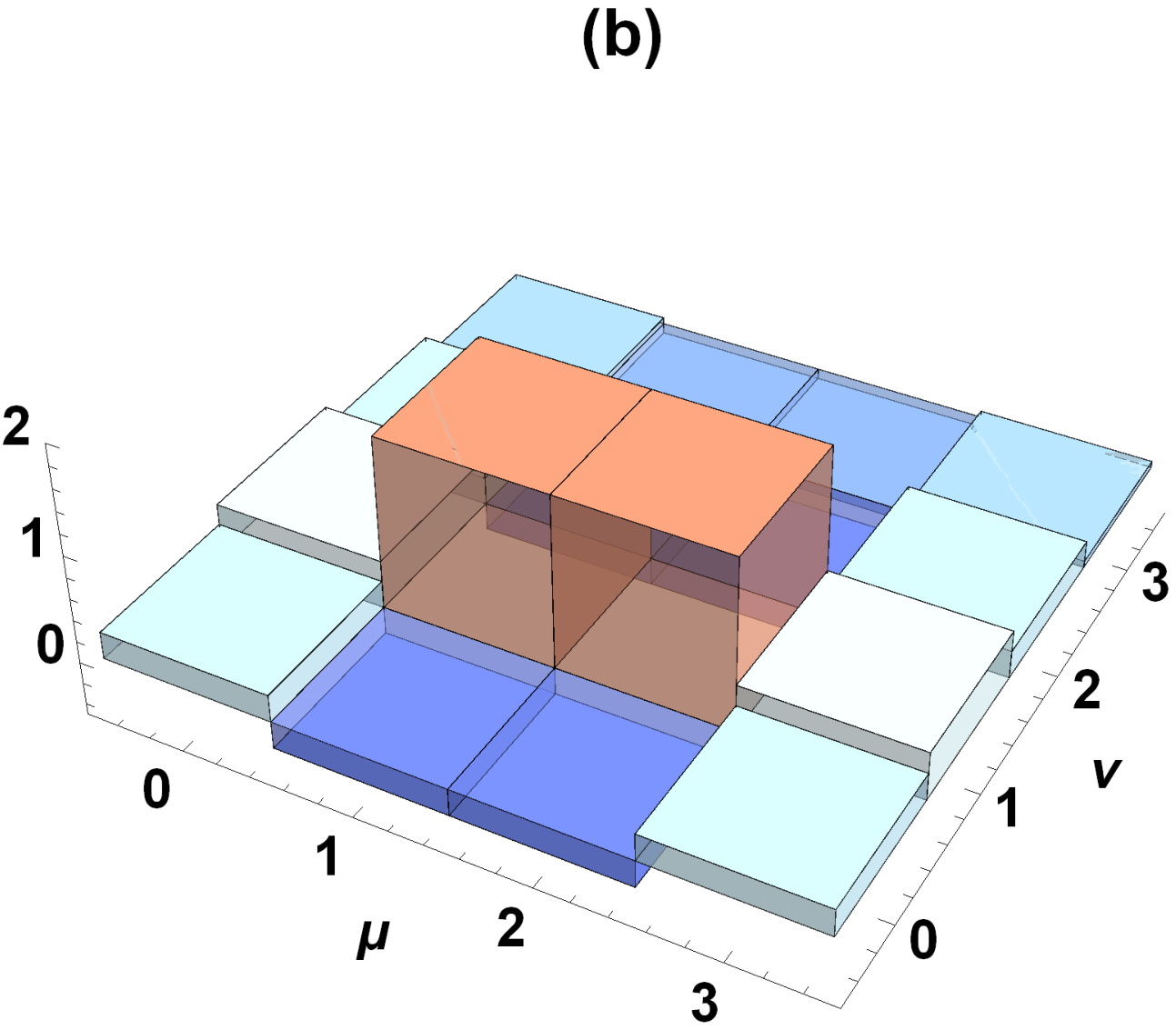}
\end{minipage} \hfill
\begin{minipage}[b]{0.4\linewidth}
\includegraphics[width=\linewidth]{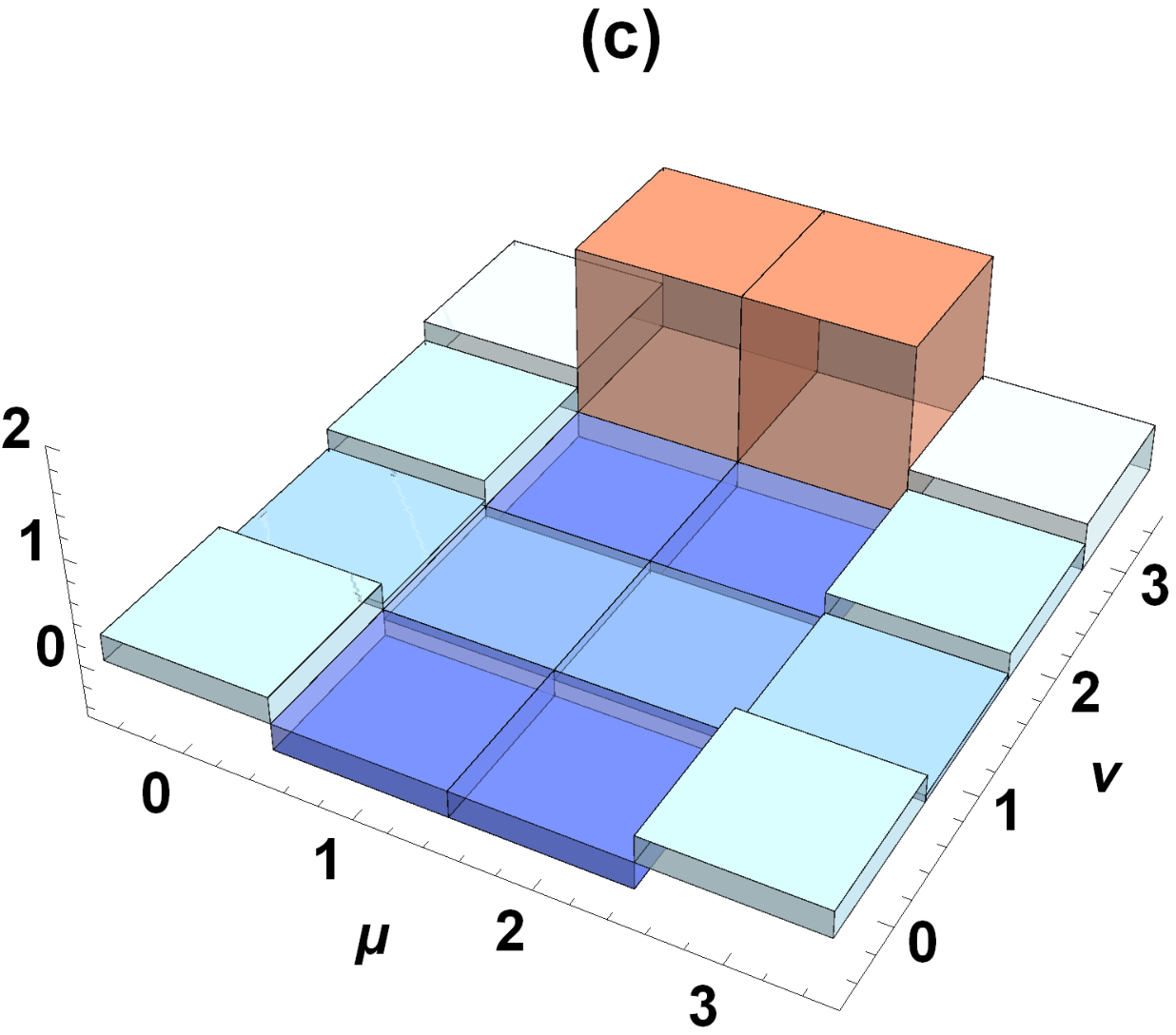}
\end{minipage} \hfill
\begin{minipage}[b]{0.4\linewidth}
\includegraphics[width=\linewidth]{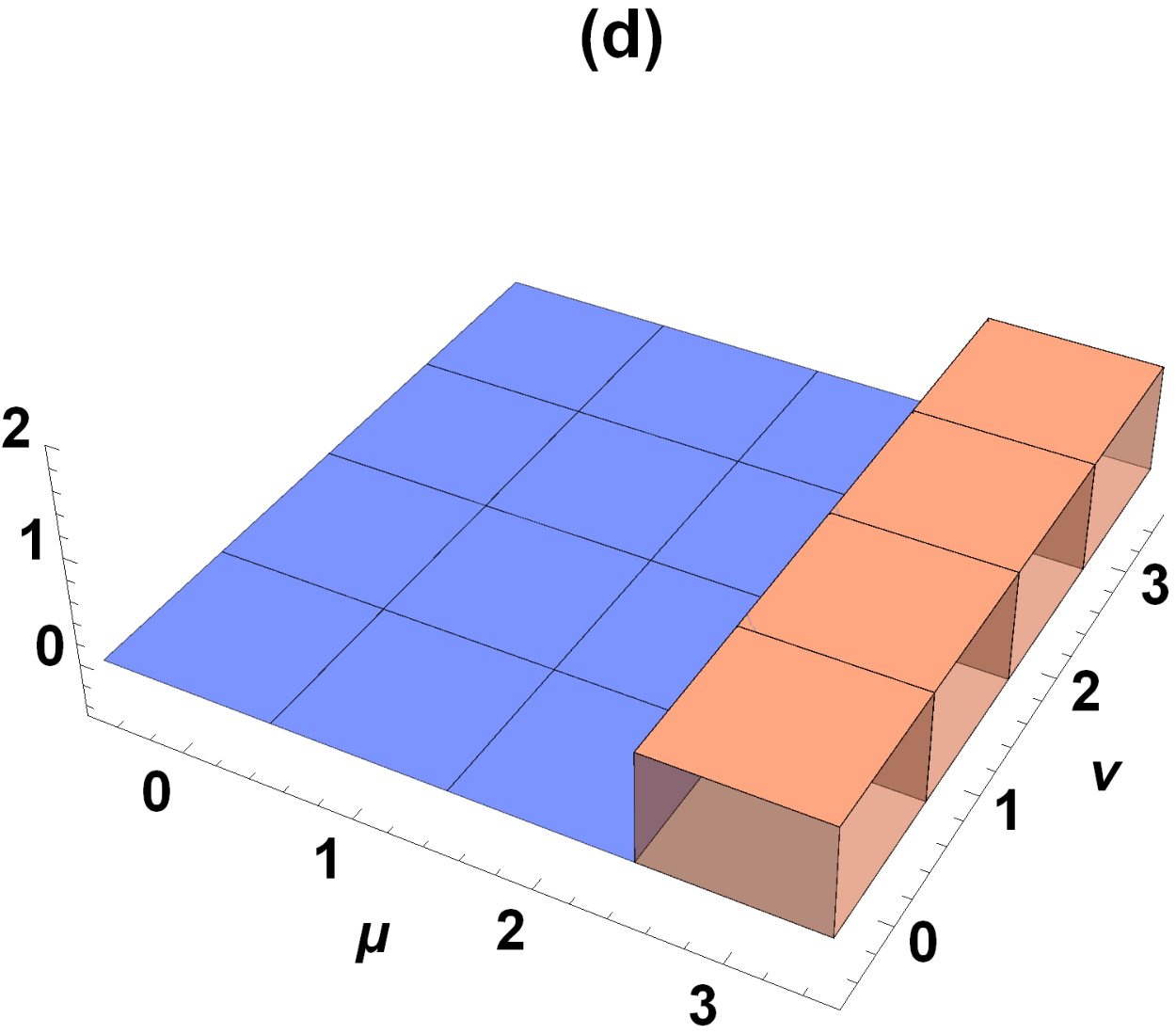}
\end{minipage}
\vspace*{13pt}
\fcaption{\lb{fig1}
Three-dimensional plots of the discrete $\mathrm{SU(4)}$ Wigner function associated with a significant part of the NMR experiment
performed by Gedik and coworkers \cite{Gedik}. Picture (a) represents the state $| 1 \rg$ and corresponds to the first step of
that experiment; (b) depicts the second step, when the state $| \psi_{1} \rg$ is achieved through the acting of discrete Fourier
operator $\hat{\mathfrak{F}}$ upon the initial state; (c) consists of applying the pulse $\hat{\mathcal{U}}_{6}$ on the state 
$| \psi_{1} \rg$ (compares with the previous picture); and finally, (d) describes the action of $\hat{\mathfrak{F}}^{\dagger}$
upon $\hat{\mathcal{U}}_{6} | \psi_{1} \rg$ and subsequent measurement -- see steps 4 and 5. As the pulse $\hat{\mathcal{U}}_{2}$
upon $| \psi_{1} \rg$ gives $- \im | \psi_{1} \rg$, this part of the experiment was not considered in the numerical calculations.}
\end{figure}

Nowadays, there are different experimental arrangements which produce ququarts of photons \cite{Nagali} as well as two entangled
qudits \cite{Kues} (each qudit encodes a ten-dimensional state), with the objective of performing information-processing tasks
related to quantum information theory. Moreover, Hu and coworkers \cite{Guo} used a path-polarization hybrid system to generate
high-dimensional entangled states (in this case, two entangled ququarts with high quality) in order to beat the channel capacity
limit for superdense coding. These different experimental frameworks represent an interesting scenario for future research in
theoretical physics where the discrete $\mathrm{SU(N)}$ Wigner function takes place.

\subsubsection{Change of basis}

Let us consider once again the Fano's prescription for the density matrix $\ro$ and its corresponding discrete Wigner function
(\ref{eq13}), to solve an apparent difficulty associated with the visualization of this function in the finite-dimensional 
discrete phase space. For such a particular task, let us adopt the isomorphic correspondence between ququart and two-qubit states,
in according to the theoretical prescription established in Refs. \cite{Resh,Manko}, namely, in this case an isomorphic
correspondence between the stationary energy states of a four-level system and the two-qubit computational basis: $| 0 \rg 
\leftrightarrow | 0_{1} 0_{2} \rg$, $| 1 \rg \leftrightarrow | 0_{1} 1_{2} \rg$, $| 2 \rg \leftrightarrow | 1_{1} 0_{2} \rg$, and
$| 3 \rg \leftrightarrow | 1 _{1} 1_{2} \rg$. Therefore, the task consists in establishing a connection between Eqs. (\ref{eq2})
for $N=4$ and (\ref{eq9}) under specific conditions; otherwise, this correspondence $\ro \leftrightarrow \hat{\varrho}$ must be
clearly stated.

In this sense, let us perform a change of basis in (\ref{eq9}) through the auxiliary results
\brr
& & \hat{\sigma}_{x}^{(1)} \otimes \hat{\mathds{I}}_{2}^{(2)} = \hat{g}_{4} + \hat{g}_{11} , \quad
\hat{\sigma}_{y}^{(1)} \otimes \hat{\mathds{I}}_{2}^{(2)} = \hat{g}_{5} + \hat{g}_{12} , \quad
\hat{\sigma}_{z}^{(1)} \otimes \hat{\mathds{I}}_{2}^{(2)} = \frac{2}{\sqrt{3}} \hat{g}_{8} + \frac{2}{\sqrt{6}} \hat{g}_{15} , 
\nn \\
& & \hat{\mathds{I}}_{2}^{(1)} \otimes \hat{\sigma}_{x}^{(2)} = \hat{g}_{1} + \hat{g}_{13} , \quad
\hat{\mathds{I}}_{2}^{(1)} \otimes \hat{\sigma}_{y}^{(2)} = \hat{g}_{2} + \hat{g}_{14} , \quad 
\hat{\mathds{I}}_{2}^{(1)} \otimes \hat{\sigma}_{z}^{(2)} = \hat{g}_{3} - \frac{1}{\sqrt{3}} \hat{g}_{8} + \frac{2}{\sqrt{6}}
\hat{g}_{15} , \nn \\
& & \hat{\sigma}_{x}^{(1)} \otimes \hat{\sigma}_{x}^{(2)} = \hat{g}_{6} + \hat{g}_{9} , \quad
\hat{\sigma}_{x}^{(1)} \otimes \hat{\sigma}_{y}^{(2)} = - \hat{g}_{7} + \hat{g}_{10} , \quad
\hat{\sigma}_{x}^{(1)} \otimes \hat{\sigma}_{z}^{(2)} = \hat{g}_{4} - \hat{g}_{11} , \nn \\
& & \hat{\sigma}_{y}^{(1)} \otimes \hat{\sigma}_{x}^{(2)} = \hat{g}_{7} + \hat{g}_{10} , \quad
\hat{\sigma}_{y}^{(1)} \otimes \hat{\sigma}_{y}^{(2)} = \hat{g}_{6} - \hat{g}_{9} , \quad
\hat{\sigma}_{y}^{(1)} \otimes \hat{\sigma}_{z}^{(2)} = \hat{g}_{5} - \hat{g}_{12} , \nn \\
& & \hat{\sigma}_{z}^{(1)} \otimes \hat{\sigma}_{x}^{(2)} = \hat{g}_{1} - \hat{g}_{13} , \quad
\hat{\sigma}_{z}^{(1)} \otimes \hat{\sigma}_{y}^{(2)} = \hat{g}_{2} - \hat{g}_{14} , \quad
\hat{\sigma}_{z}^{(1)} \otimes \hat{\sigma}_{z}^{(2)} = \hat{g}_{3} + \frac{1}{\sqrt{3}} \hat{g}_{8} - \frac{2}{\sqrt{6}}
\hat{g}_{15} , \nn
\err
which were obtained with the help of Eq. (\ref{eq1}). In particular, these results permit us to rewrite (\ref{eq9}) in the compact
form
\be
\lb{eq26}
\ro = \frac{1}{4} \lpar \hat{\mathds{I}}_{4} + \sum_{i} \mathtt{C}_{i} \, \hat{g}_{i} \rpar ,
\ee
whose coefficients are given by
\brr
& & \mathtt{C}_{1} = b_{x} + c_{zx} , \; \mathtt{C}_{2} = b_{y} + c_{zy} , \; \mathtt{C}_{3} = b_{z} + c_{zz} , \;
\mathtt{C}_{4} = a_{x} + c_{xz} , \; \mathtt{C}_{5} = a_{y} + c_{yz} , \nn \\
& & \mathtt{C}_{6} = c_{xx} + c_{yy} , \; \mathtt{C}_{7} = - c_{xy} + c_{yx} , \; \mathtt{C}_{8} = \frac{2}{\sqrt{3}} a_{z} -
\frac{1}{\sqrt{3}} b_{z} + \frac{1}{\sqrt{3}} c_{zz} , \nn \\ 
& & \mathtt{C}_{9} = c_{xx} - c_{yy} , \; \mathtt{C}_{10} = c_{xy} + c_{yx} , \; \mathtt{C}_{11} = a_{x} - c_{xz} , \;
\mathtt{C}_{12} = a_{y} - c_{yz} , \nn \\
& & \mathtt{C}_{13} = b_{x} - c_{zx} , \; \mathtt{C}_{14} = b_{y} - c_{zy} , \; \mathtt{C}_{15} = \frac{2}{\sqrt{6}} a_{z} +
\frac{2}{\sqrt{6}} b_{z} - \frac{2}{\sqrt{6}} c_{zz} . \nn
\err
As expected, the matrix elements in the computational basis reproduce exactly those obtained in Eq. (\ref{eq15}), which leads us
to establish the correspondence with (\ref{four-level}).

\subsubsection{Revisiting two-qubit Bell and Werner states}
\begin{figure}[t]
\begin{minipage}[b]{0.4\linewidth}
\includegraphics[width=\linewidth]{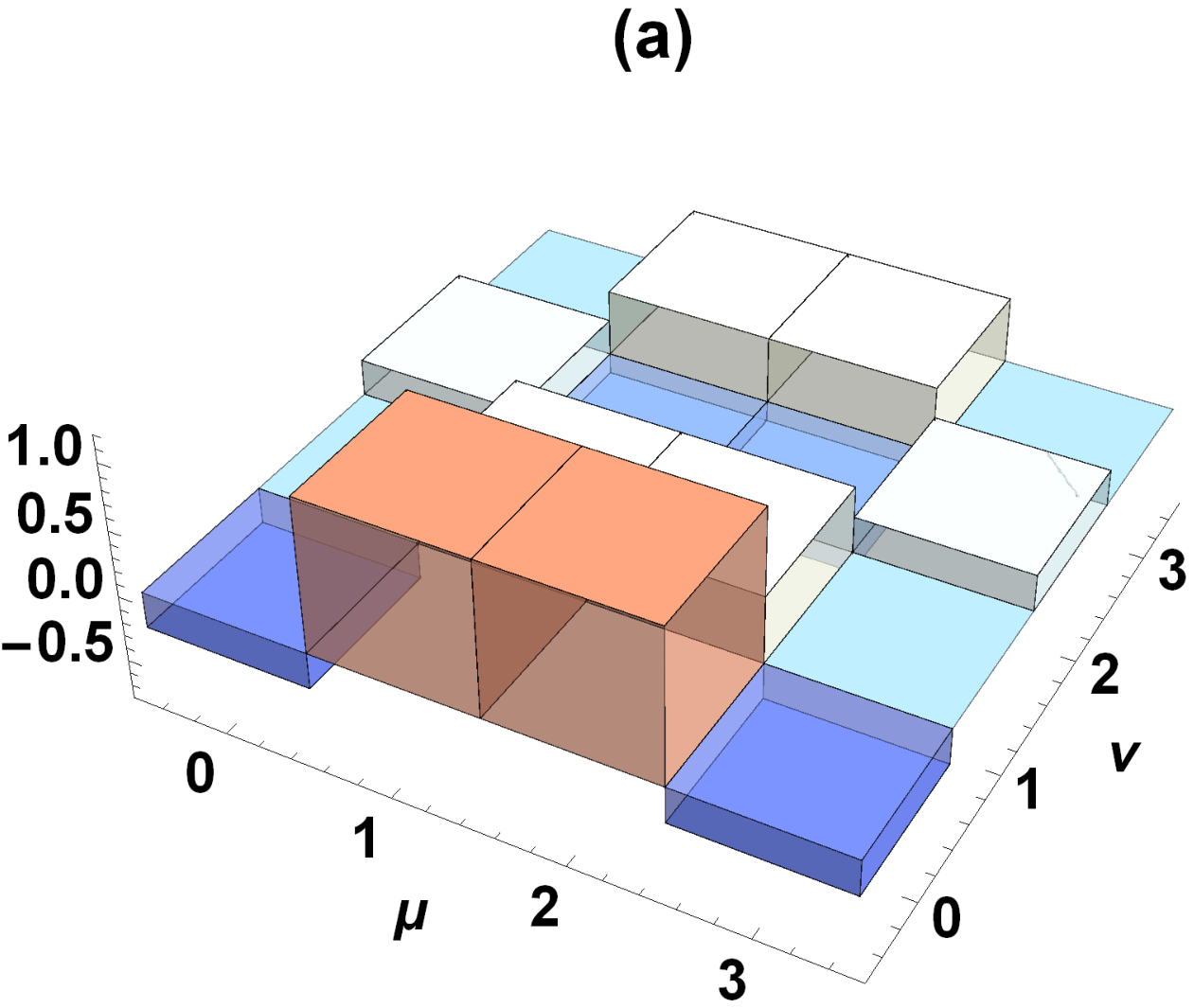}
\end{minipage} \hfill
\begin{minipage}[b]{0.4\linewidth}
\includegraphics[width=\linewidth]{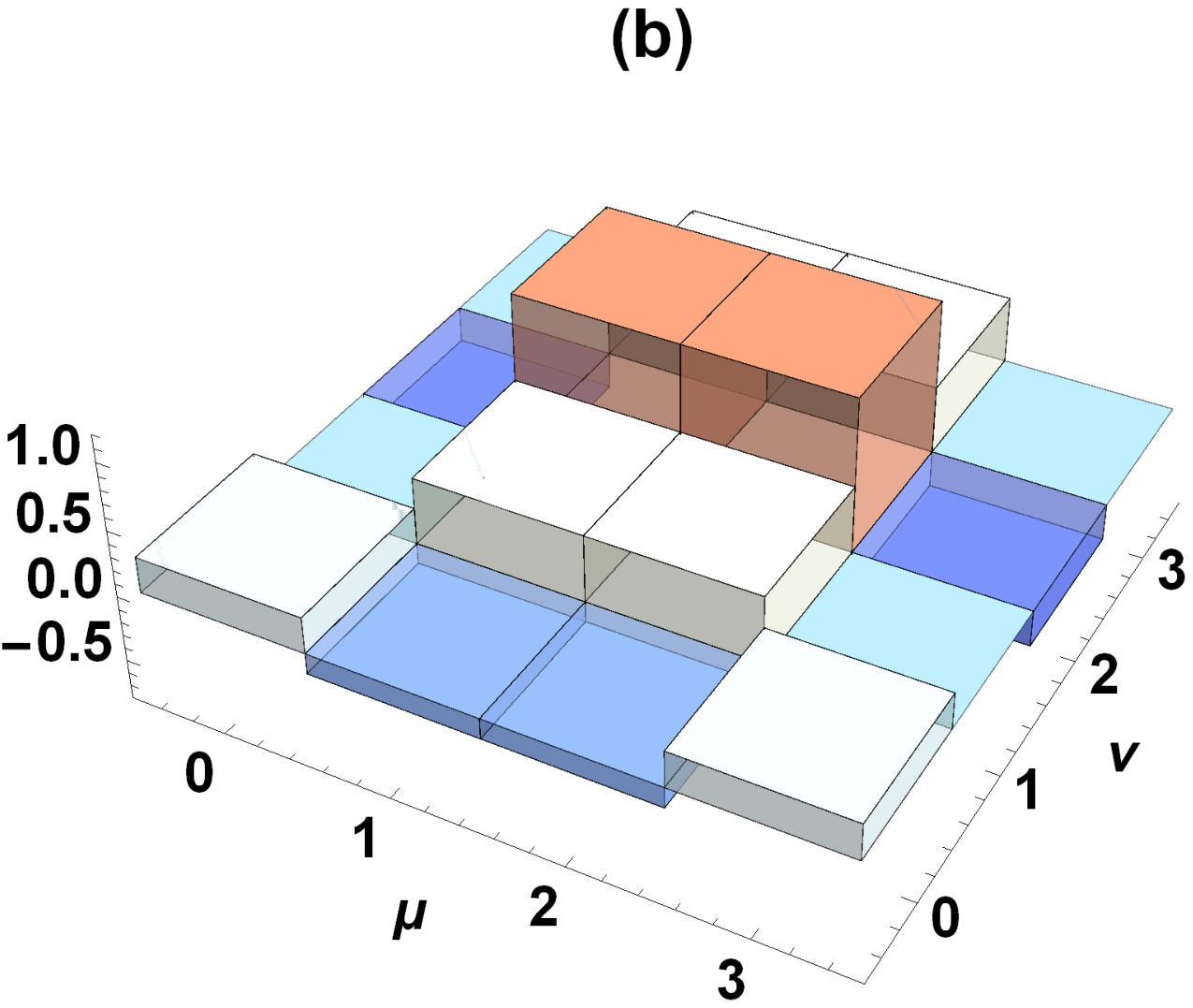}
\end{minipage} \hfill
\begin{minipage}[b]{0.4\linewidth}
\includegraphics[width=\linewidth]{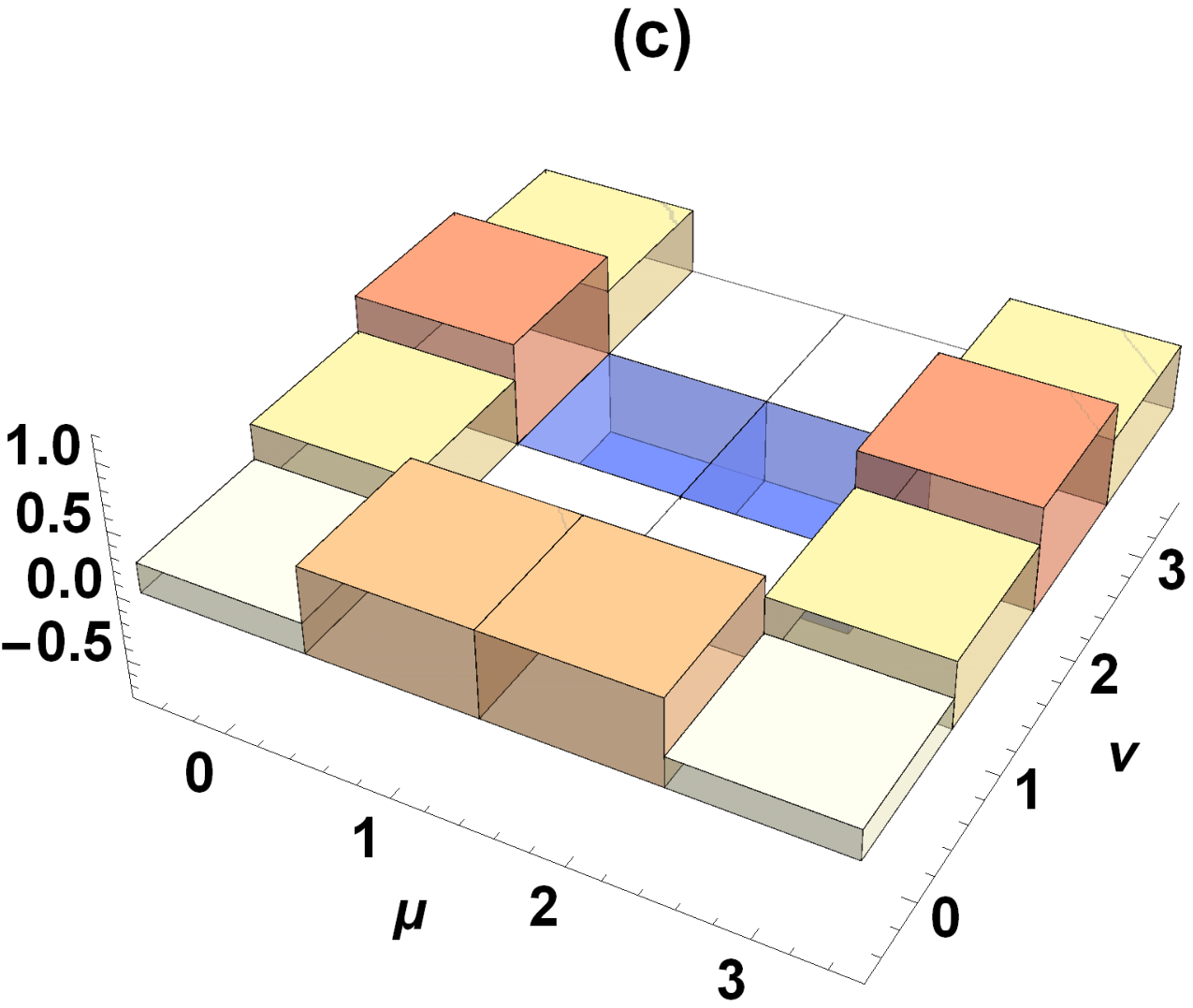}
\end{minipage} \hfill
\begin{minipage}[b]{0.4\linewidth}
\includegraphics[width=\linewidth]{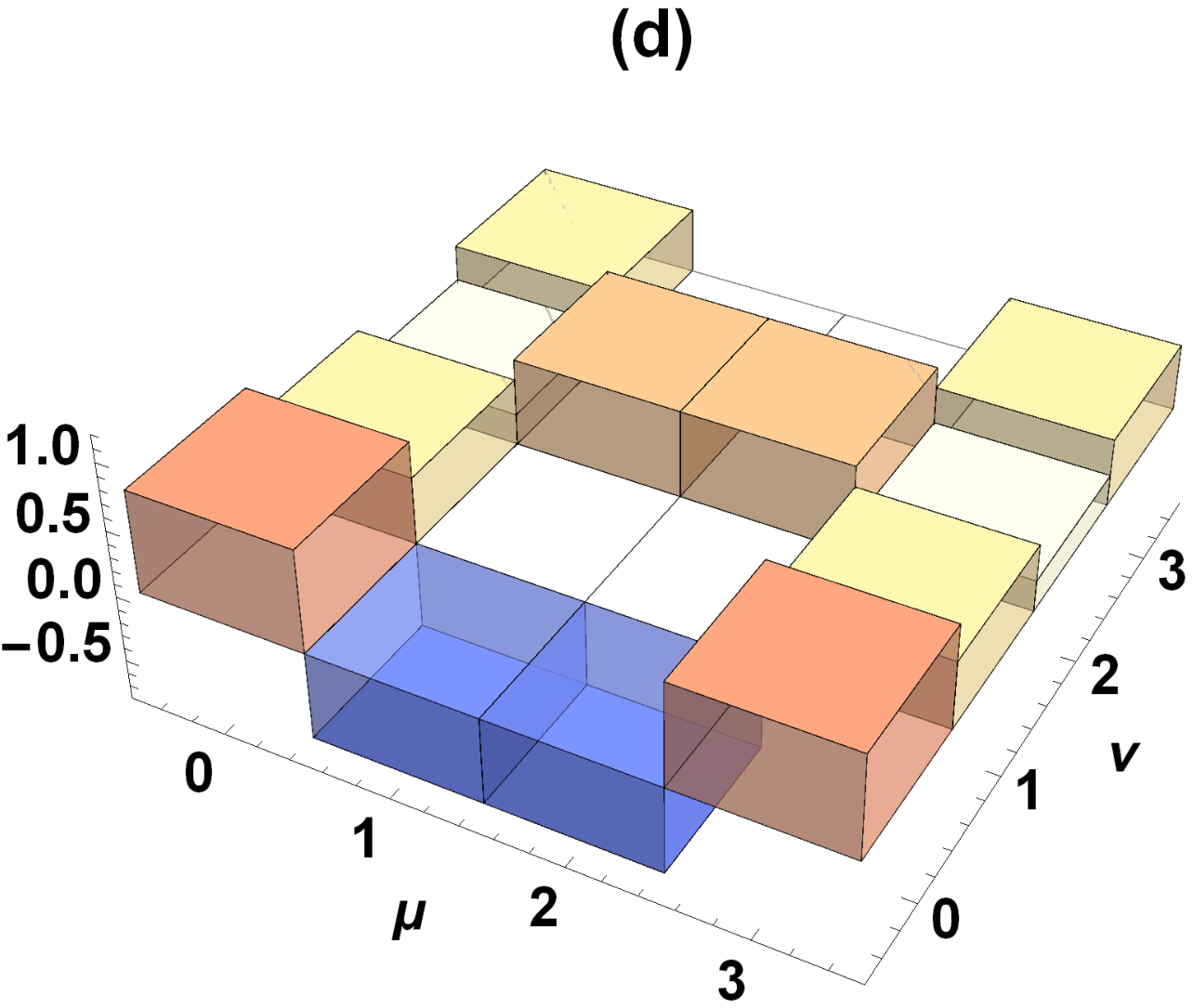}
\end{minipage}
\vspace*{13pt}
\fcaption{\lb{fig2}
Three-dimensional plots of the discrete $\mathrm{SU(4)}$ Wigner functions related to the Bell states $\Psi_{\pm}$ and $\Phi_{\pm}$
for $0 \leq \mu,\nu \leq 3$, namely, (a) and (b) describe the respective functions $W_{\Psi_{+}}$ and $W_{\Psi_{-}}$, while (c)
and (d) correspond to $W_{\Phi_{+}}$ and $W_{\Phi_{-}}$. Note that the quantum fluctuations observed in $\Psi_{\pm}$ and 
$\Phi_{\pm}$ present null contributions in all cases, which reinforce the quantum properties of being maximally entangled pure 
states.}
\end{figure}
As a first application, let us consider once again the density matrices (\ref{eq17}) and (\ref{eq18}) associated with the Bell
states $\Psi_{\pm}$ and $\Phi_{\pm}$, respectively. So, the corresponding discrete Wigner functions in these cases have the
forms 
\be
\lb{eq27}
W_{\Psi_{\pm}}(\mu,\nu) = \frac{1}{4} - \frac{1}{4} \lpar \delta_{\mu,0}^{[4]} - \delta_{\mu,1}^{[4]} - \delta_{\mu,2}^{[4]} +
\delta_{\mu,3}^{[4]} \rpar \pm \frac{1}{4} \frac{\sin \lbk \lpar \mu - \frac{3}{2} \rpar \pi \rbk}{\sin \lbk \lpar \mu - 
\frac{3}{2} \rpar \frac{\pi}{4} \rbk} \cos \lpar \frac{\nu \pi}{2} \rpar
\ee
and
\be
\lb{eq28}
W_{\Phi_{\pm}}(\mu,\nu) = \frac{1}{4} + \frac{1}{4} \lpar \delta_{\mu,0}^{[4]} - \delta_{\mu,1}^{[4]} - \delta_{\mu,2}^{[4]} +
\delta_{\mu,3}^{[4]} \rpar \pm \frac{1}{4} \frac{\sin \lbk \lpar \mu - \frac{3}{2} \rpar \pi \rbk}{\sin \lbk \lpar \mu - 
\frac{3}{2} \rpar \frac{\pi}{4} \rbk} (-1)^{\nu} \cos \lpar \frac{\nu \pi}{2} \rpar ,
\ee
whose three-dimensional representations in finite-dimensional discrete phase space were depicted by Fig. \ref{fig2} as follows:
(a) $W_{\Psi_{+}}(\mu,\nu)$, (b) $W_{\Psi_{-}}(\mu,\nu)$, (c) $W_{\Phi_{+}}(\mu,\nu)$, and (d) $W_{\Phi_{-}}(\mu,\nu)$. To start
with the analysis of the numerical results sketched in these pictures, let us consider the aforementioned correspondence between
two-qubit states and four-level systems with emphasis on two-qubit Bell states written in terms of the computational basis of a
ququart, 
\brr
& & \ro_{\Psi_{\pm}} = \half \lpar | 1 \rg \lg 1 | \pm | 1 \rg \lg 2 | \pm | 2 \rg \lg 1 | + | 2 \rg \lg 2 | \rpar \Rightarrow 
\half \lpar \begin{array}{cccc}
0 & 0     & 0     & 0 \\
0 & 1     & \pm 1 & 0 \\
0 & \pm 1 & 1     & 0 \\
0 & 0     & 0     & 0 \\
\end{array} \rpar , \nn \\
& & \ro_{\Phi_{\pm}} = \half \lpar | 0 \rg \lg 0 | \pm | 0 \rg \lg 3 | \pm | 3 \rg \lg 0 | + | 3 \rg \lg 3 | \rpar \Rightarrow
\half \lpar \begin{array}{cccc}
1     & 0 & 0 & \pm 1 \\
0     & 0 & 0 & 0 \\
0     & 0 & 0 & 0 \\
\pm 1 & 0 & 0 & 1 \\
\end{array} \rpar . \nn
\err
In this description, the states $\Psi_{\pm}$ correspond to a four-level system where only two levels are accessed: in such a case, 
the states $| 1 \rg$ and $| 2 \rg$ are equally populated in the ratio $\half$ with the same transition rates. So, the maximum
values observed in pictures (a) and (b) are coincident and equal to $\half + \half \sqrt{\frac{2+\sqrt{2}}{2}} \approx 1.153$,
while the minimum values assume $- \half \sqrt{\frac{2-\sqrt{2}}{2}} \approx - 0.271$. Table \ref{tab-su4} allows us to show that
contributions associated with $\mu=0,3$ fixed and $\nu=0,1,2,3$ have zero sum, namely, the non-accessed states $| 0 \rg$ and
$| 3 \rg$ present only quantum fluctuations; in addition, the contributions for $\mu=1,2$ fixed and $\nu=0,1,2,3$ exhibit non-zero
sum because they are directly connected in this case with the population and transition rates of the states $| 1 \rg$ and 
$| 2 \rg$. The negative signal present in $\Psi_{-}$ stands for an interchange between $\nu=0$ and $\nu=2$ for any $\mu=0,1,2,3$.
Similar analysis can also be applied in pictures (c) and (d), where now only the states $| 0 \rg$ and $| 3 \rg$ are accessed, with
the maximum and minimum values being given for $W_{\Phi_{\pm}}(\mu,\nu)$ by $\half + \half \sqrt{\frac{2-\sqrt{2}}{2}} \approx
0.771$ and $- \half \sqrt{\frac{2+\sqrt{2}}{2}} \approx - 0.653$. It is worth to observe the pronounced quantum fluctuations when
we deal with the non-accessed states $| 1 \rg$ and $| 2 \rg$ for $\mu=1,2$ fixed and $\nu=0,1,2,3$.

Let us now consider the Werner states (\ref{eq21}) written in terms of the $\mathrm{SU(4)}$ computational basis as follows:
\bd
\ro_{\mathtt{W}} = \frac{1-\mathcal{F}}{3} \lpar | 0 \rg \lg 0 | + | 3 \rg \lg 3 | \rpar + \frac{1+2\mathcal{F}}{6} \lpar | 1 \rg
\lg 1 | + | 2 \rg \lg 2 | \rpar + \frac{1-4\mathcal{F}}{6} \lpar | 1 \rg \lg 2 | + | 2 \rg \lg 1 | \rpar ,
\ed
whose matrix representation assumes the form
\bd
\ro_{\mathtt{W}} = \frac{1}{6} \lpar \begin{array}{cccc}
2-2\mathcal{F} & 0              & 0              & 0 \\
0              & 1+2\mathcal{F} & 1-4\mathcal{F} & 0 \\
0              & 1-4\mathcal{F} & 1+2\mathcal{F} & 0 \\
0              & 0              & 0              & 2-2\mathcal{F} \\
\end{array} \rpar .
\ed
In this particular four-level system, we have the states $| 0 \rg$ and $| 3 \rg$ having the same population rate of
$\frac{1-\mathcal{F}}{3}$, the states $| 1 \rg$ and $| 2 \rg$ with $\frac{1+2\mathcal{F}}{6}$, and both the transitions 
$1 \leftrightarrow 2$ sharing the rate $\frac{1-4\mathcal{F}}{6}$. The corresponding discrete $\mathrm{SU(4)}$ Wigner function
for $\mathcal{F} \in [0,1]$ is given by
\be
\lb{eq29}
W_{\mathtt{W}}(\mu,\nu) = \frac{1}{4} + \frac{1-4\mathcal{F}}{12} \lbk \delta_{\mu,0}^{[4]} - \delta_{\mu,1}^{[4]} - 
\delta_{\mu,2}^{[4]} + \delta_{\mu,3}^{[4]} + \frac{\sin \lbk \lpar \mu - \frac{3}{2} \rpar \pi \rbk}{\sin \lbk \lpar \mu - 
\frac{3}{2} \rpar \frac{\pi}{4} \rbk} \cos \lpar \frac{\nu \pi}{2} \rpar \rbk ,
\ee
such that $\mathcal{F}=1$ implies in $W_{\mathtt{W}}(\mu,\nu) = W_{\Psi_{-}}(\mu,\nu)$. Thus, the analysis on this four-level 
system is very similar to that already performed in the previous example: the basic difference is the presence of contributions
related to the states $|0 \rg$ and $| 3 \rg$ with equal weights. Figure \ref{fig3} exhibits the 3D plots of Eq. (\ref{eq29}) for
different values of $\mathcal{F} \in [0,1]$, where (a) $\mathcal{F}=0.35$, (b) $\mathcal{F}=0.50$, (c) $\mathcal{F}=0.75$, and
(d) $\mathcal{F}=1$ were considered. The quantum fluctuation contributions observed in (d) for $\mu=0,3$ fixed and $\nu=0,1,2,3$
have null sum since only quantum effects inherent to the states $| 1 \rg$ and $| 2 \rg$ prevail. 
\begin{figure}[t]
\begin{minipage}[b]{0.4\linewidth}
\includegraphics[width=\linewidth]{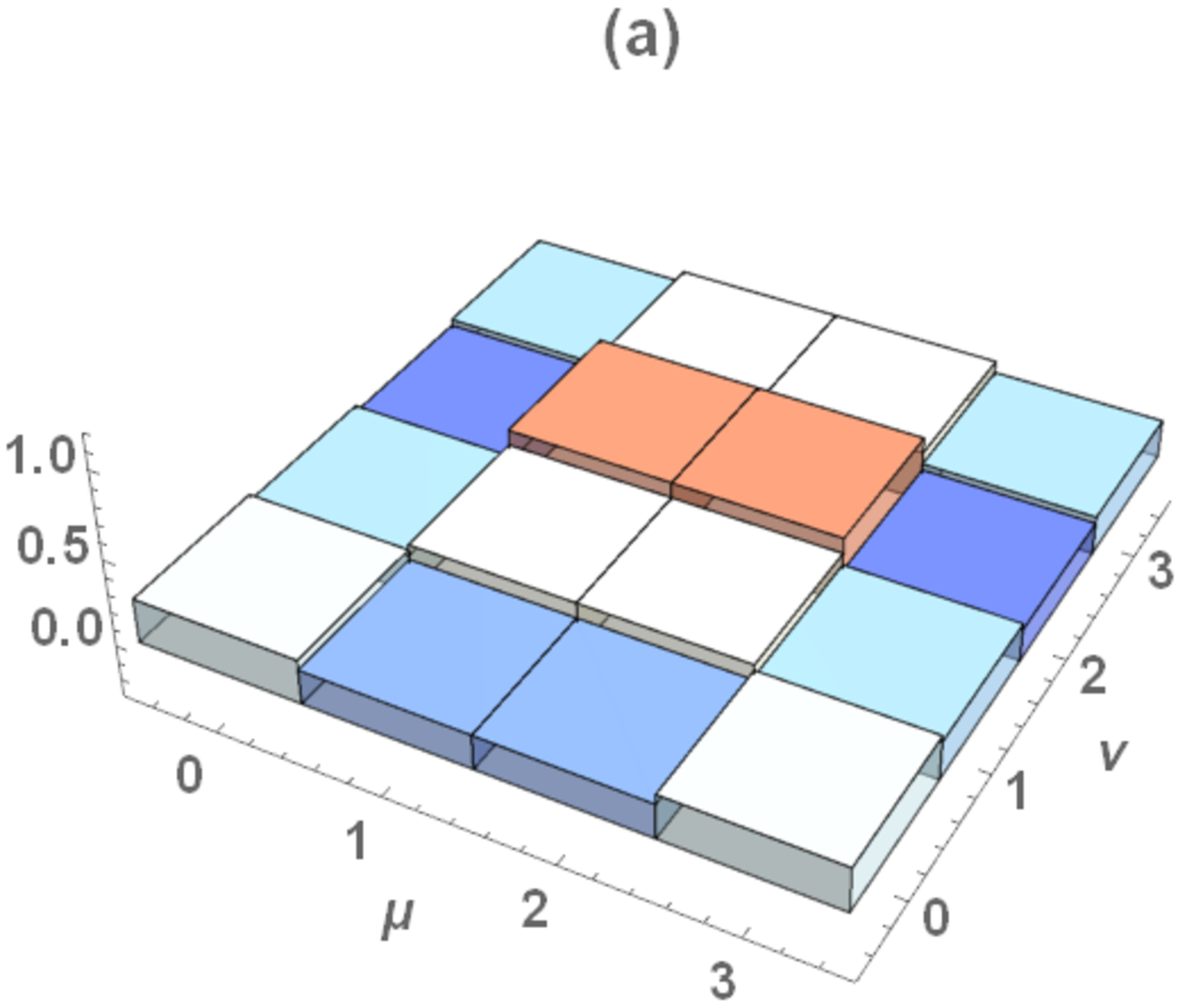}
\end{minipage} \hfill
\begin{minipage}[b]{0.4\linewidth}
\includegraphics[width=\linewidth]{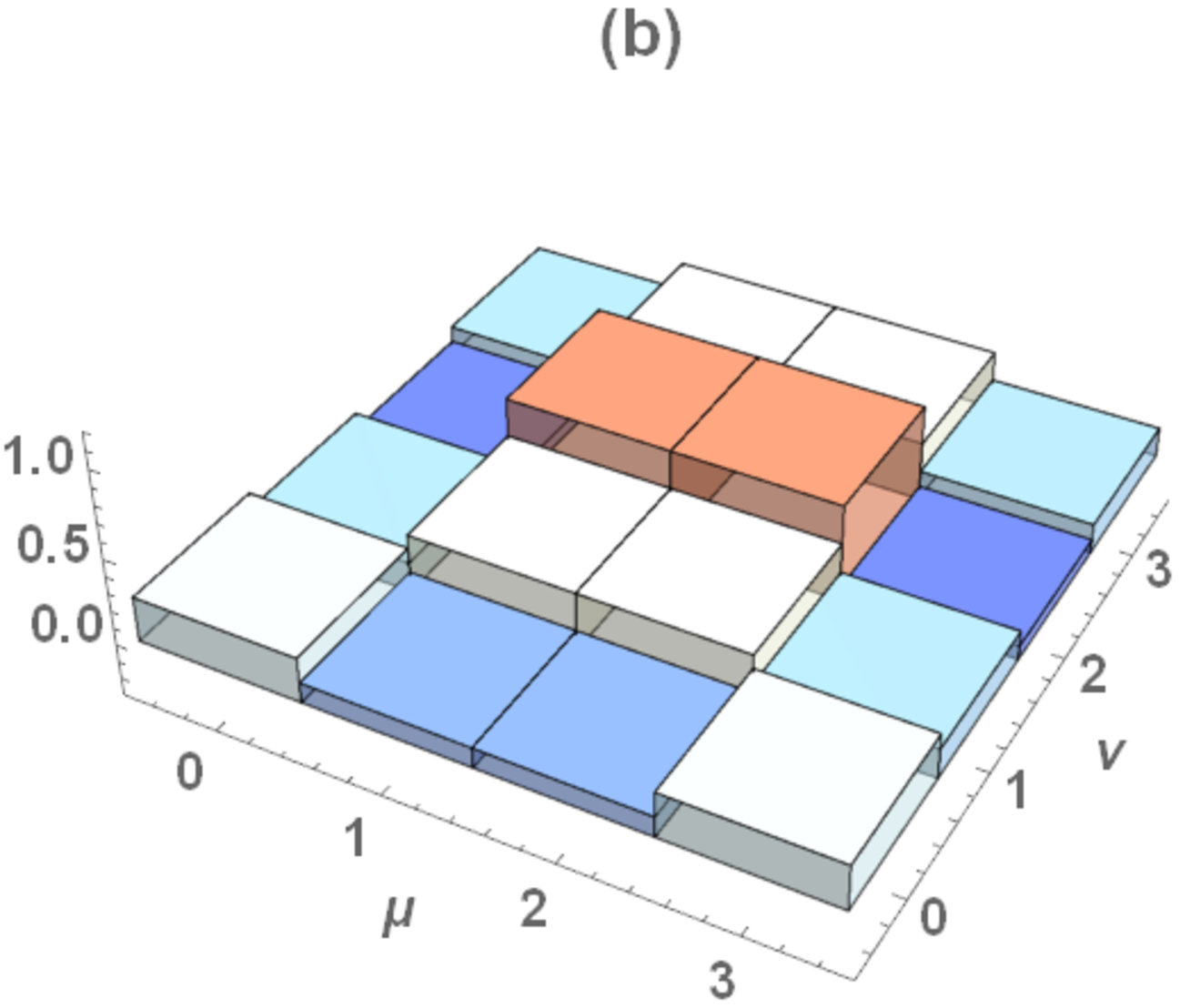}
\end{minipage} \hfill
\begin{minipage}[b]{0.4\linewidth}
\includegraphics[width=\linewidth]{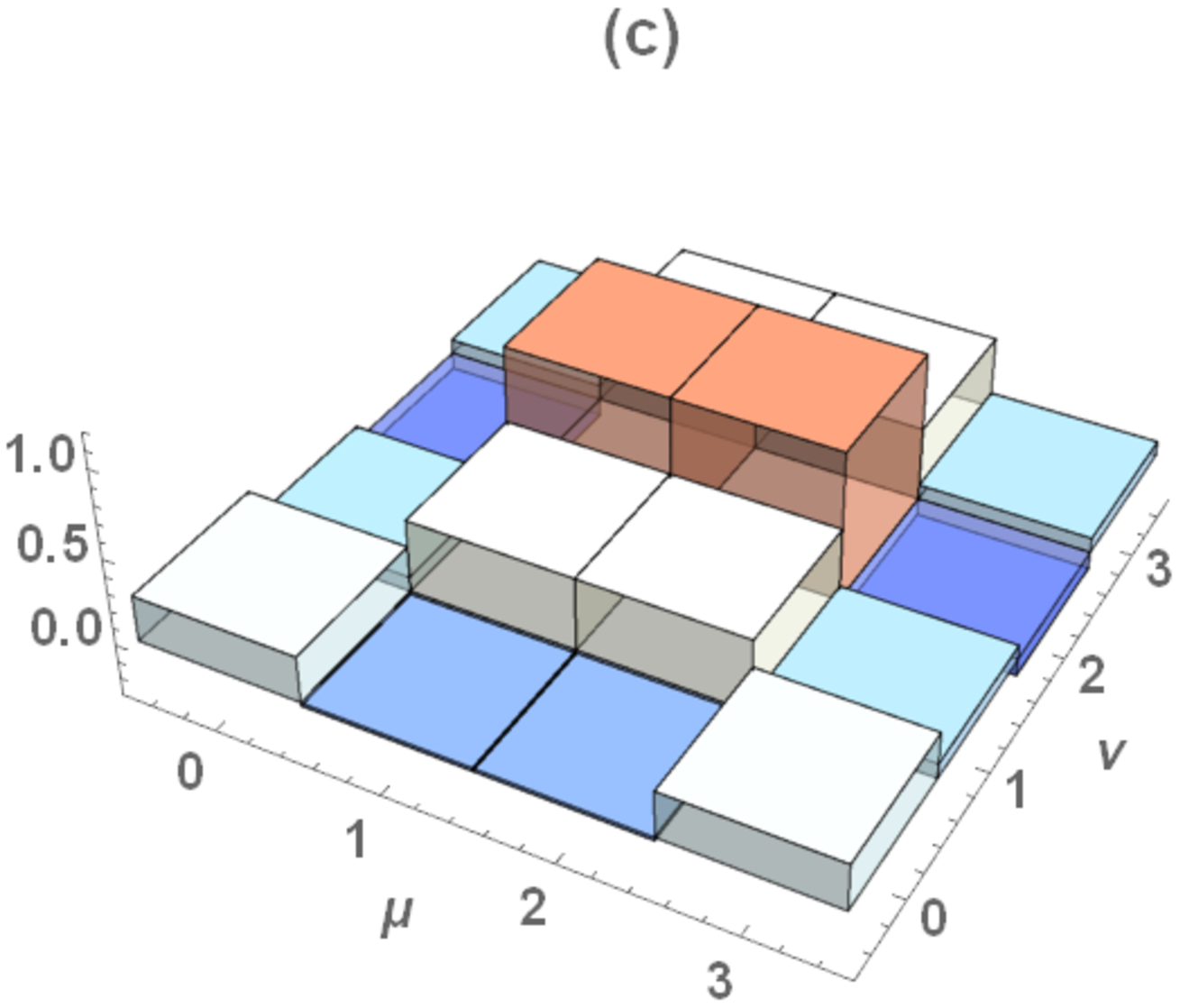}
\end{minipage} \hfill
\begin{minipage}[b]{0.4\linewidth}
\includegraphics[width=\linewidth]{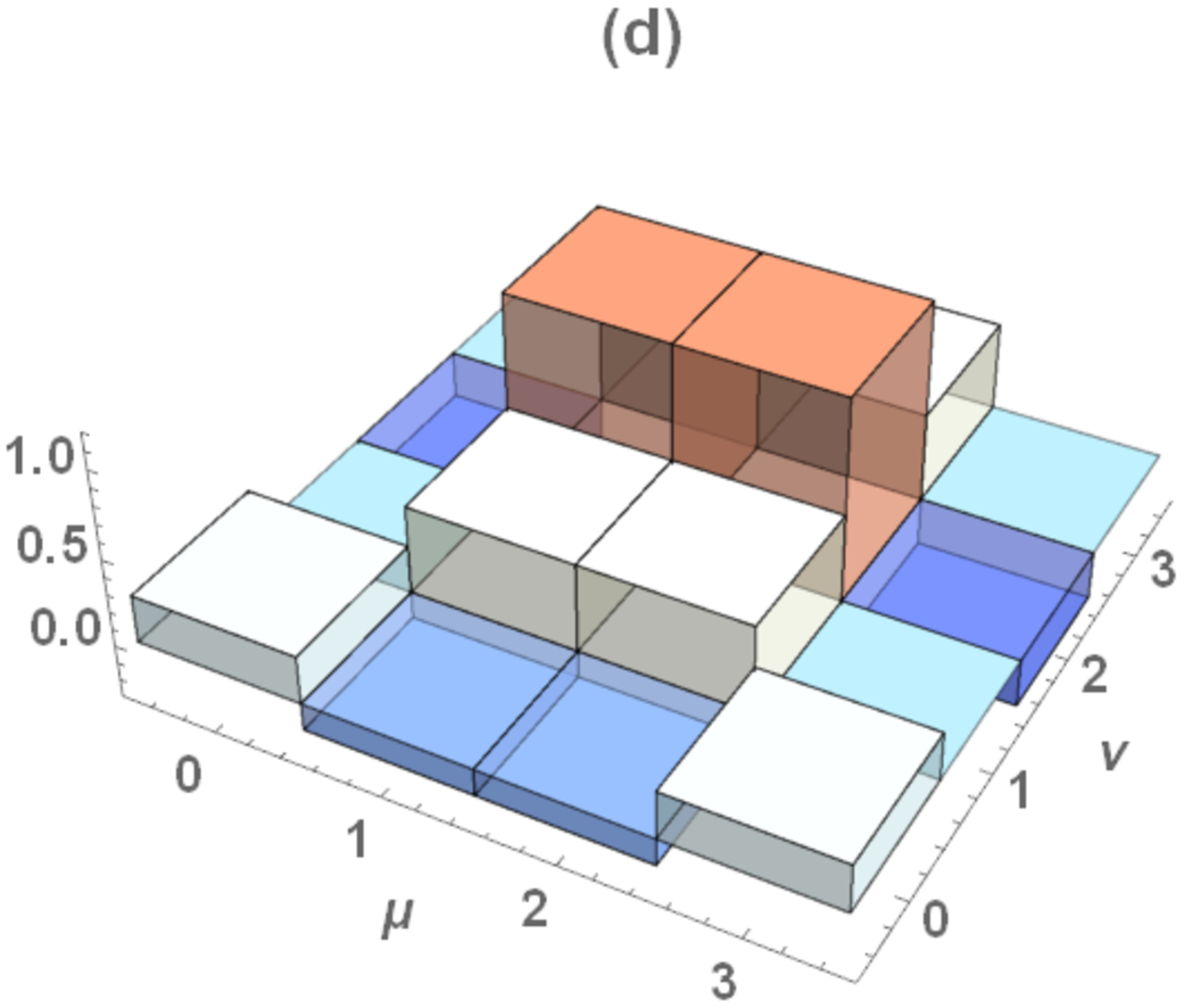}
\end{minipage}
\vspace*{13pt}
\fcaption{\lb{fig3}
Three-dimensional plots of the discrete $\mathrm{SU(4)}$ Wigner function (\ref{eq29}) related to the Werner states as a function
of $0 \leq \mu,\nu \leq 3$ and for different values of $\mathcal{F}$: (a) $\mathcal{F}=0.35$, (b) $\mathcal{F}=0.50$, (c) 
$\mathcal{F}=0.75$, and finally, (d) $\mathcal{F}=1$. Note that as $\mathcal{F}$ increases, the quantum effects underlying states
$| 1 \rg$ and $| 2 \rg$ become more evident: indeed, for $\mathcal{F}=1$ only the subspace associated with these states turns
relevant, since $\ro_{\mathtt{W}}$ coincides with $\ro_{\Psi_{-}}$ in such a case -- see picture \ref{fig2}(b).}
\end{figure}

The solution established for the visualization difficulty related to the discrete $\mathrm{SU(2)} \otimes \mathrm{SU(2)}$ Wigner
functions has, in the correspondence with four-level systems, an effective mathematical tool in the study of maximally entangled
mixed states \cite{IH-2,VAM2001,MJWK2001,Wei2003} where, in particular, the two-qubit X-states take a special place 
\cite{Mend-1,Yuri,Norek,Nandi,Hed-1,Hed-2,Yu2019}. Next, we will explore this fact to characterize the maximally entangled mixed 
two-qubit X-states through the discrete $\mathrm{SU(4)}$ Wigner functions.

\section{Two-qubit X-states}
\lb{s4}

Two-qubit X-states are an important family of quantum states belonging to a four-dimensional Hilbert space $\mathcal{H}_{4}$
characterized by a unique property: basically, they do not mix the subspaces $\mathsf{S}_{1} = \mathtt{Span} \lpar | 0_{1} 0_{2}
\rg, | 1_{1} 1_{2} \rg \rpar$ and $\mathsf{S}_{2} = \mathtt{Span} \lpar | 0_{1} 1_{2} \rg, | 1_{1} 0_{2} \rg \rpar$. Then, if one
considers the associated computational basis, these states have potentially nonzero density-matrix elements located on the main
diagonal and antidiagonal as follows \cite{Yu2007}:\ftn{Note the resemblance of Eq. (\ref{eq30}) with the alphabet letter
\texttt{X} justifying, in this way, the nomenclature \texttt{X-state}. This particular characteristic can be extended in order to 
encompass more general situations, namely, every density matrix possessing nonzero terms along the main diagonal and antidiagonal 
refers to the \texttt{X}-form.}
\be
\lb{eq30}
\ro_{\mathtt{X}} = \lpar \begin{array}{cccc}
\rho_{11}        & 0                & 0         & \rho_{14} \\
0                & \rho_{22}        & \rho_{23} & 0 \\
0                & \rho_{23}^{\ast} & \rho_{33} & 0 \\
\rho_{14}^{\ast} & 0                & 0         & \rho_{44} \\
\end{array} \rpar .
\ee
It is worth stressing that two-qubit Bell and Werner states are particular examples of X-states. Morever, an important form of
universality property with respect to two-qubit entanglement was properly established in \cite{Mend-1} through a set of 
transcendental parameters inherent to the matrix elements present in (\ref{eq30}): ``for every two-qubit state, there is a
corresponding two-qubit X-state of same spectrum and entanglement in accordance with three different entanglement measurements, 
that are concurrence, negativity, and relative entropy of entanglement." Hence, there exists an entanglement-preserving unitary
(EPU) transformation $\hat{\mathsf{U}}_{\mathrm{EPU}}$ that preserves the entanglement of the input state, that is
$\ro_{\mathtt{X}} = \hat{\mathsf{U}}_{\mathrm{EPU}} \ro \, \hat{\mathsf{U}}^{\dagger}_{\mathrm{EPU}}$, this property being termed
by `EPU equivalence'. After this, Hedemann \cite{Hed-2} established a definitive prove on the existence of such transformations
obtaining, in this way, a compact implicit solution for them; furthermore, the author also provided an explicit form for the 
corresponding two-qubit X-state family. Now, let us also mention that such X-states were recently used in the study of certain 
spin chains with emphasis on the thermal entanglement properties and quantum discord related to these models \cite{Wang2006, 
Yu2017,Park2019}.

Following, it is worth stressing that $\ro_{\mathtt{X}}$ represents a particular case of those general two-qubit states studied
until the present moment, once we have now $\rho_{12}=\rho_{13}=\rho_{24}=\rho_{34}=0$. So, in order to establish the discrete 
$\mathrm{SU(2)} \otimes \mathrm{SU(2)}$ Wigner function for two-qubit X-states, let us rewrite Eq. (\ref{eq16}) as follows:
\brr
\lb{eq31}
& & W_{\mathtt{X}}(\mu_{1},\nu_{1},\mu_{2},\nu_{2}) = \frac{1}{4} \lbr 1 + \Gamma_{11}(\mu_{1},\mu_{2}) + 
\Gamma_{22}(\mu_{1},\mu_{2}) + \Gamma_{33}(\mu_{1},\mu_{2}) + \Gamma_{44}(\mu_{1},\mu_{2}) \right. \nn \\
& & \hspace*{3cm} + \left. 2 (-1)^{\nu_{1}+\nu_{2}} \lbk \Gamma_{14}(\mu_{1},\mu_{2}) + \Gamma_{23}(\mu_{1},\mu_{2}) \rbk \rbr 
\err
where the $\Gamma$'s functions have already been previously defined. Therefore, the discrete Wigner functions 
$\mathcal{W}_{\mathtt{R}}(\mu_{1},\nu_{1})$ and $\mathcal{W}_{\mathtt{R}}(\mu_{2},\nu_{2})$ assume in this context the simplified
forms
\be
\lb{eq32}
\mathcal{W}_{\mathtt{R,X}}(\mu_{1},\nu_{1}) = \half \lbk 1 + (-1)^{\mu_{1}} \lpar \rho_{11} + \rho_{22} - \rho_{33} - \rho_{44}
\rpar \rbk
\ee
and
\be
\lb{eq33}
\mathcal{W}_{\mathtt{R,X}}(\mu_{2},\nu_{2}) = \half \lbk 1 + (-1)^{\mu_{2}} \lpar \rho_{11} - \rho_{22} + \rho_{33} - \rho_{44}
\rpar \rbk , 
\ee
which do not depend on the discrete variables $\nu_{1}$ and $\nu_{2}$, respectively. 

However, if one considers the discrete $\mathrm{SU(4)}$ Wigner function, it can be promptly obtained from Eq. (\ref{eq25}) through
a similar mathematical procedure, that is $\varrho_{12}= \varrho_{13}=\varrho_{24}=\varrho_{34}=0$,
\brr
\lb{eq34}
W_{\mathtt{X}}(\mu,\nu) &=& \frac{1}{4} + \frac{1}{4} \lpar 3 \delta_{\mu,0}^{[4]} - \delta_{\mu,1}^{[4]} - \delta_{\mu,2}^{[4]} -
\delta_{\mu,3}^{[4]} \rpar \varrho_{11} - \frac{1}{4} \lpar \delta_{\mu,0}^{[4]} - 3 \delta_{\mu,1}^{[4]} + \delta_{\mu,2}^{[4]} +
\delta_{\mu,3}^{[4]} \rpar \varrho_{22} \nn \\
& & - \, \frac{1}{4} \lpar \delta_{\mu,0}^{[4]} + \delta_{\mu,1}^{[4]} - 3 \delta_{\mu,2}^{[4]} + \delta_{\mu,3}^{[4]} \rpar 
\varrho_{33} - \frac{1}{4} \lpar \delta_{\mu,0}^{[4]} + \delta_{\mu,1}^{[4]} + \delta_{\mu,2}^{[4]} - 3 \delta_{\mu,3}^{[4]} \rpar
\varrho_{44} \nn \\
& & + \, \half \frac{\sin \lbk \lpar \mu - \frac{3}{2} \rpar \pi \rbk}{\sin \lbk \lpar \mu - \frac{3}{2} \rpar \frac{\pi}{4} \rbk}
\lbr \cos \lpar \frac{\nu \pi}{2} \rpar \lbk \re ( \varrho_{23} ) + (-1)^{\nu} \re ( \varrho_{14} ) \rbk \right. \nn \\
& & - \left. \sin \lpar \frac{\nu \pi}{2} \rpar \lbk \ima ( \varrho_{23} ) + (-1)^{\nu} \ima ( \varrho_{14} ) \rbk \rbr .  
\err
The discrete marginal distribution functions
\bd
Q_{\mathtt{X}}(\mu) \col \half \sum_{\nu} W_{\mathtt{X}}(\mu,\nu) \quad \mbox{and} \quad
R_{\mathtt{X}}(\nu) \col \half \sum_{\mu} W_{\mathtt{X}}(\mu,\nu)
\ed
complete our description of discrete Wigner functions related to the two-qubit X-states, when it is possible to show that
\brr
\lb{eq35}
& & Q_{\mathtt{X}}(\mu) = \half \lbk 1 + \lpar 3 \delta_{\mu,0}^{[4]} - \delta_{\mu,1}^{[4]} - \delta_{\mu,2}^{[4]} -
\delta_{\mu,3}^{[4]} \rpar \varrho_{11} - \lpar \delta_{\mu,0}^{[4]} - 3 \delta_{\mu,1}^{[4]} + \delta_{\mu,2}^{[4]} +
\delta_{\mu,3}^{[4]} \rpar \varrho_{22} \right. \nn \\
& & \hspace*{1.4cm} - \left. \lpar \delta_{\mu,0}^{[4]} + \delta_{\mu,1}^{[4]} - 3 \delta_{\mu,2}^{[4]} + \delta_{\mu,3}^{[4]} 
\rpar \varrho_{33} - \lpar \delta_{\mu,0}^{[4]} + \delta_{\mu,1}^{[4]} + \delta_{\mu,2}^{[4]} - 3 \delta_{\mu,3}^{[4]} \rpar
\varrho_{44} \rbk
\err
and
\brr
\lb{eq36}
& & R_{\mathtt{X}}(\nu) = \half + \frac{\sqrt{2 - \sqrt{2}}}{2} (-1)^{\nu} \lbr \cos \lpar \frac{\nu \pi}{2} \rpar \lbk 
\re ( \varrho_{14} ) + (-1)^{\nu} \re ( \varrho_{23} ) \rbk \right. \nn \\ 
& & \hspace*{1.4cm} - \left. \sin \lpar \frac{\nu \pi}{2} \rpar \lbk \ima ( \varrho_{14} ) + (-1)^{\nu} \ima ( \varrho_{23} ) \rbk 
\rbr
\err
satisfy the relations
\bd
\half \sum_{\mu} Q_{\mathtt{X}}(\mu) = \half \sum_{\nu} R_{\mathtt{X}}(\nu) = 1 .
\ed
It is important to stress that $Q_{\mathtt{X}}(\mu)$ only depends on the matrix elements of the main diagonal, while
$R_{\mathtt{X}}(\nu)$ brings information on the antidiagonal matrix elements. This important property associated with X-states
can help us to comprehend the preexisting quantum correlations in these states by means of the difference $\Delta_{\mathtt{X}}
(\mu,\nu) \col W_{\mathtt{X}}(\mu,\nu) - Q_{\mathtt{X}}(\mu) R_{\mathtt{X}}(\nu)$, this function being responsible for 
distinguishing the effects of the main-diagonal and antidiagonal matrix elements on the aforementioned quantum correlations -- see
Table \ref{tab-X-state} for calculational details.
\vspace*{4pt}
\begin{table}[!t]
\tcaption{All possible values of the discrete Wigner function $W_{\mathtt{X}}(\mu,\nu)$, the product of discrete marginal
distribution functions $Q_{\mathtt{X}}(\mu) R_{\mathtt{X}}(\nu)$, and the difference $\Delta_{\mathtt{X}}(\mu,\nu) \col
W_{\mathtt{X}}(\mu,\nu) - Q_{\mathtt{X}}(\mu) R_{\mathtt{X}}(\nu)$ between the previous functions for each cell of the 
finite-dimensional discrete phase space characterized by a specific pair $(\mu,\nu)$ with respect to the X-state (\ref{eq30}),
where the correspondence $\ro_{\mathtt{X}} \leftrightarrow \hat{\varrho}_{\mathtt{X}}$ was previously established. In particular,
the function $\Delta_{\mathtt{X}}(\mu,\nu)$ measures the preexisting quantum correlations in the X-state.}
\centerline{\footnotesize\smalllineskip
\begin{tabular}{l c c c c c}
$\mu$ & $\nu$ & $W_{\mathtt{X}}(\mu,\nu)$ & $Q_{\mathtt{X}}(\mu) R_{\mathtt{X}}(\nu)$ & $\Delta_{\mathtt{X}}(\mu,\nu)$ \\
\hline
$0$ & $0$ & $\varrho_{11} - \sqrt{\frac{2-\sqrt{2}}{2}} \, \re ( \varrho_{14} + \varrho_{23} )$ & $\varrho_{11} + 
\sqrt{2-\sqrt{2}} \, \varrho_{11} \re ( \varrho_{14} + \varrho_{23} )$ & $- \sqrt{\frac{2-\sqrt{2}}{2}} \lpar 1 + \sqrt{2}
\varrho_{11} \rpar \re ( \varrho_{14} + \varrho_{23} )$ \\
$0$ & $1$ & $\varrho_{11} - \sqrt{\frac{2-\sqrt{2}}{2}} \, \ima ( \varrho_{14} - \varrho_{23} )$ & $\varrho_{11} + 
\sqrt{2-\sqrt{2}} \, \varrho_{11} \ima ( \varrho_{14} - \varrho_{23} )$ & $- \sqrt{\frac{2-\sqrt{2}}{2}} \lpar 1 + \sqrt{2}
\varrho_{11} \rpar \ima ( \varrho_{14} - \varrho_{23} )$ \\
$0$ & $2$ & $\varrho_{11} + \sqrt{\frac{2-\sqrt{2}}{2}} \, \re ( \varrho_{14} + \varrho_{23} )$ & $\varrho_{11} - 
\sqrt{2-\sqrt{2}} \, \varrho_{11} \re ( \varrho_{14} + \varrho_{23} )$ & $\sqrt{\frac{2-\sqrt{2}}{2}} \lpar 1 + \sqrt{2}
\varrho_{11} \rpar \re ( \varrho_{14} + \varrho_{23} )$ \\
$0$ & $3$ & $\varrho_{11} + \sqrt{\frac{2-\sqrt{2}}{2}} \, \ima ( \varrho_{14} - \varrho_{23} )$ & $\varrho_{11} - 
\sqrt{2-\sqrt{2}} \, \varrho_{11} \ima ( \varrho_{14} - \varrho_{23} )$ & $\sqrt{\frac{2-\sqrt{2}}{2}} \lpar 1 + \sqrt{2}
\varrho_{11} \rpar \ima ( \varrho_{14} - \varrho_{23} )$ \\
$1$ & $0$ & $\varrho_{22} + \sqrt{\frac{2+\sqrt{2}}{2}} \, \re ( \varrho_{14} + \varrho_{23} )$ & $\varrho_{22} + 
\sqrt{2-\sqrt{2}} \, \varrho_{22} \re ( \varrho_{14} + \varrho_{23} )$ & $\sqrt{\frac{2+\sqrt{2}}{2}} \lbk 1 - ( 2-\sqrt{2} \,)
\varrho_{22} \rbk \re ( \varrho_{14} + \varrho_{23} )$ \\
$1$ & $1$ & $\varrho_{22} + \sqrt{\frac{2+\sqrt{2}}{2}} \, \ima ( \varrho_{14} - \varrho_{23} )$ & $\varrho_{22} + 
\sqrt{2-\sqrt{2}} \, \varrho_{22} \ima ( \varrho_{14} - \varrho_{23} )$ & $\sqrt{\frac{2+\sqrt{2}}{2}} \lbk 1 - ( 2-\sqrt{2} \,)
\varrho_{22} \rbk \ima ( \varrho_{14} - \varrho_{23} )$ \\
$1$ & $2$ & $\varrho_{22} - \sqrt{\frac{2+\sqrt{2}}{2}} \, \re ( \varrho_{14} + \varrho_{23} )$ & $\varrho_{22} - 
\sqrt{2-\sqrt{2}} \, \varrho_{22} \re ( \varrho_{14} + \varrho_{23} )$ & $-\sqrt{\frac{2+\sqrt{2}}{2}} \lbk 1 - ( 2-\sqrt{2} \,)
\varrho_{22} \rbk \re ( \varrho_{14} + \varrho_{23} )$ \\
$1$ & $3$ & $\varrho_{22} - \sqrt{\frac{2+\sqrt{2}}{2}} \, \ima ( \varrho_{14} - \varrho_{23} )$ & $\varrho_{22} - 
\sqrt{2-\sqrt{2}} \, \varrho_{22} \ima ( \varrho_{14} - \varrho_{23} )$ & $-\sqrt{\frac{2+\sqrt{2}}{2}} \lbk 1 - ( 2-\sqrt{2} \,)
\varrho_{22} \rbk \ima ( \varrho_{14} - \varrho_{23} )$ \\
$2$ & $0$ & $\varrho_{33} + \sqrt{\frac{2+\sqrt{2}}{2}} \, \re ( \varrho_{14} + \varrho_{23} )$ & $\varrho_{33} + 
\sqrt{2-\sqrt{2}} \, \varrho_{33} \re ( \varrho_{14} + \varrho_{23} )$ & $\sqrt{\frac{2+\sqrt{2}}{2}} \lbk 1 - ( 2-\sqrt{2} \,)
\varrho_{33} \rbk \re ( \varrho_{14} + \varrho_{23} )$ \\
$2$ & $1$ & $\varrho_{33} + \sqrt{\frac{2+\sqrt{2}}{2}} \, \ima ( \varrho_{14} - \varrho_{23} )$ & $\varrho_{33} + 
\sqrt{2-\sqrt{2}} \, \varrho_{33} \ima ( \varrho_{14} - \varrho_{23} )$ & $\sqrt{\frac{2+\sqrt{2}}{2}} \lbk 1 - ( 2-\sqrt{2} \,)
\varrho_{33} \rbk \ima ( \varrho_{14} - \varrho_{23} )$ \\
$2$ & $2$ & $\varrho_{33} - \sqrt{\frac{2+\sqrt{2}}{2}} \, \re ( \varrho_{14} + \varrho_{23} )$ & $\varrho_{33} - 
\sqrt{2-\sqrt{2}} \, \varrho_{33} \re ( \varrho_{14} + \varrho_{23} )$ & $-\sqrt{\frac{2+\sqrt{2}}{2}} \lbk 1 - ( 2-\sqrt{2} \,)
\varrho_{33} \rbk \re ( \varrho_{14} + \varrho_{23} )$ \\
$2$ & $3$ & $\varrho_{33} - \sqrt{\frac{2+\sqrt{2}}{2}} \, \ima ( \varrho_{14} - \varrho_{23} )$ & $\varrho_{33} - 
\sqrt{2-\sqrt{2}} \, \varrho_{33} \ima ( \varrho_{14} - \varrho_{23} )$ & $-\sqrt{\frac{2+\sqrt{2}}{2}} \lbk 1 - ( 2-\sqrt{2} \,)
\varrho_{33} \rbk \ima ( \varrho_{14} - \varrho_{23} )$ \\
$3$ & $0$ & $\varrho_{44} - \sqrt{\frac{2-\sqrt{2}}{2}} \, \re ( \varrho_{14} + \varrho_{23} )$ & $\varrho_{44} + 
\sqrt{2-\sqrt{2}} \, \varrho_{44} \re ( \varrho_{14} + \varrho_{23} )$ & $- \sqrt{\frac{2-\sqrt{2}}{2}} \lpar 1 + \sqrt{2}
\varrho_{44} \rpar \re ( \varrho_{14} + \varrho_{23} )$ \\
$3$ & $1$ & $\varrho_{44} - \sqrt{\frac{2-\sqrt{2}}{2}} \, \ima ( \varrho_{14} - \varrho_{23} )$ & $\varrho_{44} + 
\sqrt{2-\sqrt{2}} \, \varrho_{44} \ima ( \varrho_{14} - \varrho_{23} )$ & $- \sqrt{\frac{2-\sqrt{2}}{2}} \lpar 1 + \sqrt{2}
\varrho_{44} \rpar \ima ( \varrho_{14} - \varrho_{23} )$ \\
$3$ & $2$ & $\varrho_{44} + \sqrt{\frac{2-\sqrt{2}}{2}} \, \re ( \varrho_{14} + \varrho_{23} )$ & $\varrho_{44} - 
\sqrt{2-\sqrt{2}} \, \varrho_{44} \re ( \varrho_{14} + \varrho_{23} )$ & $\sqrt{\frac{2-\sqrt{2}}{2}} \lpar 1 + \sqrt{2}
\varrho_{44} \rpar \re ( \varrho_{14} + \varrho_{23} )$ \\
$3$ & $3$ & $\varrho_{44} + \sqrt{\frac{2-\sqrt{2}}{2}} \, \ima ( \varrho_{14} - \varrho_{23} )$ & $\varrho_{44} - 
\sqrt{2-\sqrt{2}} \, \varrho_{44} \ima ( \varrho_{14} - \varrho_{23} )$ & $\sqrt{\frac{2-\sqrt{2}}{2}} \lpar 1 + \sqrt{2}
\varrho_{44} \rpar \ima ( \varrho_{14} - \varrho_{23} )$ \\
\hline \\
\end{tabular}}
\lb{tab-X-state}
\end{table}

To illustrate these results, let us now consider the two-qubit X-states introduced by Munro \textit{et al}. \cite{Munro2001} with
maximal concurrence for a given fixed purity $\gam$,
\be
\lb{eq37}
\ro_{\mathtt{X}} = \lpar \begin{array}{cccc}
g(\gam)        & 0          & 0 & \gam / 2 \\
0              & 1-2g(\gam) & 0 & 0 \\
0              & 0          & 0 & 0 \\
\gam / 2       & 0          & 0 & g(\gam) \\
\end{array} \rpar \; \mbox{with} \;\; 
g(\gam) = \left\{ \begin{array}{rcl}
\gam / 2 & \mbox{if}   & 2/3 \leq \gam \leq 1 , \\
1/3      & \mbox{when} & 0 \leq \gam < 2/3 .
\end{array} \right.
\ee
With respect to the correspondence $\ro \leftrightarrow \hat{\varrho}$ (see appendix A), Eq. (\ref{eq37}) can be written as
\bd
\hat{\varrho}_{\mathtt{X}} = g(\gam) \lpar | 0 \rg \lg 0 | + | 3 \rg \lg 3 | \rpar + \frac{\gamma}{2} \lpar | 0 \rg \lg 3 | +
| 3 \rg \lg 0 | \rpar + [1-2g(\gam)] | 1 \rg \lg 1 | ,
\ed
where it is clear the physical rules employed by the coefficients $g(\gam)$ and $\frac{\gam}{2}$: for instance, the case $\gam=1$
deals with maximally entangled pure states, since $\ro_{\mathtt{X}}$ coincides with the Bell state $\ro_{\Phi_{+}}$; otherwise,
(\ref{eq37}) describes maximally entangled mixed states. For this particular example under scrutiny, the discrete Wigner function
(\ref{eq34}) assumes the compact form
\brr
\lb{eq38}
W_{\mathtt{X}}(\mu,\nu;\gam) &=& \frac{1}{4} - \frac{1}{4} \lpar \delta_{\mu,0}^{[4]} - 3 \delta_{\mu,1}^{[4]} + 
\delta_{\mu,2}^{[4]} + \delta_{\mu,3}^{[4]} \rpar + \lpar \delta_{\mu,0}^{[4]} - 2 \delta_{\mu,1}^{[4]} + \delta_{\mu,3}^{[4]}
\rpar g(\gam) \nn \\
& & + \, \frac{\gam}{4} \frac{\sin \lbk \lpar \mu - \frac{3}{2} \rpar \pi \rbk}{\sin \lbk \lpar \mu - \frac{3}{2} \rpar 
\frac{\pi}{4} \rbk} (-1)^{\nu} \cos \lpar \frac{\nu \pi}{2} \rpar ,  
\err
while the discrete marginal distribution functions are given by
\brr
& & Q_{\mathtt{X}}(\mu;\gam) = \half \lbk 1 - \lpar \delta_{\mu,0}^{[4]} - 3 \delta_{\mu,1}^{[4]} + \delta_{\mu,2}^{[4]} +
\delta_{\mu,3}^{[4]} \rpar + 4 \lpar \delta_{\mu,0}^{[4]} - 2 \delta_{\mu,1}^{[4]} + \delta_{\mu,3}^{[4]} \rpar g(\gam) 
\rbk , \nn \\
& & R_{\mathtt{X}}(\nu;\gam) = \half + \frac{\sqrt{2 - \sqrt{2}}}{2} \frac{\gam}{2} (-1)^{\nu} \cos \lpar \frac{\nu \pi}{2} 
\rpar . \nn
\err
Figure \ref{fig4} shows the three-dimensional plots of discrete Wigner function (\ref{eq38}) and the difference function 
$\Delta_{\mathtt{X}}(\mu,\nu) = W_{\mathtt{X}} (\mu,\nu) - Q_{\mathtt{X}}(\mu) R_{\mathtt{X}}(\nu)$ as a function of $0 \leq \mu,
\nu \leq 3$ for $\gam = \frac{3}{4}$ and $\half$. Pictures (a) $W_{\mathtt{X}}(\mu,\nu;\frac{3}{4})$ and (c) $W_{\mathtt{X}}
(\mu,\nu;\half)$ describe maximally entangled mixed states with different values of purity $\gam$, where two physical effects can
be promptly verified: as $\gam$ increases, the quantum fluctuations due to the states $| 1 \rg$ and $| 2 \rg$ -- see $\mu=1,2$ and
$\nu=0,1,2,3$ -- are more visible, while for $\gam=1$ these fluctuations exhibit equally contributions since we are now describing 
maximally entangled pure states (in such a case, the Bell state $| \Phi_{+} \rg$); on the other hand, the quantum correlations 
present in (b) $\Delta_{\mathtt{X}}(\mu,\nu;\frac{3}{4})$ and (d) $\Delta_{\mathtt{X}}(\mu,\nu;\half)$ become more prominent and 
symmetric as $\gam$ goes to 1. Indeed, for $\gam = \half, \frac{3}{4}, 1$ fixed, we obtain
\begin{itemize}
\item $\Delta_{\mathtt{X}}(1,0;\half) \approx 0.26$, $\Delta_{\mathtt{X}}(1,2;\half) \approx - 0.26$, 
$\Delta_{\mathtt{X}}(2,0;\half) \approx 0.33$, $\Delta_{\mathtt{X}}(2,2;\half) \approx - 0.33$,

\item $\Delta_{\mathtt{X}}(1,0;\frac{3}{4}) \approx 0.42$, $\Delta_{\mathtt{X}}(1,2;\frac{3}{4}) \approx - 0.42$,
$\Delta_{\mathtt{X}}(2,0;\frac{3}{4}) \approx 0.49$, $\Delta_{\mathtt{X}}(2,2;\frac{3}{4}) \approx - 0.49$,

\item $\Delta_{\mathtt{X}}(1,0;1) \approx 0.65$, $\Delta_{\mathtt{X}}(1,2;1) \approx - 0.65$, 
$\Delta_{\mathtt{X}}(2,0;1) \approx 0.65$, $\Delta_{\mathtt{X}}(2,2;1) \approx - 0.65$,
\end{itemize}
which corroborate our considerations.

Finally, let us say some few words about the potential use of discrete Wigner functions in experiments involving the two-qubit
and ququart states: the complete algebraic framework here developed for discrete $\mathrm{SU(4)}$ Wigner functions really works 
well, as expected, in detecting genuinelly quantum effects (for example, entanglement, among others); besides, experiments
associated with NMR techniques, where the matrix elements of the density matrix are tomographycally reconstructed, can be 
considered as the best scenario to implement this important mathematical tool.
\begin{figure}[t]
\begin{minipage}[b]{0.4\linewidth}
\includegraphics[width=\linewidth]{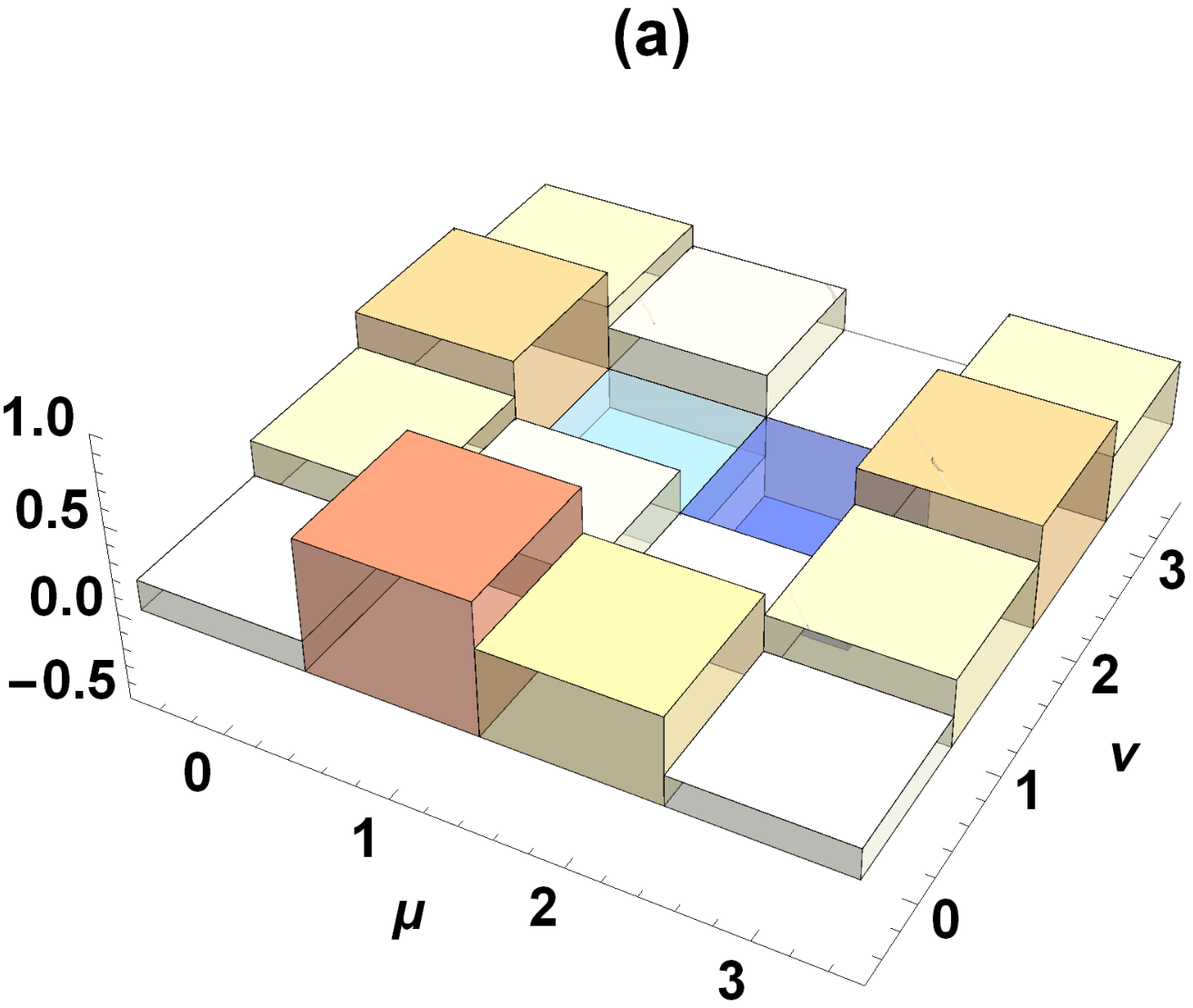}
\end{minipage} \hfill
\begin{minipage}[b]{0.4\linewidth}
\includegraphics[width=\linewidth]{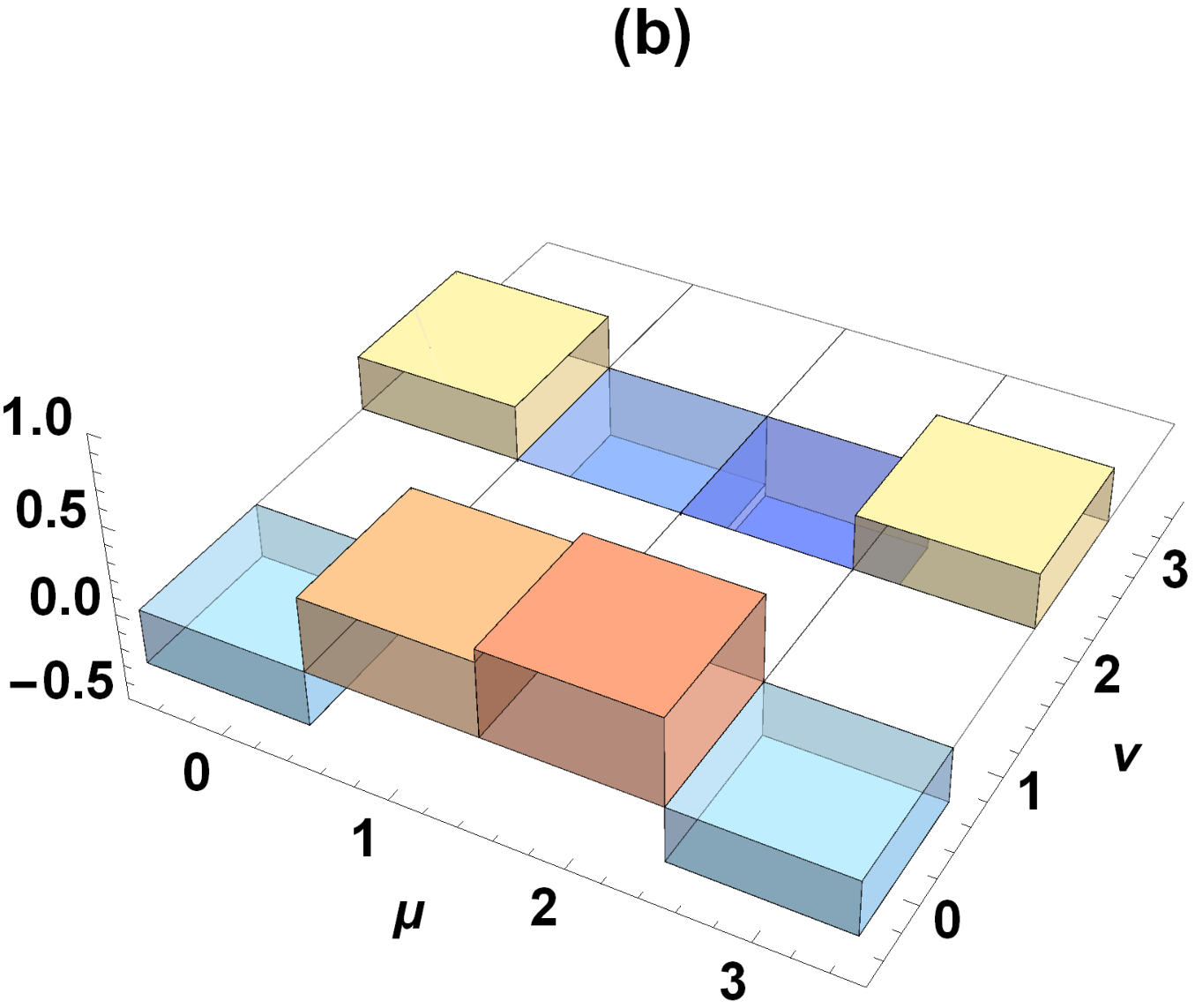}
\end{minipage} \hfill
\begin{minipage}[b]{0.4\linewidth}
\includegraphics[width=\linewidth]{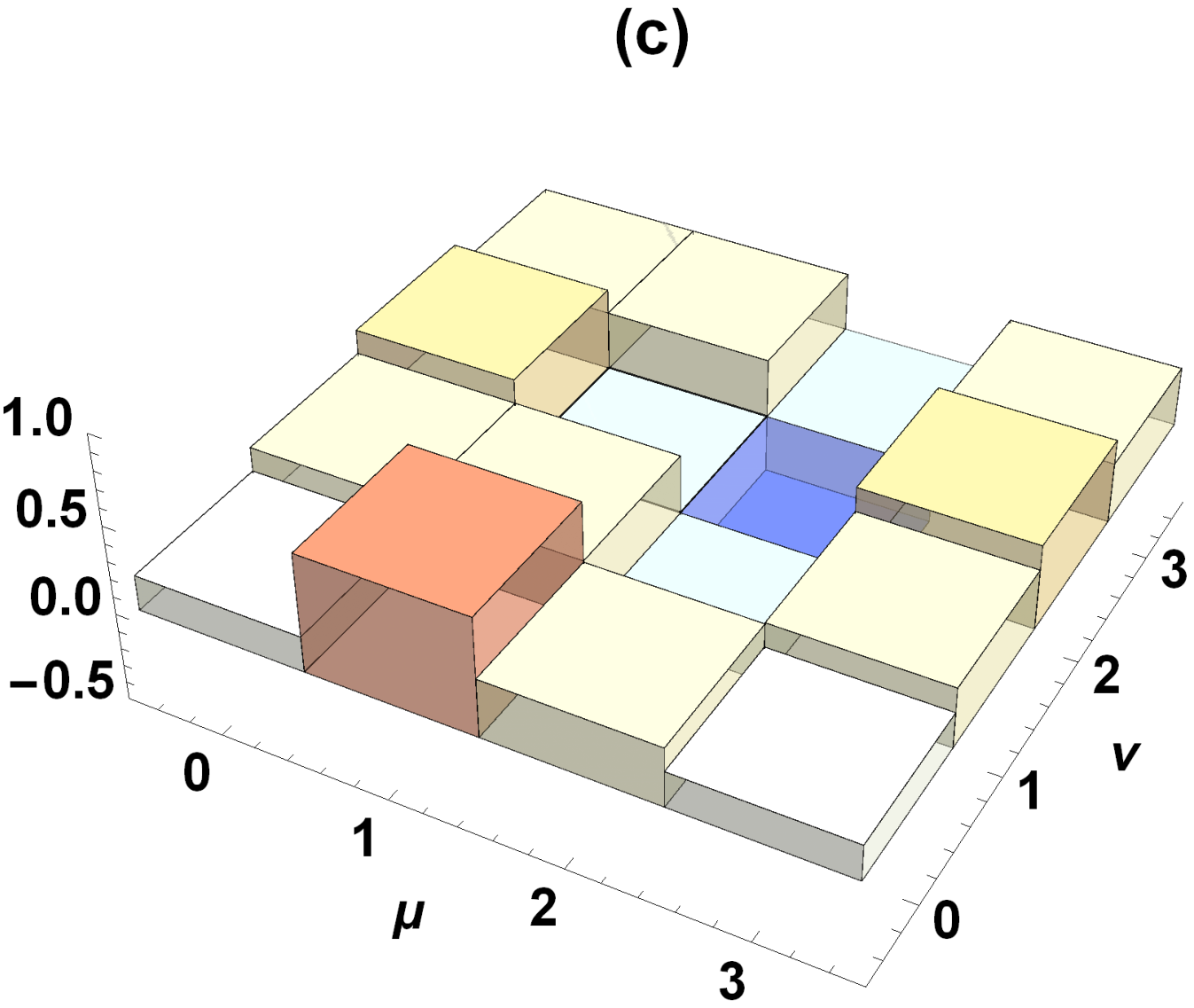}
\end{minipage} \hfill
\begin{minipage}[b]{0.4\linewidth}
\includegraphics[width=\linewidth]{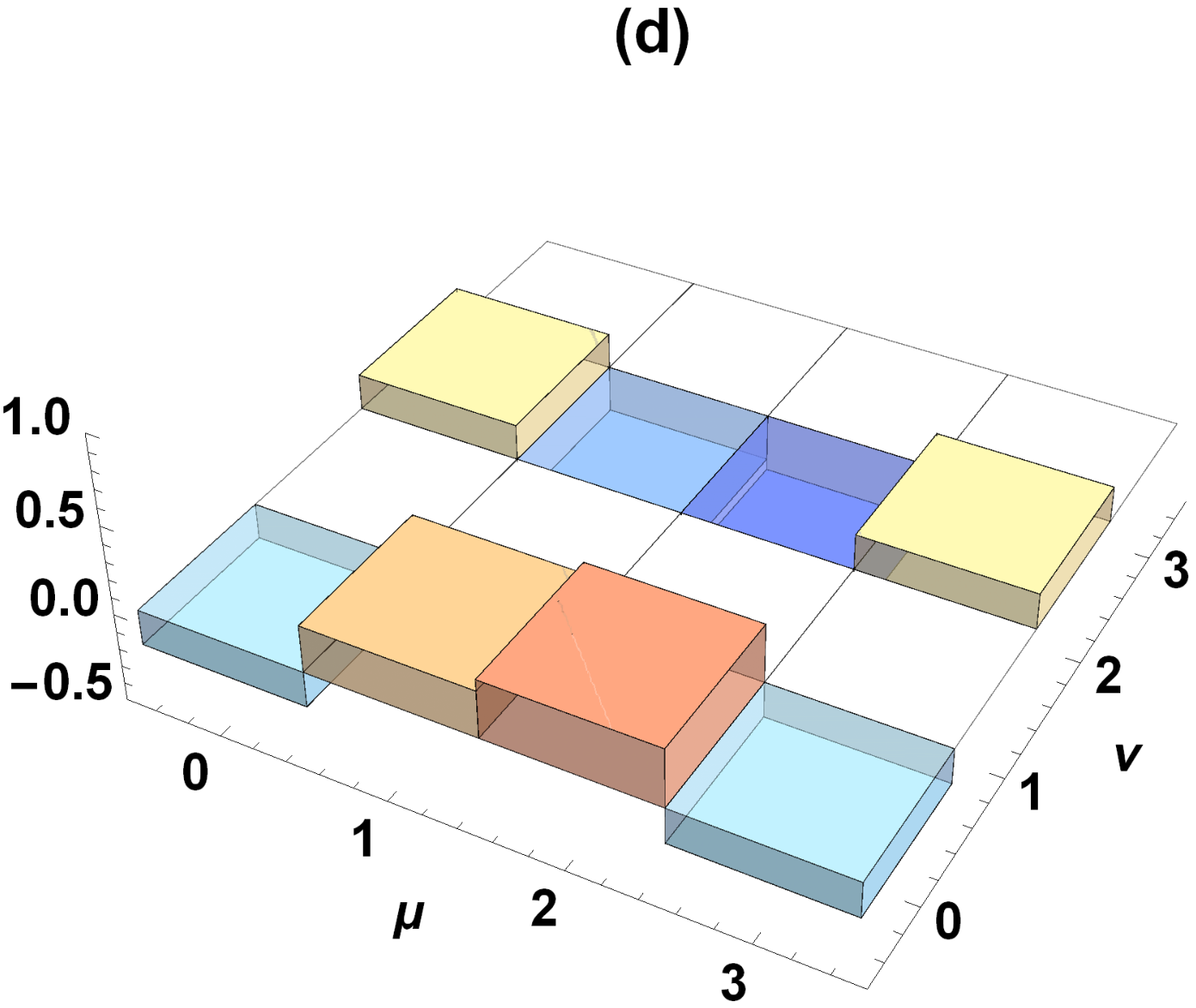}
\end{minipage}
\vspace*{13pt}
\fcaption{\lb{fig4}
Three-dimensional plots of $W_{\mathtt{X}}(\mu,\nu;\gam)$ and $\Delta_{\mathtt{X}}(\mu,\nu;\gam)$ versus $0 \leq \mu,\nu \leq 3$
for two distinct values of $\gam$: (a) $W_{\mathtt{X}}(\mu,\nu;\frac{3}{4})$, (b) $\Delta_{\mathtt{X}}(\mu,\nu;\frac{3}{4})$, (c)
$W_{\mathtt{X}}(\mu,\nu;\half)$, and (d) $\Delta_{\mathtt{X}}(\mu,\nu;\half)$. In both cases, the quantum correlations associated 
with the non-accessed state $| 2 \rg$ are more prominent, as we can see from pictures (b) and (d) for $\mu=2$ fixed and any value
of $\nu$, when one compares with those due to the state $| 1 \rg$ (in this situation, for all $\nu=0,1,2,3$ and $\mu=1$ fixed) --
see Table \ref{tab-X-state} for numerical estimates.}
\end{figure}

\section{Concluding remarks}
\lb{s5}

In this work, we have established an algebraic approach that allows us to describe in a general way both the two two-level
and four-level quantum-mechanical systems through their respective discrete Wigner functions. For this specific task, we have 
employed the connection between $\mathrm{SU(N)}$ generators and Schwinger unitary operators that, in particular, paves the way 
to introduce a genuinely discrete finite-dimensional phase space \cite{MG2019}. So, the discrete Wigner function framework emerged
from this approach is completely general since it allows, among other things, to describe arbitrary two-qubit and ququart states.
Furthermore, experimental researches dealing with NMR techniques (or even dealing with different experimental arrangements) have, 
in our results on discrete Wigner functions, a new solid mathematical tool for searching on entanglement in analogous or even more 
complex systems \cite{NC2019,Gedik,Nagali,Kues,Guo,Resh}. Next, we will discuss on effective gains and future perspectives derived
from this manuscript.

The correspondence between discrete $\mathrm{SU(2)} \otimes \mathrm{SU(2)}$ and $\mathrm{SU(4)}$ Wigner functions not only solved
the difficulty of visualizing general two-qubit states in finite-dimensional discrete phase spaces, but also introduced a new 
mathematical tool that provides qualitative information on the entanglement effects associated with two-qubit X-states 
\cite{Mend-1} through the function\ftn{In particular, for the Bell states $\ro_{\Psi_{\pm}}$ and $\ro_{\Phi_{\pm}}$, the
corresponding functions $\Delta_{\Psi_{\pm}}(\mu,\nu)$ and $\Delta_{\Phi_{\pm}}(\mu,\nu)$ are restricted to the symmetric
intervals
\bd
- \frac{1}{4} \sqrt{2+\sqrt{2}} \, \leq \Delta_{\Psi_{\pm}}(\mu,\nu) \leq \frac{1}{4} \sqrt{2+\sqrt{2}} \quad \mbox{and} \;\,
- \half \sqrt{\frac{2+\sqrt{2}}{2}} \, \leq \Delta_{\Phi_{\pm}}(\mu,\nu) \leq \half \sqrt{\frac{2+\sqrt{2}}{2}} \; ,
\ed
whose expressions are numerically equivalent to $| \Delta_{\Psi_{\pm}}(\mu,\nu) | \leq 0.46$ and $| \Delta_{\Phi_{\pm}}(\mu,\nu) |
\leq 0.65$. Such results will be our guidelines for subsequent comparisons with different entangled X-states.}
\bd
\Delta_{\mathtt{X}}(\mu,\nu) = W_{\mathtt{X}}(\mu,\nu) - Q_{\mathtt{X}}(\mu) R_{\mathtt{X}}(\nu) ,
\ed
where the discrete marginal distribution functions $Q_{\mathtt{X}}(\mu)$ and $R_{\mathtt{X}}(\nu)$ are then responsible for the
main diagonal and antidiagonal matrix elements of $\ro_{\mathtt{X}}$, respectively. It will be quite interesting to apply this 
result in different two-qubit X-states under kinematical and dynamical perspectives: for example, let us initially consider 
certain static states as those introduced by Peres-Horodecki (\texttt{PH}) \cite{Peres1996,HHH1996}
\be
\lb{eq39}
\hat{\varrho}_{\mathtt{PH}} = (1-x) | 0 \rg \lg 0 | + \frac{x}{2} \lpar | 1 \rg \lg 1 | + | 2 \rg \lg 2 | \rpar - \frac{x}{2}
\lpar | 1 \rg \lg 2 | + | 2 \rg \lg 1 | \rpar 
\ee
for $x \in (0,1]$ (it is separable in $x=0$), as well as the Gisin (\texttt{G}) state \cite{Gisin1996}
\brr
\lb{eq40}
& & \hat{\varrho}_{\mathtt{G}} = \half (1-\mathrm{x}) \lpar | 0 \rg \lg 0 | + | 3 \rg \lg 3 | \rpar + \lpar a^{2} - b^{2} + 1/2
\rpar \mathrm{x} \, | 1 \rg \lg 1 | - \lpar a^{2} - b^{2} - 1/2 \rpar \mathrm{x} \, | 2 \rg \lg 2 | \nn \\
& & \qquad - \, a b \mathrm{x} \lpar | 1 \rg \lg 2 | + | 2 \rg \lg 1 | \rpar 
\err
which basically depends on three intrinsic parameters $(a,b,\mathrm{x})$ such that $a > b$ and $\mathrm{x} \in [0,1]$. By varying
the unique internal parameter present in $\ro_{\mathtt{PH}}$, it is easy to verify that $\Delta_{\mathtt{PH}}(\mu,\nu)$ is null
for $x=0$ and it attains its maximum value for $x=1$ when we get $| \Delta_{\mathtt{PH}}(\mu,\nu) | \leq 0.46$ (in such a case,
$\Delta_{\mathtt{PH}} = \Delta_{\Psi_{-}}$). Now, if one considers the Gisin state $\hat{\varrho}_{\mathtt{G}}$ with 
$a^{2} - b^{2} = \frac{\sqrt{2}}{4}$, $ab = \half$, and $\mathrm{x} = 1$ fixed, we obtain  $| \Delta_{\mathtt{G}}(\mu,\nu) | \leq
0.60$, which represents a value close to that reached by the Bell states $\Phi_{\pm}$. Therefore, these results lead us to
establish a hierarchy relation among the static two-qubit X-states through the function $\Delta_{\mathtt{X}}(\mu,\nu)$; in 
addition, it can also be extended to include dynamic states where both the continuous \cite{Braga2010,GB2019} and discrete
\cite{BR2018} time descriptions take place. 

Nowadays, it is well-known that \textsl{fidelity} corresponds to an important concept to quantum information since it provides an
effective measurement of the degree of similarity between two quantum states \cite{Paulo2019}. From the experimental point of 
view, this specific measurement allows to quantify how close the state produced in any experimental apparatus -- once this state 
is limited by imperfections and noise -- stays from that intended one. A theoretical application comes from the entanglement
quantification context, since it measures how close an entangled state is to the set of separable states. Hence, let us consider
the definition of fidelity initially introduced in Ref. \cite{Paulo2008} as
\be
\lb{eq41}
\mathcal{F}_{\mathrm{N}}(\ro,\hat{\sigma}) \col \tr [ \ro \hat{\sigma} ] + \sqrt{1 - \tr [ \ro^{2} ]} \, 
\sqrt{1 - \tr [ \hat{\sigma}^{2} ]}
\ee
and later studied independently in \cite{Horo2009} by the name of super-fidelity, where $\ro$ and $\hat{\sigma}$ represent two
density matrices that belong to $\mathscr{L}_{+,1}(\mathcal{H}_{N})$. With respect to $\mathcal{F}_{\mathrm{N}}(\ro,\hat{\sigma})$
the connection with the discrete Wigner functions $W_{\rho}(\mu,\nu)$ and $W_{\sigma}(\mu,\nu)$ can be promptly established as
follows:
\be
\lb{eq42}
\tr [ \ro \hat{\sigma} ] = \frac{1}{N} + \half \sum_{i=1}^{N^{2}-1} \lg \hat{g}_{i} \rg_{\rho} \lg \hat{g}_{i} \rg_{\sigma} =
\frac{1}{N} \sum_{\mu,\nu=0}^{N-1} W_{\rho}(\mu,\nu) W_{\sigma}(\mu,\nu) \; .
\ee
For $N=4$, the fidelity (\ref{eq41}) perfectly matches with the results established in this paper.

Finally, let us discuss on a possible extension of the mathematical framework exposed here, in order to include the discrete 
Husimi and Glauber-Sudarshan distribution functions \cite{RMG,MRG}. As mentioned in Ref. \cite{MG2019}, the change 
$\hat{G}(\mu,\nu) \rightarrow \hat{T}^{(s)}(\mu,\nu)$ of operator bases in Eq. (\ref{eq4}) permits to include a wide range of
possibilities in what concerns the quasiprobability distribution functions defined over a finite-dimensional discrete phase
space: indeed, for $s=-1,0,+1$ the parametrized function $F^{(s)}(\mu,\nu) = \tr [ \hat{T}^{(s)}(\mu,\nu) \ro ]$ recovers the 
discrete Husimi, Wigner, and Glauber-Sudarshan distribution functions, respectively. To conclude, an interesting scenario of
possible applications for discrete Wigner functions refers to the study of certain spin chains where two-qubit X-states have a 
key role \cite{Wang2006,Yu2017,Park2019}, as well as the study on the separability of multi-qubit states \cite{HHK2019}.

\nonumsection{References}


\appendix

The mathematical prescription adopted to obtain the generators of $\mathrm{SU(4)}$ basically follows that outlined in
\cite{HE1981}, and subsequently adapted to describe finite-dimensional discrete phase spaces by means of the connection between 
these generators and the Schwinger unitary operators \cite{MG2019}. In this case, the computational basis $\{ | 0 \rg, | 1 \rg, | 
2 \rg, | 3 \rg \}$ is made to coincide with $\{ | u_{0} \rg, | u_{1} \rg, | u_{2} \rg, | u_{3} \rg \}$ in a one-to-one 
correspondence, where $\{ | u_{\sigma} \rg \}$ represent the eigenvectors of the unitary operator $\hat{U}$ with eigenvalues 
$\im^{\sigma}$ for $\sigma = 0,\ldots,3$ \cite{GM1996}. Table \ref{tab-ap} exhibits all the generators $\{ \hat{g}_{i} \}_{i=1,
\ldots,15}$ expressed in terms of the transition/projection operators and also as a function of the Schwinger unitary operators. 
For the sake of completeness, we also present below the matrix representations of these generators -- see Ref. \cite{Pfeifer} for 
further details.

\begin{center}
\textbf{Generators for $\mathbf{SU(4)}$}
\end{center}
\bd
\hat{g}_{1} = \lpar \begin{array}{cccc}
0 & 1 & 0 & 0 \\
1 & 0 & 0 & 0 \\
0 & 0 & 0 & 0 \\
0 & 0 & 0 & 0 \\
\end{array} \rpar , \qquad 
\hat{g}_{2} = \lpar \begin{array}{cccc}
0   & -\im & 0 & 0 \\
\im & 0    & 0 & 0 \\
0   & 0    & 0 & 0 \\
0   & 0    & 0 & 0 \\
\end{array} \rpar , 
\ed
\bd
\hat{g}_{3} = \lpar \begin{array}{cccc}
1 & 0  & 0 & 0 \\
0 & -1 & 0 & 0 \\
0 & 0  & 0 & 0 \\
0 & 0  & 0 & 0 \\
\end{array} \rpar , \qquad
\hat{g}_{4} = \lpar \begin{array}{cccc}
0 & 0 & 1 & 0 \\
0 & 0 & 0 & 0 \\
1 & 0 & 0 & 0 \\
0 & 0 & 0 & 0 \\
\end{array} \rpar , 
\ed
\bd
\hat{g}_{5} = \lpar \begin{array}{cccc}
0   & 0 & -\im & 0 \\
0   & 0 & 0    & 0 \\
\im & 0 & 0    & 0 \\
0   & 0 & 0    & 0 \\
\end{array} \rpar , \qquad 
\hat{g}_{6} = \lpar \begin{array}{cccc}
0 & 0 & 0 & 0 \\
0 & 0 & 1 & 0 \\
0 & 1 & 0 & 0 \\
0 & 0 & 0 & 0 \\
\end{array} \rpar , 
\ed
\bd
\hat{g}_{7} = \lpar \begin{array}{cccc}
0 & 0   & 0    & 0 \\
0 & 0   & -\im & 0 \\
0 & \im & 0    & 0 \\
0 & 0   & 0    & 0 \\
\end{array} \rpar , \qquad 
\hat{g}_{8} = \frac{1}{\sqrt{3}} \lpar \begin{array}{cccc}
1 & 0 & 0  & 0 \\
0 & 1 & 0  & 0 \\
0 & 0 & -2 & 0 \\
0 & 0 & 0  & 0 \\
\end{array} \rpar , 
\ed
\bd
\hat{g}_{9} = \lpar \begin{array}{cccc}
0 & 0 & 0 & 1 \\
0 & 0 & 0 & 0 \\
0 & 0 & 0 & 0 \\
1 & 0 & 0 & 0 \\
\end{array} \rpar , \qquad
\hat{g}_{10} = \lpar \begin{array}{cccc}
0   & 0 & 0 & -\im \\
0   & 0 & 0 & 0 \\
0   & 0 & 0 & 0 \\
\im & 0 & 0 & 0 \\
\end{array} \rpar ,  
\ed
\bd
\hat{g}_{11} = \lpar \begin{array}{cccc}
0 & 0 & 0 & 0 \\
0 & 0 & 0 & 1 \\
0 & 0 & 0 & 0 \\
0 & 1 & 0 & 0 \\
\end{array} \rpar , \qquad
\hat{g}_{12} = \lpar \begin{array}{cccc}
0 & 0   & 0 & 0 \\
0 & 0   & 0 & -\im \\
0 & 0   & 0 & 0 \\
0 & \im & 0 & 0 \\
\end{array} \rpar ,
\ed
\bd
\hat{g}_{13} = \lpar \begin{array}{cccc}
0 & 0 & 0 & 0 \\
0 & 0 & 0 & 0 \\
0 & 0 & 0 & 1 \\
0 & 0 & 1 & 0 \\
\end{array} \rpar , \qquad 
\hat{g}_{14} = \lpar \begin{array}{cccc}
0 & 0 & 0   & 0 \\
0 & 0 & 0   & 0 \\
0 & 0 & 0   & -\im \\
0 & 0 & \im & 0 \\
\end{array} \rpar ,
\ed
\be
\lb{su4-generators}
\hat{g}_{15} = \frac{1}{\sqrt{6}} \lpar \begin{array}{cccc}
1 & 0 & 0 & 0 \\
0 & 1 & 0 & 0 \\
0 & 0 & 1 & 0 \\
0 & 0 & 0 & -3 \\
\end{array} \rpar .
\ee
\vspace*{4pt}
\begin{table}[!t]
\tcaption{Generators of $\mathrm{SU(4)}$ in terms of the transition/projection operators and the Schwinger unitary operators. In
particular, the transition operators $\hat{\mathscr{P}}_{\alf,\bet} = | u_{\alf} \rg \lg u_{\bet} |$ with $0 \leq \alf < \bet 
\leq 3$ and the projection operators $\hat{\mathscr{P}}_{\sigma,\sigma} = | u_{\sigma} \rg \lg u_{\sigma} |$ for $0 \leq \sigma
\leq 3$, jointly represent a complete orthonormal operator basis constituted by the elements $\{ \hat{g}_{i} \}_{i=1,\ldots,15}$.
In such a case, $\hat{U}$ and $\hat{V}$ describe a pair of unitary operators defined in a four-dimensional state vector space,
whose respective orthonormal eigenvectors $| u_{\sigma} \rg$ and $| v_{\eps} \rg$ are related through the inner product 
$\lg u_{\sigma} | v_{\eps} \rg = \half \im^{\sigma \eps}$ -- see also Ref. \cite{GM1996}.}
\centerline{\footnotesize\smalllineskip
\begin{tabular}{l c c c}
Generators & Transition/Projection Operators & Schwinger Unitary Operators \\
\hline
$\hat{g}_{1}$ & $\hat{\mathscr{P}}_{0,1} + \hat{\mathscr{P}}_{1,0}$ & $\frac{1}{4} \bigl( \hat{V} + \hat{V}^{3} + \hat{U} \hat{V} 
+ \hat{U}^{2} \hat{V} + \hat{U}^{3} \hat{V} - \im \hat{U} \hat{V}^{3} - \hat{U}^{2} \hat{V}^{3} + \im \hat{U}^{3} \hat{V}^{3} 
\bigr)$ \\
$\hat{g}_{2}$ & $- \im \bigl( \hat{\mathscr{P}}_{0,1} - \hat{\mathscr{P}}_{1,0} \bigr)$ & $- \frac{\im}{4} \bigl( \hat{V} - 
\hat{V}^{3} + \hat{U} \hat{V} + \hat{U}^{2} \hat{V} + \hat{U}^{3} \hat{V} + \im \hat{U} \hat{V}^{3} + \hat{U}^{2} \hat{V}^{3} -
\im \hat{U}^{3} \hat{V}^{3} \bigr)$ \\
$\hat{g}_{3}$ & $\hat{\mathscr{P}}_{0,0} - \hat{\mathscr{P}}_{1,1}$ & $\frac{1}{4} \bigl[ (1+\im) \hat{U} + 2 \hat{U}^{2} + 
(1-\im) \hat{U}^{3} \bigr]$ \\
$\hat{g}_{4}$ & $\hat{\mathscr{P}}_{0,2} + \hat{\mathscr{P}}_{2,0}$ & $\half \bigl( \hat{V}^{2} + \hat{U}^{2} \hat{V}^{2} 
\bigr)$ \\
$\hat{g}_{5}$ & $- \im \bigl( \hat{\mathscr{P}}_{0,2} - \hat{\mathscr{P}}_{2,0} \bigr)$ & $- \frac{\im}{2} \bigl( \hat{U} 
\hat{V}^{2} + \hat{U}^{3} \hat{V}^{2} \bigr)$ \\
$\hat{g}_{6}$ & $\hat{\mathscr{P}}_{1,2} + \hat{\mathscr{P}}_{2,1}$ & $\frac{1}{4} \bigl( \hat{V} + \hat{V}^{3} - \im \hat{U}
\hat{V} - \hat{U}^{2} \hat{V} + \im \hat{U}^{3} \hat{V} - \hat{U} \hat{V}^{3} + \hat{U}^{2} \hat{V}^{3} - \hat{U}^{3} \hat{V}^{3}
\bigr)$ \\
$\hat{g}_{7}$ & $- \im \bigl( \hat{\mathscr{P}}_{1,2} - \hat{\mathscr{P}}_{2,1} \bigr)$ & $- \frac{\im}{4} \bigl( \hat{V} - 
\hat{V}^{3} - \im \hat{U} \hat{V} - \hat{U}^{2} \hat{V} + \im \hat{U}^{3} \hat{V} + \hat{U} \hat{V}^{3} - \hat{U}^{2} \hat{V}^{3}
+ \hat{U}^{3} \hat{V}^{3} \bigr)$ \\
$\hat{g}_{8}$ & $\frac{1}{\sqrt{3}} \bigl( \hat{\mathscr{P}}_{0,0} + \hat{\mathscr{P}}_{1,1} - \hat{\mathscr{P}}_{2,2} \bigr)$ & 
$\frac{1}{4 \sqrt{3}} \bigl[ (3-\im) \hat{U} - 2 \hat{U}^{2} + (3+\im) \hat{U}^{3} \bigr]$ \\
$\hat{g}_{9}$ & $\hat{\mathscr{P}}_{0,3} + \hat{\mathscr{P}}_{3,0}$ & $\frac{1}{4} \bigl( \hat{V} + \hat{V}^{3} + \im \hat{U}
\hat{V} - \hat{U}^{2} \hat{V} - \im \hat{U}^{3} \hat{V} + \hat{U} \hat{V}^{3} + \hat{U}^{2} \hat{V}^{3} + \hat{U}^{3} \hat{V}^{3}
\bigr)$ \\
$\hat{g}_{10}$ & $- \im \bigl( \hat{\mathscr{P}}_{0,3} - \hat{\mathscr{P}}_{3,0} \bigr)$ & $\frac{\im}{4} \bigl( \hat{V} - 
\hat{V}^{3} + \im \hat{U} \hat{V} - \hat{U}^{2} \hat{V} - \im \hat{U}^{3} \hat{V} - \hat{U} \hat{V}^{3} - \hat{U}^{2} \hat{V}^{3}
- \hat{U}^{3} \hat{V}^{3} \bigr)$ \\
$\hat{g}_{11}$ & $\hat{\mathscr{P}}_{1,3} + \hat{\mathscr{P}}_{3,1}$ & $\half \bigl( \hat{V}^{2} - \hat{U}^{2} \hat{V}^{2} 
\bigr)$ \\
$\hat{g}_{12}$ & $- \im \bigl( \hat{\mathscr{P}}_{1,3} - \hat{\mathscr{P}}_{3,1} \bigr)$ & $- \half \bigl( \hat{U} \hat{V}^{2} -
\hat{U}^{3} \hat{V}^{2} \bigr)$ \\
$\hat{g}_{13}$ & $\hat{\mathscr{P}}_{2,3} + \hat{\mathscr{P}}_{3,2}$ & $\frac{1}{4} \bigl( \hat{V} + \hat{V}^{3} - \hat{U} \hat{V}
+ \hat{U}^{2} \hat{V} - \hat{U}^{3} \hat{V} + \im \hat{U} \hat{V}^{3} - \hat{U}^{2} \hat{V}^{3} - \im \hat{U}^{3} \hat{V}^{3}
\bigr)$ \\
$\hat{g}_{14}$ & $- \im \bigl( \hat{\mathscr{P}}_{2,3} - \hat{\mathscr{P}}_{3,2} \bigr)$ & $- \frac{\im}{4} \bigl( \hat{V} -
\hat{V}^{3} - \hat{U} \hat{V} + \hat{U}^{2} \hat{V} - \hat{U}^{3} \hat{V} - \im \hat{U} \hat{V}^{3} + \hat{U}^{2} \hat{V}^{3} +
\im \hat{U}^{3} \hat{V}^{3} \bigr)$ \\
$\hat{g}_{15}$ & $\frac{1}{\sqrt{6}} \bigl( \hat{\mathscr{P}}_{0,0} + \hat{\mathscr{P}}_{1,1} + \hat{\mathscr{P}}_{2,2} -
3 \hat{\mathscr{P}}_{3,3} \bigr)$ & $- \frac{\im}{\sqrt{6}} \bigl( \hat{U} + \im \hat{U}^{2} - \hat{U}^{3} \bigr)$ \\
\hline \\
\end{tabular}}
\lb{tab-ap}
\end{table}

The next step consists in determining the mean values $\lg \hat{g}_{i} \rg \col \tr [ \vro \hat{g}_{i} ]$ for a given density
matrix
\be
\lb{four-level}
\vro = \left( \begin{array}{cccc}
\varrho_{11}        & \varrho_{12}        & \varrho_{13}        & \varrho_{14} \\
\varrho_{12}^{\ast} & \varrho_{22}        & \varrho_{23}        & \varrho_{24} \\
\varrho_{13}^{\ast} & \varrho_{23}^{\ast} & \varrho_{33}        & \varrho_{34} \\
\varrho_{14}^{\ast} & \varrho_{24}^{\ast} & \varrho_{34}^{\ast} & \varrho_{44}
\end{array} \right) \in \mathscr{L}_{+,1}(\mathcal{H}_{4})
\ee
in the computational basis and associated with a single four-level quantum system \cite{Gedik}. Here, we do not enter in technical
details w.r.t. the calculations involved in the mean values, since only the final results are necessary -- see the list below. In
fact, these results correspond to the components of the generalized Bloch vector $\mathbf{g} = \lpar \lg \hat{g}_{1} \rg,\ldots,
\lg \hat{g}_{15} \rg \rpar \in \mathbb{R}^{15}$.

\begin{center}
\textbf{Mean Values}
\end{center}
\brr
\lb{valores-medios}
& & \lg \hat{g}_{1} \rg = 2 \, \re \lpar \varrho_{12} \rpar , \;\; \lg \hat{g}_{2} \rg = - 2 \, \ima \lpar \varrho_{12} \rpar , 
\nn \\
& & \lg \hat{g}_{3} \rg = \varrho_{11} - \varrho_{22} , \;\; \lg \hat{g}_{4} \rg = 2 \, \re \lpar \varrho_{13} \rpar , \nn \\
& & \lg \hat{g}_{5} \rg = - 2 \, \ima \lpar \varrho_{13} \rpar , \;\; \lg \hat{g}_{6} \rg = 2 \, \re \lpar \varrho_{23} \rpar , 
\nn \\
& & \lg \hat{g}_{7} \rg = - 2 \, \ima \lpar \varrho_{23} \rpar , \;\; \lg \hat{g}_{8} \rg = \frac{1}{\sqrt{3}} \lpar \varrho_{11} 
+ \varrho_{22} - 2 \varrho_{33} \rpar , \nn \\
& & \lg \hat{g}_{9} \rg = 2 \, \re \lpar \varrho_{14} \rpar , \;\; \lg \hat{g}_{10} \rg = - 2 \, \ima \lpar \varrho_{14} \rpar , 
\nn \\
& & \lg \hat{g}_{11} \rg = 2 \, \re \lpar \varrho_{24} \rpar , \;\; \lg \hat{g}_{12} \rg = - 2 \, \ima \lpar \varrho_{24} \rpar , 
\nn \\
& & \lg \hat{g}_{13} \rg = 2 \, \re \lpar \varrho_{34} \rpar , \;\; \lg \hat{g}_{14} \rg = - 2 \, \ima \lpar \varrho_{34} \rpar , 
\nn \\
& & \lg \hat{g}_{15} \rg = \frac{1}{\sqrt{6}} \lpar \varrho_{11} + \varrho_{22} + \varrho_{33} - 3 \varrho_{44} \rpar .
\err

Finally, let us determine the representatives in the finite-dimensional discrete phase space of all these generators through the
expression $\lpar \hat{g}_{i} \rpar \! (\mu,\nu) = \tr [ \hat{G}^{\dagger}(\mu,\nu) \hat{g}_{i} ]$. For such a task, we adopt the
theoretical framework described in Ref. \cite{MG2019} for discrete $\mathrm{SU(N)}$ Wigner function, as well as the results 
obtained in Table \ref{tab-ap} which describe the $\mathrm{SU(4)}$ generators in terms of the Schwinger unitary operators. This
important link leads us to establish the expressions below for the aforementioned representatives.\ftn{Note that $\lpar 
\hat{g}_{5} \rpar \! (\mu,\nu)$ and $\lpar \hat{g}_{12} \rpar \! (\mu,\nu)$ for $0 \leq \mu,\nu \leq 3$ do not present any
contribution in the calculation of the discrete Wigner function (\ref{eq24}).}

\begin{center}
\textbf{Mapped expressions of the $\mathbf{SU(4)}$ generators}
\end{center}
%
\bd
\lpar \hat{g}_{1} \rpar \! (\mu,\nu) = \half \cos \lpar \frac{\nu \pi}{2} \rpar \frac{\sin \lbk \lpar \mu - \frac{1}{2} \rpar
\pi \rbk}{\sin \lbk \lpar \mu - \frac{1}{2} \rpar \frac{\pi}{4} \rbk} \, ,
\ed
\bd
\lpar \hat{g}_{2} \rpar \! (\mu,\nu) = \half \sin \lpar \frac{\nu \pi}{2} \rpar \frac{\sin \lbk \lpar \mu - \frac{1}{2} \rpar
\pi \rbk}{\sin \lbk \lpar \mu - \frac{1}{2} \rpar \frac{\pi}{4} \rbk} \, ,
\ed
\bd
\lpar \hat{g}_{3} \rpar \! (\mu,\nu) = \delta_{\mu,0}^{[4]} - \delta_{\mu,1}^{[4]} \, , 
\ed
\bd
\lpar \hat{g}_{4} \rpar \! (\mu,\nu) = 2 \cos \lpar \nu \pi \rpar \delta_{\mu,1}^{[4]} \, , \;\; \lpar \hat{g}_{5} \rpar \! 
(\mu,\nu) = 2 \sin \lpar \nu \pi \rpar \delta_{\mu,1}^{[4]} \, ,
\ed
\bd
\lpar \hat{g}_{6} \rpar \! (\mu,\nu) = \half \cos \lpar \frac{\nu \pi}{2} \rpar \frac{\sin \lbk \lpar \mu - \frac{3}{2} \rpar
\pi \rbk}{\sin \lbk \lpar \mu - \frac{3}{2} \rpar \frac{\pi}{4} \rbk} \, ,
\ed
\bd
\lpar \hat{g}_{7} \rpar \! (\mu,\nu) = \half \sin \lpar \frac{\nu \pi}{2} \rpar \frac{\sin \lbk \lpar \mu - \frac{3}{2} \rpar
\pi \rbk}{\sin \lbk \lpar \mu - \frac{3}{2} \rpar \frac{\pi}{4} \rbk} \, ,
\ed
\bd
\lpar \hat{g}_{8} \rpar \! (\mu,\nu) = \frac{1}{\sqrt{3}} \lpar \delta_{\mu,0}^{[4]} + \delta_{\mu,1}^{[4]} - 2 
\delta_{\mu,2}^{[4]} \rpar \, ,
\ed
\bd
\lpar \hat{g}_{9} \rpar \! (\mu,\nu) = \half (-1)^{\nu} \cos \lpar \frac{\nu \pi}{2} \rpar \frac{\sin \lbk \lpar \mu - 
\frac{3}{2} \rpar \pi \rbk}{\sin \lbk \lpar \mu - \frac{3}{2} \rpar \frac{\pi}{4} \rbk} \, ,
\ed
\bd
\lpar \hat{g}_{10} \rpar \! (\mu,\nu) = \half (-1)^{\nu} \sin \lpar \frac{\nu \pi}{2} \rpar \frac{\sin \lbk \lpar \mu - 
\frac{3}{2} \rpar \pi \rbk}{\sin \lbk \lpar \mu - \frac{3}{2} \rpar \frac{\pi}{4} \rbk} \, ,
\ed
\bd
\lpar \hat{g}_{11} \rpar \! (\mu,\nu) = 2 \cos \lpar \nu \pi \rpar \delta_{\mu,2}^{[4]} \, , \;\; \lpar \hat{g}_{12} \rpar \!
(\mu,\nu) = 2 \sin \lpar \nu \pi \rpar \delta_{\mu,2}^{[4]} \, ,
\ed
\bd
\lpar \hat{g}_{13} \rpar \! (\mu,\nu) = \half \cos \lpar \frac{\nu \pi}{2} \rpar \frac{\sin \lbk \lpar \mu - \frac{5}{2} \rpar
\pi \rbk}{\sin \lbk \lpar \mu - \frac{5}{2} \rpar \frac{\pi}{4} \rbk} \, ,
\ed
\bd
\lpar \hat{g}_{14} \rpar \! (\mu,\nu) = \half \sin \lpar \frac{\nu \pi}{2} \rpar \frac{\sin \lbk \lpar \mu - \frac{5}{2} \rpar
\pi \rbk}{\sin \lbk \lpar \mu - \frac{5}{2} \rpar \frac{\pi}{4} \rbk} \, ,
\ed
\be
\lb{mapa-geradores}
\lpar \hat{g}_{15} \rpar \! (\mu,\nu) = \frac{1}{\sqrt{6}} \lpar \delta_{\mu,0}^{[4]} + \delta_{\mu,1}^{[4]} + 
\delta_{\mu,2}^{[4]} - 3 \delta_{\mu,3}^{[4]} \rpar \, .
\ee

To conclude, let us briefly mention that $\{ \lg \hat{g}_{i} \rg \}$ and $\{ \lpar \hat{g}_{i} \rpar \! (\mu,\nu) \}$ for
$i=1,\ldots,15$ provide completely general expressions for the discrete $\mathrm{SU(4)}$ Wigner function, which allows us to deal
with at least four-level systems and their intrinsic quantum properties from a different perspective.

\end{document}